\def\Mode{2} 
\def\InsertAll{0} 
\def\IRAS{{\it IRAS }}			
\def\um{$\mu$m}		        
\def\pwr #1 {$10^{#1}$}			
\def\pwrx #1 #2 {$#1 \times$\pwr #2 }	
\def\EA{et al}			
\def\arcm{$^{\prime\ \ }$}				
\def\deg{$^{\circ}$}				

\def\13CO{$^{\rm 13}$CO}
\def\12CO{$^{\rm 12}$CO}

\def\gtapp
{\mathrel{\hbox{\raise0.3ex\hbox{$>$}\kern-0.8em\lower0.8ex\hbox{$\sim$}}}}
\def\ltapp
{\mathrel{\hbox{\raise0.3ex\hbox{$<$}\kern-0.75em\lower0.8ex\hbox{$\sim$}}}}
\def\app{$\approx$}

\def\IRAS{{\it IRAS }}			
\def\um{$\mu$m}		        
\def\pwr #1 {$10^{#1}$}			
\def\pwrx #1 #2 {$#1 \times$\pwr #2 }	
\def\arcm{$^{\prime\ \ }$}				
\def\deg{$^{\circ}$}				

\def\BGS{BGS$_1+$BGS$_2$}
\ifnum\Mode=0
\documentclass[manuscript]{aastex}
\fi
\ifnum\Mode=1
\documentclass[preprint]{aastex}
\fi
\ifnum\Mode=2
\documentclass[preprint2]{aastex}
\fi
\received{}
\accepted{}
\journalid{}{}
\articleid{}{}
\slugcomment{\small Accepted 12 May 2003 for publication in the {\it Astronomical Journal} (August issue).}

\lefthead{Sanders et al.}
\righthead{The \IRAS Revised Bright Galaxy Sample (RBGS)}

\begin{document}



\title{{\small Accepted 12 May 2003 for publication in the {\it Astronomical Journal}, scheduled for August 2003 issue.}\\
       \vskip 0.3truein
       The \IRAS Revised Bright Galaxy Sample (RBGS)
} 

\author{D. B. Sanders,\altaffilmark{1,2} J. M. Mazzarella,\altaffilmark{3} 
D.-C. Kim,\altaffilmark{3,4} J. A. Surace,\altaffilmark{5} B. T. Soifer\altaffilmark{5,6}}

\altaffiltext{1}{Institute for Astronomy, University of Hawaii, 2680
Woodlawn Drive, Honolulu, HI 96822;\ E-mail: sanders@ifa.hawaii.edu}

\altaffiltext{2}{Max-Planck Institut fur Extraterrestrische Physik, D-85740, 
Garching, Germany}

\altaffiltext{3}{IPAC, MS 100-22, California Institute of Technology, Jet
Propulsion Laboratory, Pasadena, CA 91125;\ Email: mazz@ipac.caltech.edu}

\altaffiltext{4}{Current Address:\ School of Earth and Environmental Sciences (BK21), 
Seoul National University, Seoul, Korea; Email: dckim@astro.snu.ac.kr}

\altaffiltext{5}{SIRTF Science Center, MS 314-6, California Institute of Technology, Pasadena,
CA 91125;\  Email: jason@ipac.caltech.edu, bts@ipac.caltech.edu}

\altaffiltext{6}{Division of Physics, Math and Astronomy, Downes Lab, MS 320-47, California
Institute of Technology, Pasadena, CA 91125;\ Email: bts@mop.caltech.edu}

\def\ts{\thinspace}

\begin{abstract}
{\small

\IRAS flux densities, redshifts, and infrared luminosities are reported for all
sources identified in the \IRAS Revised Bright Galaxy Sample (RBGS), a complete
flux-limited survey of all extragalactic objects with total 60{\ts}\um\
flux density greater than 5.24{\ts}Jy, covering the entire sky surveyed
by \IRAS at Galactic latitude $\vert b \vert > 5^\circ$.  The RBGS includes
629 objects, with a median (mean) sample redshift of 0.0082 (0.0126) and a
maximum redshift of 0.0876.  The RBGS supersedes the previous two-part \IRAS
Bright Galaxy Samples (hereafter \BGS), which were compiled before the final
(\lq\lq Pass 3\rq\rq) calibration of the \IRAS Level{\ts}1 Archive in May 1990.
The RBGS also makes use of more accurate and consistent automated methods
to measure the flux of objects with extended emission.  The RBGS contains
39 objects which were not present in the \BGS, and 28 objects from the \BGS\
have been dropped from RBGS because their revised 60{\ts}\um\ flux densities
are not greater than 5.24{\ts}Jy.  Comparison of revised flux measurements
for sources in both surveys shows that most flux differences are in the
range $\sim 5-25$\%, although some faint sources at 12{\ts}\um\ and 25{\ts}\um\
differ by as much as a factor of 2. Basic properties of the RBGS sources
are summarized, including estimated total infrared luminosities, as well as
updates to cross-identifications with sources from optical galaxy catalogs
established using the NASA/IPAC Extragalactic Database (NED).  In addition,
an atlas of images from the Digitized Sky Survey with overlays of the \IRAS
position uncertainty ellipse and annotated scale bars is provided for ease in
visualizing the optical morphology in context with the angular and metric size
of each object.  The revised bolometric infrared luminosity function, $\phi
(L_{\rm ir})$, for infrared bright galaxies in the local Universe remains best
fit by a double power law, $\phi (L) \propto L^\alpha$, with $\alpha = -0.6\
(\pm 0.1), {\rm and}\ \alpha = -2.2\ (\pm 0.1)$ below and above the \lq\lq
characteristic\rq\rq\ infrared luminosity $L_{\rm ir}^\ast \sim 10^{10.5}
L_\odot$, respectively.  A companion paper provides \IRAS High
Resolution (HIRES) processing of over 100 RBGS sources where improved
spatial resolution often provides better \IRAS source
positions or allows for deconvolution of close galaxy pairs.

}
\end{abstract}
\keywords{galaxies:{\ts}general --- infrared:{\ts}general --- infrared:sources}

\def\figcapOne{
\footnotesize 
DSS1 images of each RBGS object. Horizontal bars on the bottom
of each plot show the angular scale labeled in arcminutes, and vertical
bars on the lower right side of each plot show the metric scale labeled in
kiloparsecs. The ellipses represent 3-sigma uncertainties in the \IRAS positions.
(Note: For the LMC and SMC, no ellipse is shown because there is no
FSC or PSC source corresponding to the ``center'' of these large, diffuse galaxies.)
Pairs and groups for which improved 2-D spatial resolution has been attempted
using HIRES processing of the IRAS data (Surace, Sanders \& Mazzarella 2003) are indicated by
``(H)'' following the object name.
{\bf NOTE: Due to astro-ph size limits, only 1 of 26 pages in Figure 1 is included
here, and it is degraded by bit-mapping. A complete, full resolution version of this 
figure will be published in the
{\it Astronomical Journal} and is also available at
{\it http://nedwww.ipac.caltech.edu/level5/March03/IRAS\_RBGS/Figures/} (Fig1p*.ps.gz).}}
\def\figcapTwo{
\footnotesize 
The ratio of new total flux density
measurements to original estimates versus the base 10 log of the new total flux density
at 12{\ts}\um, 25{\ts}\um, 60{\ts}\um\ and 100{\ts}\um.
}
\def\figcapThree{
\footnotesize 
Ratio of the best estimate of the total flux density to the peak value in the
coadded scan vs the base 10 log of the new total flux density at 12{\ts}\um, 25{\ts}\um, 60{\ts}\um\ and 100{\ts}\um. 
Only a few objects with very large ratios are outside the plot region; the selected
range in the ratio was chosen to show details for the largest number of points.
}
\def\figcapFour{
\footnotesize 
Integral and differential number counts 
vs. flux density (${\rm log}N - {\rm log}S$) for objects in the RBGS.
Nineteen objects with upper limits are not included in the 12{\ts}\um\ plot.
The bins have width $d({\rm log} S) = 0.1$, and the thin vertical 
bars represent statistical uncertainties (i.e., $\sqrt N$).
}
\def\figcapFive{
\footnotesize 
Aitoff projection in Galactic coordinates
of all sources in the RBGS. The dark shaded region represents the Galactic
plane region excluded from the survey ($\vert b\vert < 5^\circ$).  The three small lightly
shaded regions denote areas with extensive
contamination from nearby Galactic molecular clouds and the LMC (see text
for detailed boundaries). The total survey area is 37,658 deg$^2$, or 91.3\%
of the sky. {\tiny \bf NOTE: This figure is degraded by bit-mapping to meet the astro-ph 
size limit. Full resolution version available at 
{\it http://nedwww.ipac.caltech.edu/level5/March03/IRAS\_RBGS/Figures/hammer.ps.gz}.}
}

\def\figcapSix{
\footnotesize 
Surface density of RBGS objects versus
Galactic latitude ($b$). The shaded region denotes the omitted survey region
of $\pm 5^\circ$ around the Galactic plane.  The line indicates the median value of
log$_{10}$(\# source/deg$^2$)~$= -1.80$, or 0.016 sources per square degree for the whole
sample. This plot illustrates the well-known excess of \IRAS galaxies in
the northern compared to the southern Galactic hemisphere, due primarily
to the Virgo cluster centered near $b = +74^\circ$ as well as the local supercluster .
}
\def\figcapSeven{
\footnotesize 
Distributions of total \IRAS flux density ratios:
(a) $S_\nu (12\mu m) / S_\nu (25\mu m)$, 
(b) $S_\nu (12\mu m) / S_\nu (60\mu m)$, 
(c) $S_\nu (25\mu m) / S_\nu (60\mu m)$, and 
(d) $S_\nu (60\mu m) / S_\nu (100\mu m)$.
}
\def\figcapEight{
\footnotesize 
\IRAS flux density ratio correlations using the revised total flux estimates in Table 1.
}
\def\figcapNine{
\footnotesize 
Distribution of heliocentric radial velocities ($c*z$) for the RBGS.
}
\def\figcapTen{
\footnotesize 
Distribution of estimated distances (Mpc) for sources in the RBGS.
}
\def\figcapEleven{
\footnotesize 
Distribution of the base ten logarithm of the total infrared luminosity in Solar units.
}
\def\figcapTwelve{
\footnotesize 
The infrared luminosity function for the RBGS, computed using the $1/V_{\rm max}$ method.
The space densities and uncertainties plotted are those listed in Table 6;
the points represent the center of each luminosity bin, 
and each bin has a uniform width of 0.5
in units of log{\ts}($L_{\rm ir}/L_{\odot}$).
The solid lines are linear least-square fits to the data points below and above
the \lq\lq characteristic\rq\rq\ infrared luminosity $L_{\rm ir}^\ast \sim
10^{10.5} L_\odot$, respectively. The corresponding power laws,
$\phi (L) \propto L^\alpha$, have $\alpha = -0.6\ (\pm 0.1)\ {\rm and}\ \alpha = -2.2\ (\pm 0.1)$.
}
\def\figcapThirteen{
\footnotesize 
Histogram illustrating the distribution of the ratio
$f_{\nu}(t)/f_{\nu}(template)$ at 60{\ts}\um. The entire range of this parameter
is not shown in order to highlight details of the distribution for objects
that have a ratio near unity.  The solid line is a Gaussian fit to the upper
envelop of the galaxy distribution, constructed by rejecting objects outside
the range 0.95 -- 1.05 from the fit. The asterisks show counts in the same
bins for a sample of 121 stars selected from the \IRAS Point Source Catalog
(PSC) with $f_{\nu}(60\mu m) > 5.24~Jy$ and positional associations with objects
in various star catalogs; cross-identification of these \IRAS sources with
stars was confirmed using SIMBAD. The dotted line is a Gaussian fit to the
distribution for the stars, again omitting objects outside the range 0.95 -
1.05 from the fit.  The dashed line is a Gaussian distribution with the same
mean and standard deviation as the fit to the star distribution (dotted line), but
scaled to the peak of the galaxy distribution (solid line).
Horizontal line segments have been drawn in the bins with 
$f_{\nu}(t)/f_{\nu}(template)$ values between 1.0 and 1.10; these represent a
``reflection'' of the bins with $f_{\nu}(t)/f_{\nu}(template)$ values between 0.90 and 1.00
for the RBGS objects, indicating the expected counts if differences between
$f_{\nu}(t)$ and $f_{\nu}(template)$ were due only to noise and not real excess,
extended emission detected in the $f_{\nu}(t)$ measurement.
The small plots inset around the top and right sides of the figure
illustrate representative coadded 60{\ts}\um\ scan profiles from SCANPI; the object
name and corresponding value of $f_{\nu}(t)/f_{\nu}(template)$ at 60{\ts}\um\ 
is listed above each plot. The larger inset plot on the left illustrates
how dips in the background noise cause $f_{\nu}(t)$ to be less than the
template fit (point spread function) value for some objects; in such cases,
$f_{\nu}(template)$ was chosen over $f_{\nu}(t)$.
This diagram was used to establish the threshold requirement of
$f_{\nu}(t)/f_{\nu}(template) > 1.05$ for selecting $f_{\nu}(t)$ as a
confident measurement of at least marginally extended emission in excess 
of the value measured by the point source template fit.
}
\def\figcapFourteen{
\footnotesize 
Coadded \IRAS scan profiles that illustrate source size codes
listed in  columns (8) -- (11) of Table 1 (the ``S" in ``SMF").
Panel (a) R - resolved source NGC{\ts}1961.
Panel (b) M - marginally resolved source (called ``U+" in \BGS) NGC{\ts}625.
Panel (c) U - unresolved source NGC{\ts}34.
In this figure, as well as in Figs. 15  and 16, the solid points are the median
coadded \IRAS scan data (SCANPI coadd method 1002),
the dashed lines are the baseline fits to
the background noise, and the solid lines are the point source
template fits (point-spread function). The vertical bars below the
fitted baseline show the integration range used for the total
flux density estimation, $f_\nu (t)$, which is measured within a defined region;
the default SCANPI ranges for $f_\nu (t)$ were used, which are
$\pm 2, \pm 2, \pm 2.5$ and $\pm 4$ arcminutes at 12{\ts}\um, 25{\ts}\um, 60{\ts}\um\ and 100{\ts}\um.
}
\def\figcapFifteen{
\footnotesize 
Coadded \IRAS scan profiles that illustrate flux density
estimator methods listed as codes in columns (8) -- (11) of Table 1 (the ``M" in
``SMF"). Panel (a): Z - total flux density estimated from integration of the 
averaged scan between the zero crossings; called ``$f_\nu (z)$" in SCANPI output.
Panel (b): I - total flux density estimated from integration of the 
coadded scan between fixed points defining an integration range; 
called ``$f_\nu (t)$" in SCANPI output. Note the slightly raised point source
template fit, which is another indicator of small, but statistically
significant excess emission compared to the case of a pure point source.
The default SCANPI ranges for $f_\nu (t)$ were used, which are
$\pm 2, \pm 2, \pm 2.5$ and $\pm 4$ arcminutes at 12{\ts}\um, 25{\ts}\um, 60{\ts}\um\ and 100{\ts}\um.
Panel (c): T - flux density estimated from the best-fitting 
point source template; called 'template amplitude' in the SCANPI output.
This is the method most often selected by the software for unresolved
sources with no confusion. Note that the point source template fit
is not raised with respect to the baseline fit of the background
emission; this is a visual confirmation of the result found
by the automated method selection---  within the statistical noise,
$f_\nu (t)$ is not larger than the template fit.
Panel (d): P - the maximum flux density measured within the signal range;
called ``peak" in SCANPI output. This method was used when emission from
a nearby source or cirrus confused the template fit 
and contaminated the $f_\nu (t)$ and $f_\nu (z)$ methods.
Panel (e): S - result of SCLEAN point source subtraction (see Appendix); 
this method was used to estimate the flux density of components of
some pairs, for comparison with generally more reliable HIRES results 
(Surace, Sanders \& Mazzarella 2003).  The first four codes are referred to
as ``peak", ``temp", ``tot" and ``zc", respectively, in the header of Table 7.
Another value of the measurement method code listed in Table 1 is ``R"; 
this indicates that the total flux estimate
from Rice (1993) or Rice et al. (1988) was used because the SCANPI 1-D scan
coaddition method does not work well for objects larger than $\sim25$ arcminutes. 
See text for details.
}
\def\figcapSixteen{
\footnotesize 
Coadded \IRAS scan profiles that illustrate uncertainty codes 
listed in columns (8) -- (11) of Table 1 (the ``F" in ``SMF"). These codes
identify the origin of large uncertainty flagged generally by a colon (``:")
following the associated flux density measurement in Table 1.
Panel (a): g - a nearby {\bf companion galaxy} influenced the choice of flux
estimator. Panel (b): b - emission from two or more galaxies is {\bf blended};
the components are unresolved by \IRAS at the indicated wavelength.
Panel (c): c - prominent Galactic {\bf cirrus} taints the measurement.
Panel (d): n - excessive {\bf noise} or source confusion prevented a reliable
flux density estimate.
}

\def\tableTwo{
\begin{deluxetable}{lll}
\tabletypesize{\scriptsize} 
\tablenum{2}
\tablewidth{0pt}
\tablecaption{${\rm BGS_1+BGS_2}$ Objects Omitted Due To Revised
$S_\nu(60\mu m) \leq 5.24$ Jy}
\tablehead{
\colhead{Name} & 
\colhead{Revised $\rm S_\nu(60\mu m)~Jy$} &
\colhead{~~~Status}
}
\startdata

NGC 0578               & ~~~$4.64\pm0.04$ & ~~~${\rm BGS_1~had~S_\nu(60\mu m)=4.64~Jy}$\tablenotemark{a}\\
NGC 1012               & ~~~$5.19\pm0.04$ & ~~~${\rm BGS_2~had~S_\nu(60\mu m)=5.37~Jy}$ \\
NGC 1141/2             & ~~~$5.06\pm0.02$ & ~~~${\rm BGS_1~had~S_\nu(60\mu m)=5.06~Jy}$\tablenotemark{a}\\
UGC 2789               & ~~~$5.23\pm0.03$ & ~~~${\rm BGS_2~had~S_\nu(60\mu m)=5.24~Jy}$ \\
ESO 484-G036           & ~~~$5.10\pm0.04$ & ~~~${\rm BGS_1~had~S_\nu(60\mu m)=5.44~Jy}$; named~IR 0433-25 in ${\rm BGS_1}$ \\
VII Zw 019             & ~~~$5.05\pm0.03$ & ~~~${\rm BGS_2~had~S_\nu(60\mu m)=5.35~Jy}$ \\
IRAS F00555+7614       & ~~~$5.14\pm0.05$ & ~~~${\rm BGS_2~had~S_\nu(60\mu m)=5.32~Jy}$; named IR 00555+7614 in ${\rm BGS_2}$ \\
IRAS F06035-7102       & ~~~$5.18\pm0.03$ & ~~~${\rm BGS_2~had~S_\nu(60\mu m)=5.34~Jy}$; named IR 06035-7102 in ${\rm BGS_2}$ \\
NGC 2820               & ~~~$4.34\pm0.03$ & ~~~${\rm BGS_1~had~S_\nu(60\mu m)=4.23~Jy}$\tablenotemark{a} \\
NGC 2990               & ~~~$5.20\pm0.03$ & ~~~${\rm BGS_1~had~S_\nu(60\mu m)=5.49~Jy}$ \\
NGC 3353               & ~~~$5.17\pm0.05$ & ~~~${\rm BGS_1~had~S_\nu(60\mu m)=5.54~Jy}$ \\
ESO 174-G005           & ~~~$5.23\pm0.13$ & ~~~${\rm BGS_2~had~S_\nu(60\mu m)=5.29~Jy}$ \\
NGC 4395               & ~~~$4.69\pm0.06$ & ~~~${\rm BGS_1~had~S_\nu(60\mu m)=4.21~Jy}$\tablenotemark{a} \\
NGC 4438               & ~~~$3.87\pm0.03$ & ~~~${\rm BGS_1~had~S_\nu(60\mu m)=3.86~Jy}$\tablenotemark{a} \\
NGC 4594               & ~~~$4.28\pm0.03$ & ~~~${\rm BGS_1~had~S_\nu(60\mu m)=3.98~Jy}$\tablenotemark{a} \\
NGC 4618               & ~~~$5.13\pm0.03$ & ~~~${\rm BGS_1~had~S_\nu(60\mu m)=4.92~Jy}$\tablenotemark{a} \\
MCG +08-23-097         & ~~~$4.91\pm0.04$ & ~~~${\rm BGS_1~had~S_\nu(60\mu m)=4.79~Jy}$\tablenotemark{a} \\
ESO 383-G027           & ~~~$5.16\pm0.05$ & ~~~${\rm BGS_2~had~S_\nu(60\mu m)=5.31~Jy}$ \\
UGC 09668              & ~~~$5.24\pm0.03$ & ~~~${\rm BGS_1~had~S_\nu(60\mu m)=5.70~Jy}$ \\
IRAS F15335-0513       & ~~~$5.16\pm0.03$ & ~~~${\rm BGS_1~had~S_\nu(60\mu m)=5.25~Jy}$; named IR 1533-05 in ${\rm BGS_1}$\\
IRAS 18587-1653        & ~~~$5.20\pm0.12$ & ~~~${\rm BGS_2~had~S_\nu(60\mu m)=5.51~Jy}$ \\
AM 1925-724            & ~~~$5.16\pm0.03$ & ~~~${\rm BGS_2~had~S_\nu(60\mu m)=5.30~Jy}$ \\
IRAS 19279+3534        & ~~~$5.22\pm0.09$ & ~~~${\rm BGS_2~had~S_\nu(60\mu m)=5.43~Jy}$ \\
IRAS 20100-4156        & ~~~$5.19\pm0.04$ & ~~~${\rm BGS_2~had~S_\nu(60\mu m)=5.44~Jy}$ \\
IC 4946                & ~~~$5.24\pm0.03$ & ~~~${\rm BGS_2~had~S_\nu(60\mu m)=5.31~Jy}$; named A2020-44 in ${\rm BGS_2}$\\
NGC 6015               & ~~~$4.40\pm0.03$ & ~~~${\rm BGS_1~had~S_\nu(60\mu m)=4.42~Jy}$\tablenotemark{a}\\
NGC 6070               & ~~~$5.10\pm0.03$ & ~~~${\rm BGS_1~had~S_\nu(60\mu m)=5.07~Jy}$\tablenotemark{a}\\
NGC 7673               & ~~~$4.98\pm0.05$ & ~~~${\rm BGS_1~had~S_\nu(60\mu m)=4.98~Jy}$\tablenotemark{a}\\

\tablenotetext{a}{These objects were listed in Table 1(b) of ${\rm BGS_1}$
(Soifer et al. 1989) as having $\rm S_\nu(60\mu m) < 5.24 ~Jy$. However,
they were also included in the main ${\rm BGS_1}$ catalog listing [Table
1(a)], because of their presence in an earlier compilation
(Soifer et al. 1987). Since some readers may have been confused
regarding the status of these objects, these sources are listed here again
and omitted in Table 1 of this work to clarify their status.}

\enddata
\end{deluxetable}
}

\def\tableThree{

\begin{deluxetable}{lll}
\tabletypesize{\scriptsize} 
\singlespace
\tablenum{3}
\tablewidth{0pt}
\tablecaption{Catalog Cross-Identification Changes}
\tablehead{
\colhead{Name} & 
\colhead{~~~Status}
}
\startdata

MCG -02-01-051/2 (Arp 256)  & ~~~IRAS emission identified with MCG -02-01-051 in ${\rm BGS_1}$\tablenotemark{a}\\
SMC                  & ~~~Named A0051-73 in ${\rm BGS_2}$ (not recognized by NED)\\
NGC 0317B            & ~~~IRAS emission identified with pair NGC 317A/B in ${\rm BGS_2}$\tablenotemark{b}\\
CGCG 436-030         & ~~~Named MCG +02-04-035 in ${\rm BGS_1}$ (same object)\\
NGC 0716             & ~~~Named UGC 1351 in ${\rm BGS_1}$ (same object)\\
ESO 297-G011/12      & ~~~IRAS emission identified with largest galaxy in the pair (NGC 633 $=$ ESO 297-G011) in ${\rm BGS_2}$\tablenotemark{c}\\
IC 0214              & ~~~Named UGC 1720 in ${\rm BGS_1}$ (same object)\\
IRAS F03217+4023     & ~~~Named IRAS 03217+4022 in ${\rm BGS_2}$ (same object)\\
CGCG 465-012         & ~~~Named Zw 465.012 in ${\rm BGS_2}$ (same object)\\
ESO 550-IG02         & ~~~Named MCG -03-12-002 in ${\rm BGS_1}$ (same object)\\
ESO 485-G003         & ~~~Named MCG -04-12-003 in ${\rm BGS_1}$ (same object)\\
ESO 491-G020/021     & ~~~Named ESO 491-G020 in ${\rm BGS_2}$ (same object)\\
CGCG 011-076         & ~~~Named MCG +00-29-023 in ${\rm BGS_2}$ (same object)\\
NGC 4175             & ~~~Mistakenly named NGC 4174 (12h12m26.9s, +29d08m58s J2000) in ${\rm BGS_1}$\\
NGC 2798             & ~~~IRAS emission identified with NGC 2799 (SE companion) in ${\rm BGS_1}$\\
MCG +08-18-013       & ~~~IRAS emission identified with MCG +08-18-012 (SW companion) in ${\rm BGS_1}$ \\
IC 2810              & ~~~Named UGC 6436 in ${\rm BGS_1}$ (same object)\\
NGC 3994/5           & ~~~IRAS emission identified with NGC 3994 in ${\rm BGS_1}$\tablenotemark{d} \\
NGC 4568/7           & ~~~IRAS emission identified with NGC 4568 (SE component of pair) in ${\rm BGS_1}$ \\
NGC 4793             & ~~~Mistakenly named NGC 4783 (12h54m36.6s, -12d33m28s J2000) in ${\rm BGS_1}$\\
CGCG 043-099         & ~~~Named MCG +01-33-036 in ${\rm BGS_1}$ (same object)\\
MCG -02-33-098/9     & ~~~Named MCG -02-33-098 in ${\rm BGS_2}$; IRAS emission dominated by MCG -02-33-098 \\
MCG -03-34-014       & ~~~Named A1309-17 in ${\rm BGS_2}$ (not recognized by NED) \\
VV 250a              & ~~~IRAS emission identified with UGC 8335 = VV 250 in ${\rm BGS_1}$\tablenotemark{e} \\
MCG -03-34-064       & ~~~IRAS emission identified with MCG -03-34-063 in ${\rm BGS_2}$\\
NGC 5257/8           & ~~~IRAS emission identified with NGC 5257 in ${\rm BGS_1}$\tablenotemark{f} \\
CGCG 247-020         & ~~~Named Zw 247.020 in ${\rm BGS_1}$ (same object)\\
NGC 5665             & ~~~Mistakenly named NGC 5663 (14h33m56.3s, -16d34m52s J2000) in ${\rm BGS_1}$\\
VV 340a              & ~~~IRAS emission identified with UGC 9618 = VV 340 in ${\rm BGS_1}$\tablenotemark{g} \\
CGCG 049-057         & ~~~Named Zw 049.057 in ${\rm BGS_1}$ (same object)\\
VV 705               & ~~~Named I Zw 107 in ${\rm BGS_1}$ (same object)\\
NGC 5930             & ~~~IRAS emission identified with NGC 5929 in ${\rm BGS_1}$\tablenotemark{h}\\
CGCG 052-037         & ~~~Mistakenly named MCG +01-42-088 in ${\rm BGS_1}$ (no such object in NED)\\
NGC 6286             & ~~~IRAS emission identified with NGC 6285/6 in ${\rm BGS_1}$ \\
CGCG 083-025         & ~~~Named Zw 083.025 in ${\rm BGS_2}$ (same object)\\
CGCG 141-034         & ~~~Named Zw 141.034 in ${\rm BGS_2}$ (same object)\\
NGC 6621             & ~~~IRAS emission identified with NGC 6621/22 in ${\rm BGS_2}$ \\
CGCG 142-034         & ~~~Named Zw 142.034 in ${\rm BGS_2}$ (same object)\\
IC 4946              & ~~~Named A2020-44 in ${\rm BGS_2}$ ((not recognized by NED)\\
CGCG 448-020         & ~~~Named Zw 448.020 in ${\rm BGS_2}$ (same object)\\
ESO 602-G025         & ~~~Named MCG -03-57-017 in ${\rm BGS_1}$ (same object)\\
CGCG 453-062         & ~~~Named Zw 453.062 in ${\rm BGS_1}$ (same object)\\
IC 5298              & ~~~Named Zw 475.056 in ${\rm BGS_1}$ (same object)\\

\tablenotetext{a}{IRAS centroid is on the SE component of the pair, but both galaxies are
within the IRAS beam (Fig. 1); see also Surace, Sanders \& Mazzarella 2003.}
\tablenotetext{b}{IRAS centroid is clearly on the SE component of NGC 317; see Fig. 1.}
\tablenotetext{c}{IRAS centroid is clearly between the NW (NGC 633 = ESO 297-G011) and SE (ESO 297-G012) pair components; see Fig. 1.}
\tablenotetext{d}{IRAS centroid is clearly between SW (NGC 3994) and NE (NGC 3995) pair components; see Fig. 1.}
\tablenotetext{e}{IRAS centroid is clearly dominated by VV 250a = MCG +10-19-057; see Fig. 1.}
\tablenotetext{f}{IRAS centroid is clearly between NGC 5257/8; see Figure 1.}
\tablenotetext{g}{IRAS centroid is clearly dominated by VV 340a = MCG +04-35-019; see Fig. 1.}
\tablenotetext{h}{NGC 5929 is the SW companion of the dominant IRAS source NGC 5930; see Fig. 1.}

\enddata
\end{deluxetable}
}

\def\tableFour{
\begin{deluxetable}{lrl}
\tabletypesize{\scriptsize} 
\tablenum{4}
\tablewidth{0pt}
\tablecaption{Additions to the RBGS}
\tablehead{
\colhead{Name} & 
\colhead{Revised $\rm S_\nu(60\mu m)~Jy$} &
\colhead{~~~Status}
}
\startdata

NGC 0289               & ~~~5.47 & ~~~Found in IRAS PSC/FSC and confirmed with SCANPI\\
NGC 0300               & ~~~15.30 & ~~~Large optical galaxy (Rice et al. 1993)\tablenotemark{a} \\
NGC 0925               & ~~~7.82 & ~~~Large optical galaxy (Rice et al. 1988)\tablenotemark{a} \\
IC 0356                & ~~~6.77 & ~~~Found in IRAS PSC/FSC and confirmed with SCANPI\\
NGC 1532               & ~~~9.63 & ~~~Found in IRAS PSC/FSC and confirmed with SCANPI\\
CGCG 468-002           & ~~~9.66 & ~~~Orion region with high Galactic foreground confusion (see Fig. 5)\\
NGC 1819               & ~~~6.85 & ~~~Orion region with high Galactic foreground confusion (see Fig. 5)\\
IRAS F05170+0535       & ~~~14.4 & ~~~Orion region with high Galactic foreground confusion (see Fig. 5)\\
IRAS F05405+0035       & ~~~7.03 & ~~~Orion region with high Galactic foreground confusion (see Fig. 5)\\
UGCA 116               & ~~~6.57 & ~~~Orion region with high Galactic foreground confusion (see Fig. 5)\\
IRAS F06076-2139       & ~~~6.43 & ~~~Orion region with high Galactic foreground confusion (see Fig. 5)\\
IC 2163                & ~~~17.55& ~~~Orion region with high Galactic foreground confusion (see Fig. 5)\\
UGCA 127               & ~~~17.61& ~~~Orion region with high Galactic foreground confusion (see Fig. 5)\\
UGCA 128               & ~~~5.38 & ~~~Orion region with high Galactic foreground confusion (see Fig. 5)\\
NGC 2221               & ~~~6.41 & ~~~Found in IRAS PSC/FSC and confirmed with SCANPI\\
ESO 557-G002           & ~~~7.42 & ~~~Orion region with high Galactic foreground confusion (see Fig. 5)\\
AM 0702-601            & ~~~6.70 & ~~~Found in IRAS PSC/FSC and confirmed with SCANPI\\
UGCA 150               & ~~~6.03 & ~~~Found in IRAS PSC/FSC and confirmed with SCANPI\\
NGC 2992               & ~~~7.51 & ~~~Each component of NGC 2992/3 has {$\rm S_\nu(60\mu m) > 5.24$} Jy \\
IC 2522                & ~~~6.47 & ~~~Found in IRAS PSC/FSC and confirmed with SCANPI\\
NGC 3125               & ~~~5.33 & ~~~Found in IRAS PSC/FSC and confirmed with SCANPI\\
NGC 3732               & ~~~5.36 & ~~~Found in IRAS PSC/FSC and confirmed with SCANPI\\
NGC 4151               & ~~~6.46 & ~~~Found in IRAS FSC Rejects (IRAS Z12080+3940) and confirmed with SCANPI \\
NGC 4217               & ~~~11.60& ~~~Found in IRAS PSC/FSC and confirmed with SCANPI\\
NGC 4437               & ~~~7.87 &  ~~~Large optical galaxy (Rice et al. 1988)\tablenotemark{a} \\
NGC 5010               & ~~~10.29& ~~~Found in IRAS PSC/FSC and confirmed with SCANPI\\
NGC 5068               & ~~~12.50& ~~~Found in IRAS PSC/FSC and confirmed with SCANPI\\
NGC 5331               & ~~~5.86 & ~~~In Soifer et al. (1987) sample but dropped in ${\rm BGS_1}$ \\
NGC 5483               & ~~~6.30 & ~~~Found in IRAS PSC/FSC and confirmed with SCANPI\\
UGCA 394               & ~~~5.83 & ~~~Found in IRAS PSC/FSC and confirmed with SCANPI\\
IRAS 16399-0937        & ~~~8.42 & ~~~Found in IRAS PSC/FSC and confirmed with SCANPI\\
NGC 6744               & ~~~18.92& ~~~Large optical galaxy (Rice et al. 1988)\tablenotemark{a} \\
NGC 6786/UGC 11415     & ~~~7.58 & ~~~Found in IRAS PSC/FSC and confirmed with SCANPI\\
NGC 6822               & ~~~47.63& ~~~Large optical galaxy (Rice et al. 1988)\tablenotemark{a} \\
IRAS 19542+1110        & ~~~6.18 & ~~~Found in IRAS PSC/FSC and confirmed with SCANPI\\
NGC 7331               & ~~~45.00& ~~~Large optical galaxy (Rice et al. 1988)\tablenotemark{a} \\
NGC 7418               & ~~~6.67 & ~~~Found in IRAS PSC/FSC and confirmed with SCANPI\\
MCG -01-60-022         & ~~~5.39 & ~~~Found in IRAS PSC/FSC and confirmed with SCANPI\\
NGC 7752/3             & ~~~5.79 & ~~~Found in IRAS PSC/FSC and confirmed with SCANPI\\
\tablenotetext{a}{SCANPI {$\rm S_\nu(60\mu m)$} profile zero-crossing size is comparable to optical size of the galaxy,
and SCANPI {$\rm S_\nu(12\mu m)$} and {$\rm S_\nu(25\mu m)$} values are larger than or comparable to those of Rice et al. (1988) and Rice
(1993), so SCANPI zero-crossing values are adopted in RBGS for this object.}

\enddata
\end{deluxetable}
}

\def\tableFive{
\begin{deluxetable}{lrrrr}
\tabletypesize{\small} 
\singlespace
\tablenum{5}
\tablewidth{0pt}
\tablecaption{IRAS Source Detection Types\tablenotemark{\dagger}}
\tablehead{
\colhead{Detection Type} & 
\colhead{$12 \mu m$} &
\colhead{$25 \mu m$} &
\colhead{$60 \mu m$} &
\colhead{$100 \mu m$}
}
\startdata
RBGS Resolved\tablenotemark{a} - R             & 338 (54\%) & 266 (42\%) & 219 (35\%) & 81 (13\%) \\
BGS$_1$$+$BGS$_2$ Resolved - R                 & 349 (56\%) & 321 (52\%) & 195 (32\%) & 72 (12\%) \\
                                               &            &            &            &           \\
RBGS Marginally Resolved\tablenotemark{a} - M  & ~43 ~(7\%) & ~77 (12\%) & ~82 (13\%) & 105 (17\%)\\
BGS$_1$$+$BGS$_2$ Marginally Resolved - U+     & ~84 (14\%) & 112 (18\%) & ~80 (13\%) & ~71 (11\%)\\
                                               &            &            &            &           \\
RBGS Unresolved\tablenotemark{a} - U           & 229 (36\%) & 286 (46\%) & 328 (52\%) & 443 (70\%)\\
BGS$_1$$+$BGS$_2$ Unresolved - U               & 179 (29\%) & 185 (30\%) & 343 (55\%) & 471 (76\%)\\
                                               &            &            &            &           \\
RBGS Upper Limits                              & ~19 ~(3\%) &   0        &          0 &         0 \\
RBGS Uncertain fluxes\tablenotemark{b}         & ~34 ~(5\%) &  37 ~(6\%) &  40 ~(6\%) &  59~(9\%) \\

\tablenotetext{\dagger}{The number and approximate percentages of objects in each category are listed.
The RBGS contains 629 objects and BGS$_1$$+$BGS$_2$ contains 618 objects.
}

\tablenotetext{a}{IRAS scan profile size information as identified with the
size code (S) following each flux density and uncertainty listed in Table 1.
See the description of columns (8) -- (11) in Table 1, and Figure 14 (Appendix)
for examples of coadded scan profiles that illustrate size codes ``R'',
``M'', and ``U''.
}
\tablenotetext{b}{Various types of measurement uncertainties as encoded in the 
flag (F) following some flux density and uncertainty values listed in Table 1.
See the description of columns (8) -- (11) in Table 1, and Figure 16 (Appendix)
for examples of scan profiles that illustrate the uncertainty flags 
``g'', ``b'', ``c'', and ``n''.
}
\enddata
\end{deluxetable}
}

\def\tableSix{
\begin{deluxetable}{rrrr}
\tiny
\singlespace
\tablenum{6}
\tablewidth{0pt}
\tablecaption{Infrared Luminosity Function}
\tablehead{
\colhead{$L_{\rm ir}$} & 
\colhead{$N$} &
\colhead{$V/V_{\rm max}$} &
\colhead{$\rho\ ({\rm Mpc}^{-3}\ {\rm M}_{\rm ir}^{-1})$}
}
\startdata
 $7.75$ &   $3$ & $0.16 \pm 0.03$ & $11.2 \pm 6.5 \times 10^{-2}$\\
 $8.25$ &   $3$ & $0.28 \pm 0.05$ & $14.0 \pm 8.1 \times 10^{-3}$\\
 $8.75$ &   $9$ & $0.39 \pm 0.04$ & $13.2 \pm 4.3 \times 10^{-3}$\\
 $9.25$ &  $24$ & $0.37 \pm 0.02$ &  $6.7 \pm 1.3 \times 10^{-3}$\\
 $9.75$ &  $69$ & $0.56 \pm 0.02$ & $32.9 \pm 4.0 \times 10^{-4}$\\
$10.25$ & $168$ & $0.44 \pm 0.01$ & $17.0 \pm 1.3 \times 10^{-4}$\\
$10.75$ & $157$ & $0.45 \pm 0.01$ & $26.5 \pm 2.1 \times 10^{-5}$\\
$11.25$ & $122$ & $0.53 \pm 0.02$ & $38.6 \pm 3.5 \times 10^{-6}$\\
$11.75$ &  $56$ & $0.50 \pm 0.02$ & $30.1 \pm 4.0 \times 10^{-7}$\\
$12.25$ &  $18$ & $0.44 \pm 0.03$ & $14.5 \pm 3.4 \times 10^{-8}$\\
\enddata
\end{deluxetable}
}

\section{Introduction} 

This paper presents the complete list of objects contained in the \IRAS Revised
Bright Galaxy Sample (RBGS), a flux-limited sample of all extragalactic objects
brighter than 5.24{\ts}Jy at 60{\ts}\um, covering the entire sky surveyed
by the {\it Infrared Astronomical Satellite} (\IRAS; Neugebauer et al. 1984)
at Galactic latitude $\vert b \vert {\ts}>{\ts}5^\circ$. The RBGS replaces
the earlier \IRAS Bright Galaxy Samples (\BGS), which were compiled using the
\IRAS {\it Point Source Catalog} (PSC: 1988) along with other intermediate
releases of the \IRAS data products and are therefore now out of date with the
more accurate ``Pass 3" calibration \footnote{``Pass 3" refers to the final
calibration adopted for the archived \IRAS data by the Infrared Processing and
Analysis Center (IPAC) in 1990. Details of the calibration methods for each of
the \IRAS infrared bands, and the various \IRAS catalogs and atlases can be found
in the Explanatory Supplement to the \IRAS {\it Faint Source Survey} (Moshir
et al. 1992).} adopted for the final release of the \IRAS Level{\ts}1 Archive.

The RBGS objects are the brightest 60{\ts}\um\ sources in the extragalactic
infrared sky, and as such they remain the best sources for studying
the infrared emission processes in galaxies and for comparison with
extragalactic observations at other wavelengths.  In this regard the RBGS
can be considered the infrared equivalent of the 3CR survey of extragalactic
radio sources.  Most sources in the RBGS have extended \IRAS flux densities
that are underestimated by measurements in the \IRAS PSC and in the
{\it Faint Source Catalog} (FSC: Moshir et al. 1992), because
the \IRAS catalogs are the result of point-source filtering. Therefore, the
RBGS should be the reference of choice for accurate \IRAS fluxes and infrared
luminosities of galaxies in the local Universe.

In this paper we report revised 12{\ts}\um, 25{\ts}\um, 60{\ts}\um\ and 100{\ts}\um\ total flux
densities for all 629 infrared sources in the RBGS.  Section 2 describes the
methods used to select the objects. Details of the \IRAS data 
processing are deferred to the Appendix.  Section 3 presents the
summary data for the RBGS objects, including the revised \IRAS flux measurements,
source size information, and derived infrared luminosities.  Also included
is an atlas of images from the Digitized Sky Survey with overlays of the \IRAS
position uncertainty ellipse and annotated scale bars; this is intended as
a reference to visualize the optical morphology of the infrared source in
context with the angular and metric size of each object, and more importantly,
in the case of confused or double sources, to quickly see how large the \IRAS
position offset may be from the optical sources. Section 4 is a discussion
of the general properties of the RBGS which includes a comparison of the
revised measurements with those published previously, and a summary of the sky
coverage, number counts, infrared colors, and infrared luminosity function.
A companion paper (Surace, Sanders \& Mazzarella 2003) provides High Resolution
(HIRES) processing of $\sim$20\% of the RBGS sources where it was thought
that enhanced resolution might provide better source positions, or in the
case of close galaxy pairs, allow deconvolution of the individual components.

\section{Sample Selection and Data Processing}

In constructing the RBGS, the basic methods used to extract candidate
bright galaxies from the \IRAS catalogs and to compute total fluxes in each
of the four \IRAS bands using ADDSCAN/SCANPI (hereafter referred to as SCANPI;
Helou et al. 1988) were similar to procedures used in earlier compilations
of the \IRAS BGS (here referred to as BGS$_1$: Soifer \EA\ 1989, 1987, 1986)
and the \IRAS BGS--Part II (here referred to as BGS$_2$: Sanders \EA\ 1995);
the reader is referred to these papers for historical perspective as well
as for a more thorough description of original sample membership criteria.
In this paper we outline the major steps that were followed to construct
the new sample, and where appropriate, emphasize the differences between
the earlier \BGS\ procedures and the more refined RBGS processing.

The \IRAS FSC and \IRAS PSC were used as starting points for the initial search
of the \IRAS data archive. It was necessary to use the PSC because the FSC does
not include objects in confused regions of the infrared sky with Galactic
latitude $\vert b \vert < 10^\circ$.  The FSC Rejects (FSCR) \footnote{The
FSCR, along with all other \IRAS catalogs, is available for queries through
the GATOR service of the Infrared Science Archive (IRSA)
at IPAC; {\it http://irsa.ipac.caltech.edu/}. The FSCR contains objects that
have only one \IRAS hours-confirmed (HCON) observation; however, \IRAS measurements
in the FSCR are much more reliable when a high signal-to-noise detection
is combined with confirmation via positional cross-identification with a
known source detected at another wavelength.} were also examined for bright
objects at 60{\ts}\um.  In the end, the only source accepted from the FSCR is
NGC 4151 (\IRAS Z12080+3940). The reasons for its inclusion in RBGS are:
a) the SCANPI signals in all 4 \IRAS bands are very strong (see Table 1);
b) the FIR signal is positionally coincident with the optical galaxy, and
the coadded scan profile from ADDSCAN/SCANPI is consistent with the optical
size of the galaxy; c) the \IRAS SCANPI flux density values at 12{\ts}\um\ and
25{\ts}\um\ are comparable (slightly higher, as expected given some extended
flux) to independent ground-based measurements at 10.6 {\ts}\um\ and 21{\ts}\um\
in apertures centered on the nucleus (Lebofsky \& Rieke 1979).

A second step involved another search of the \IRAS PSC and FSC, this time
with a lower 60{\ts}\um\ point source threshold of 4.5{\ts}Jy, which was designed
primarily for the purpose of capturing sources that might have a reasonable
chance of having extended flux sufficient to bring their total 60{\ts}\um\ flux
densities above the RBGS threshold. To investigate how many objects with
total 60{\ts}{\ts}\um\ flux density greater than 5.24{\ts}Jy may be ``hiding''
among objects with even fainter point-source components, a few dozen galaxies
randomly selected from the FSC with $4.0~Jy < S_{\nu}(60\mu m) < 4.5~Jy$ were
also coadded with SCANPI. These objects were not found to contain sufficient
extended emission to bring their total fluxes above the RBGS flux limit.
Searches of NED for optical galaxies larger than 4 arcminutes, without regard
for their (likely underestimated) flux measurements from the \IRAS catalogs,
were also performed to obtain additional candidates.  However, objects with
total $S_{\nu}(60\mu m) > 5.42~Jy$ among these large optical galaxies objects
were also recovered in the \IRAS catalog searches noted above.  This provides
confidence that our procedures resulted in a sample with a very high level of
completeness at 60{\ts}\um. 

All candidate \IRAS sources were compared with the latest catalog
cross-correlations available in the NASA/IPAC Extragalactic Database
(NED)\footnote{NED is available at {\it http://ned.ipac.caltech.edu/}.},
which were then inspected using overlays on
the DSS1 images.  The data were then reprocessed (co-addition of all acceptable
\IRAS scans) using SCANPI, and the resulting 1-D coadded scan profiles were
visually inspected to determine the amount of extended emission, and in
the case of blended or confused sources, to determine the best method for
computing the total flux density. A computer program was written to determine
objectively and consistently whether the total flux density of each source
is better represented by the baseline zero-crossing method $f_\nu(z)$, the
value integrated within the nominal \IRAS detector size $f_\nu(t)$, the point
source template fit amplitude (``template"), or the peak flux in the profile
(``peak"). This was accomplished by checking whether the $f_\nu(t)$ value is
significantly larger than the template fit and peak values (considering also the
RMS noise in the coadded scan outside the detected signal range), and if so whether
the $f_\nu(z)$ value is significantly larger than the $f_\nu(t)$ value. This
procedure, combined with comparisons of the coadded scan profile widths at
25\% (``W25") and 50\% (``W50") of the peak flux with the nominal beam size
(point-spread function), was also used to determine whether each \IRAS source is
resolved, marginally extended or unresolved.  All measurements were performed
using SCANPI's median (1002) method of coadding the individual \IRAS scans,
and all sources with
60{\ts}\um\ point source flux greater than 5.24{\ts}Jy (the completeness limit
of the original BGS) were included in the final sample.

Appendix A gives more details regarding the various methods used for setting
thresholds to select between the flux density estimators and for
estimating the total flux densities of the extended sources, and provides
a tabular listing of all key SCANPI measurements in each \IRAS band, as well
as the ratio of the new RBGS total flux density measurements compared with
the old values reported in the \BGS\ (where available). Example coadded scan
profiles are also plotted to illustrate the extent of \IRAS emission compared to
the point spread function and to show how different flux measurement methods
were objectively selected in the automated SCANPI reprocessing of
the data for all potential RBGS sources.

Finally, \IRAS flux measurements from the \lq\lq Catalog of \IRAS Observations
of Large Optical Galaxies\rq\rq\ (Rice et al. 1988) and \lq\lq An Atlas of
High-Resolution \IRAS Maps of Nearby Galaxies\rq\rq\ (Rice 1993) were examined.
Careful comparison of flux densities integrated within the signal range between
SCANPI's baseline fit zero-crossing points, $f_\nu(z)$, to those measured
by Rice et al. from \IRAS 2-D image products showed generally good agreement
for galaxies with optical diameters smaller than about 25 arcminutes. 
For objects with optical diameters larger than $\sim$25 arcminutes it is clear
that even SCANPI's $f_\nu(z)$ method systematically underestimates the total
flux density compared to the 2-D method used by Rice.  Therefore, to maintain
consistency with other objects in the RBGS to the highest degree possible,
in the final compilation we adopted Rice et al. measurements over those from
SCANPI's $f_\nu(z)$ estimator only for objects larger than 25 arcminutes;
such objects are flagged in Table 1.

\section{The \IRAS RBGS Data}

Table 1 presents the summary data for all of the sources in the \IRAS RBGS.
The complete sample contains 629 galaxies that met our threshold criteria
of $S_{60} > 5.24${\ts}Jy.  The column entries are as follows:

({\bf 1}) {\it  Common Name} -- Common name taken in decreasing priority
order from the NGC, UGC, ESO, IC, A, MCG, CGCG, Zwicky compact galaxies,
Markarian, and \IRAS catalogs.  ``N" follows the common name for 
``new" RBGS objects which were not included in the compilation of the BGS$_1$
(Soifer et al. 1989) or BGS$_2$ (Sanders et al. 1995); ``1" flags objects
from the BGS$_1$, and ``2" flags objects from the BGS$_2$.

({\bf 2}) {\it \IRAS Name} -- Names from the \IRAS Faint Source Catalog (FSC)
are given using the standard ``F" prefix; if the source is not in the FSC, the
name from the Point Source Catalog (PSC) is given, which has no letter prefix.
\IRAS source names are based on equinox B1950.0 coordinates.  Names from the
\IRAS {\it Faint Source Catalog Rejects} (FSCR) are given using the standard
``Z" prefix. (NGC 4151 is the only FSCR object in the RBGS; see Section 2.)
\IRAS catalog names are not listed for a few very large galaxies (indicated
by `` ---") because entries in the \IRAS catalogs do not represent the total
flux and generally correspond only to the nucleus or other bright component.
Flags following the \IRAS name are as follows: ``R" indicates large galaxies
with for which total flux densities derived from \IRAS images were published by
Rice et al. (1988) or Rice (1993); the Rice et al. values are adopted here only
in cases listing an ``R" value for the Method code in columns (8) -- (11); see text
for details. `` *" identifies objects located in regions of the Milky Way,
LMC, or SMC with high source confusion.  [Note: The objects are listed in
order of increasing B1950 right ascension (R.A.), as reflected in the
source names from the \IRAS catalogs; this places some objects out 
of order in terms of their J2000 coordinates.]

({\bf 3--4}) {\it R.A., Dec} -- Equinox J2000.0 R.A.  and declination
(Dec), transformed from equinox B1950.0 coordinates published in the \IRAS FSC
(Version 2.0) or PSC (Version 2), as indicated in Col. (2).  These coordinates
correspond to the centroid of the \IRAS positional uncertainty ellipse as
superimposed on the Digitized Sky Survey fields shown in Figure 1.  Exceptions
are for the large galaxies NGC{\ts}55, SMC, NGC{\ts}300, LMC, NGC{\ts}6744,
and NGC{\ts}6822 whose positions are taken from Rice et al. (1988).

({\bf 5--6}) {\it $\ell$,{\it b}} -- Galactic latitude and longitude in
decimal degrees.

({\bf 7}) {\it HIRES code} -- \IRAS HIRES code from Surace, Sanders \& Mazzarella (2003).
A listing here indicates that the infrared source contains two or more objects
that were confused in one or more of the \IRAS detectors.  The high-resolution
image restoration algorithm HIRES (Aumann, Fowler, \& Melnyk 1990) was applied
to these sources in an attempt to resolve flux from individual galaxies. ``R"
means the source was resolved into two or more components, resulting in new
\IRAS positions and fluxes; ``*" means the source was resolved, but the \IRAS
position lies between the two galaxies as seen on the DSS; ``S" means the
source was partially resolved into two or more components, and fluxes were
estimated for each object, but no new reliable positions were derived; ``U"
means HIRES was unsuccessful at resolving separate components. See Surace
et al. (2003) for detailed HIRES results for these objects.
For consistency, the fluxes listed in Table 1 were derived using
SCANPI in the same manner as the single objects in the table. There
are known calibration problems with the HIRES data product (Surace et
al. 1993), and while the flux ratio between individual galaxies
within a system can be more accurately measured using HIRES, the absolute
calibration of the galaxy system flux is better determined
by SCANPI. Since in many cases a single galaxy dominates the
far-infrared flux of a multiple galaxy system, there are only a
handful of systems that would have not been selected in the RBGS
but which are included by virtue of having an integrated flux above the
selection limit.

({\bf 8--11}) {\it Flux Densities} -- Total \IRAS flux densities (Jy) in the
12{\ts}\um, 25{\ts}\um, 60{\ts}\um, and 100{\ts}\um\ bands, respectively.
Following each flux density value are the uncertainty (1{\ts}$\sigma$ in
mJy) followed by three code letters ({\bf SMF}) indicating the infrared
{\underline{\bf S}ize} (U = ``unresolved", M = ``marginally extended", R =
``resolved"); the {\underline{\bf M}ethod} chosen as the best flux estimate
from SCANPI (Z = ``zero crossing", I = ``in-band total", T = ``template fit",
P = ``peak value", S = ``deconvolution with SCLEAN" (see Appendix), R =
from Rice et al. (1988); and a {\underline{\bf F}lag} indicating the type of
confusion (c = ``cirrus", g = ``nearby galaxy", n = ``excessive noise", b =
``blended objects") causing large uncertainty as indicated by a colon (``:")
prefix on the flux density. Note that some large galaxies flagged with ``R"
in Col. (2) have new SCANPI zero-crossing estimates, indicated here using
``Z" for the second flag, which are judged to have more accurate calibration
than previously adopted values from Rice et al. (1988).

({\bf 12}) {\it cz} -- The Heliocentric radial velocity (km{\ts}s$^{-1}$)
of the \IRAS source computed as $c$ times the redshift $z$.  The source for the
redshift is given in Col. (2) of Table 2 as a 19 digit reference code from NED.
In cases where either a millimeter (e.g. CO) or HI 21-cm line measurement has
been reported we have chosen to adopt these redshifts given that they better
reflect the systemic velocity of the galaxy, as opposed to optical measurements
which are often biased (typically blue-shifted) due to optical depth effects.

({\bf 13}) {\it Distance} -- The estimated source distance in Mpc; metric
(``proper") distance is listed, not luminosity distance.  For most objects
this was calculated from $cz$ using the cosmic attractor model outlined in
Appendix A of Mould et al. (2000), using $H_{\rm o} = 75~{\rm km}~{\rm
s}^{-1}~{\rm Mpc}^{-1}$ and adopting a flat cosmology in which $\Omega_{\rm
M} = 0.3$  and $\Omega_{\Lambda} = 0.7$ (which corresponds to $q_0 = -0.55$).
Superscripts are defined as follows: `` P" indicates the distance is not
computed by correcting the heliocentric redshift using the cosmic attractor
model, but comes from a direct {\it primary} distance measurement; ``S"
flags a direct {\it secondary} distance measurement; ``V" indicates that
the cosmic attractor flow model places the object within the {\it Virgo}
cluster at a distance of 15.3 Mpc; ``G" indicates that the cosmic attractor
flow model places the object within the {\it Great Attractor}.  References
for the adopted primary and secondary distance indicators are Freedman et
al. (2001), Madore \& Freedman (1998), Ferrarese et al. (2000), Mould et al. (1991),
Tully \& Shaya (1984),  Aaronson \& Mould (1983), Aaronson et al. 1982,
and Aaronson, Mould \& Huchra (1980).

({\bf 14--15}) {\it Luminosities} -- The base$_{10}$ logarithm of the
far-infrared luminosity, $L_{\rm fir} = L(40-400\mu m)$, determined using
the prescription described in Appendix B of Cataloged Galaxies and Quasars
Observed in the \IRAS Survey (1989), and the infrared luminosity, $L_{\rm
ir} = L(8-1000\mu m)$, determined using the fluxes in all four \IRAS bands
(Perault 1987; see also Table 1 in Sanders \& Mirabel 1996), in units of solar
bolometric luminosity, $L_\odot = 3.83 \times 10^{33}${\ts}erg{\ts}s$^{-1}$.

({\bf 16}) {\it Rank} -- The object's sequential rank in the distribution
of $L_{\rm ir}$ values, where 1 is the most luminous source.

({\bf 17}) {\it Other Names} -- Other names from catalogs of interacting
galaxies (IG) or active galactic nuclei (AGN).  Spaces between catalog name
prefix and number (as standard in NED) are omitted here to save space.

\ifnum\Mode=0 
\vskip 0.3in
\fbox{\bf Table 1 goes here (13 pages).}
\vskip 0.3in
\else
\ifnum\Mode=2 \onecolumn \fi 
\clearpage
{
\headheight     0.0in
\headsep        0.0in
\topmargin     -1.0in 
\begin{figure}[!htb]
\includegraphics[scale=0.9,angle=180]{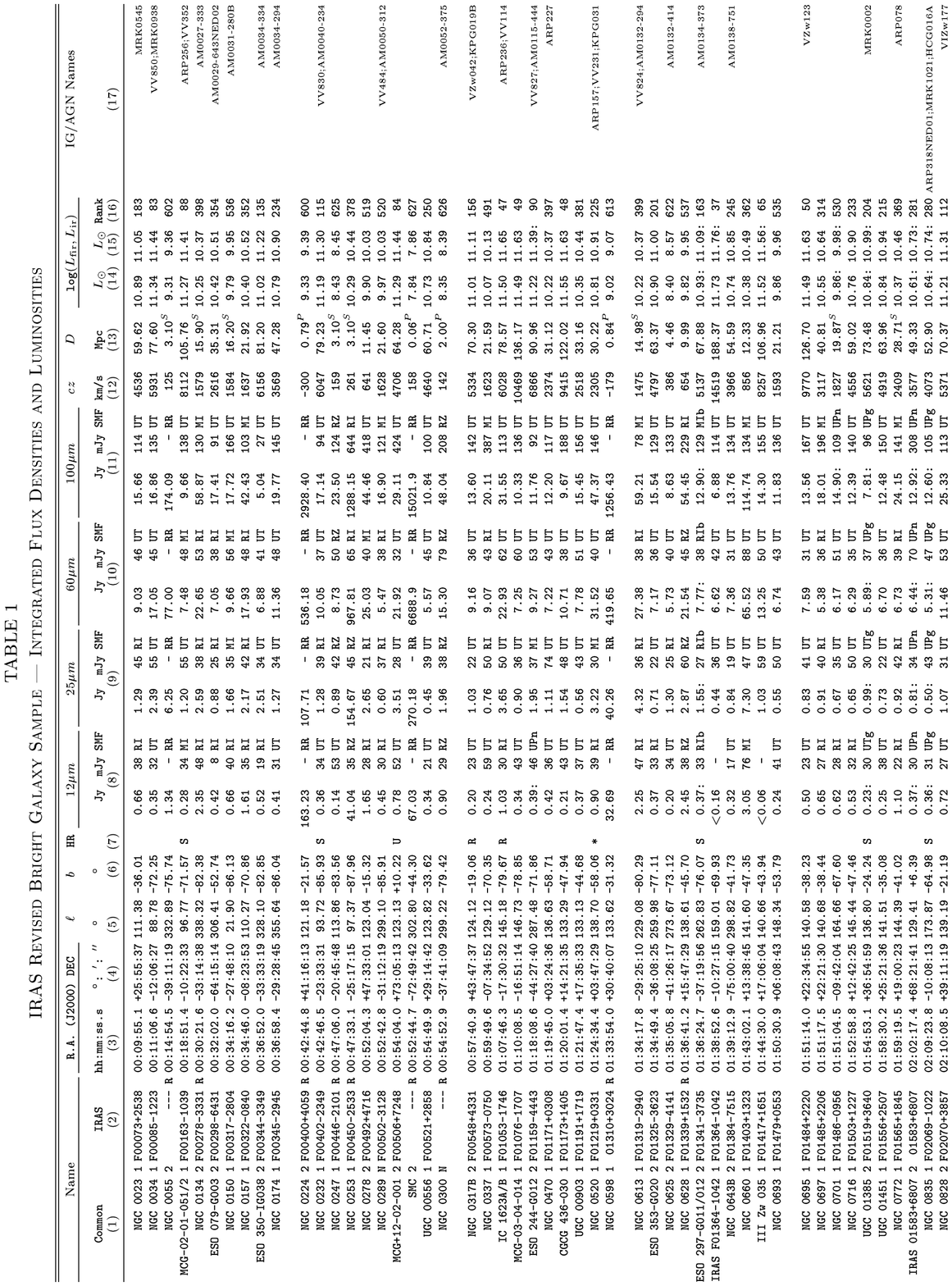}
\end{figure}
\clearpage
\begin{figure}[!t]
\includegraphics[scale=0.9,angle=180]{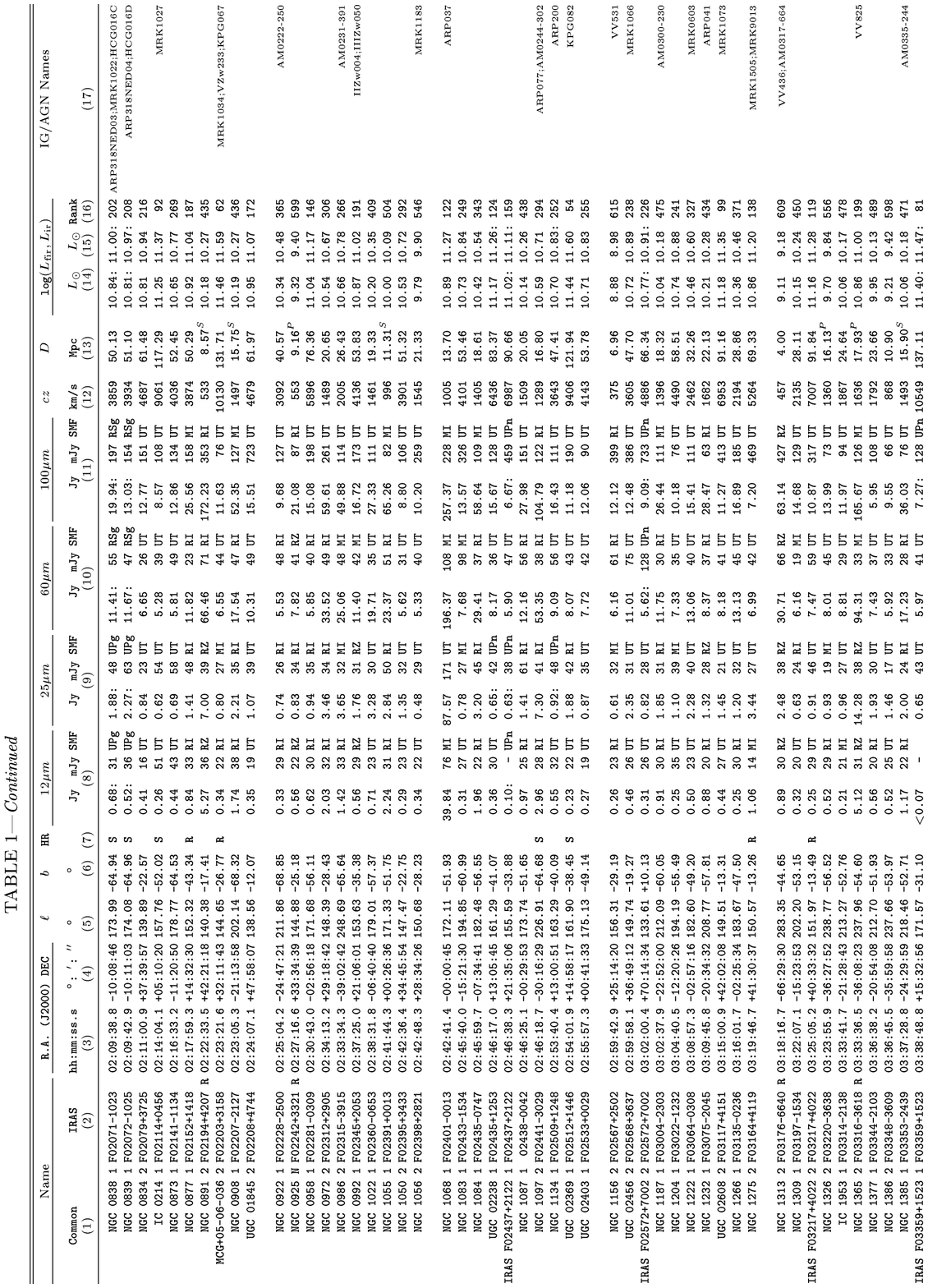}
\end{figure}
\clearpage
\begin{figure}[!t]
\includegraphics[scale=0.9,angle=180]{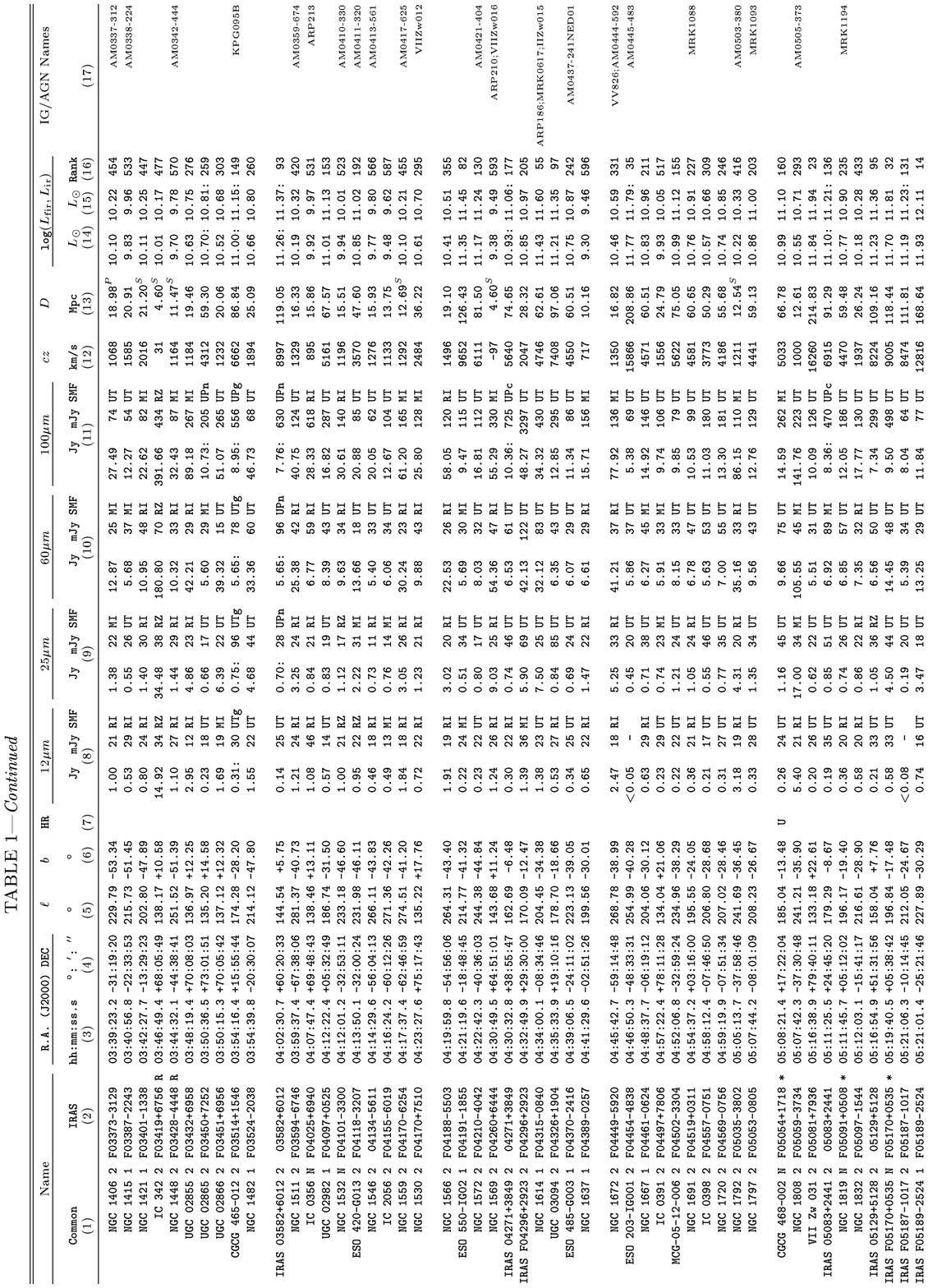}
\end{figure}
\clearpage
\begin{figure}[!t]
\includegraphics[scale=0.9,angle=180]{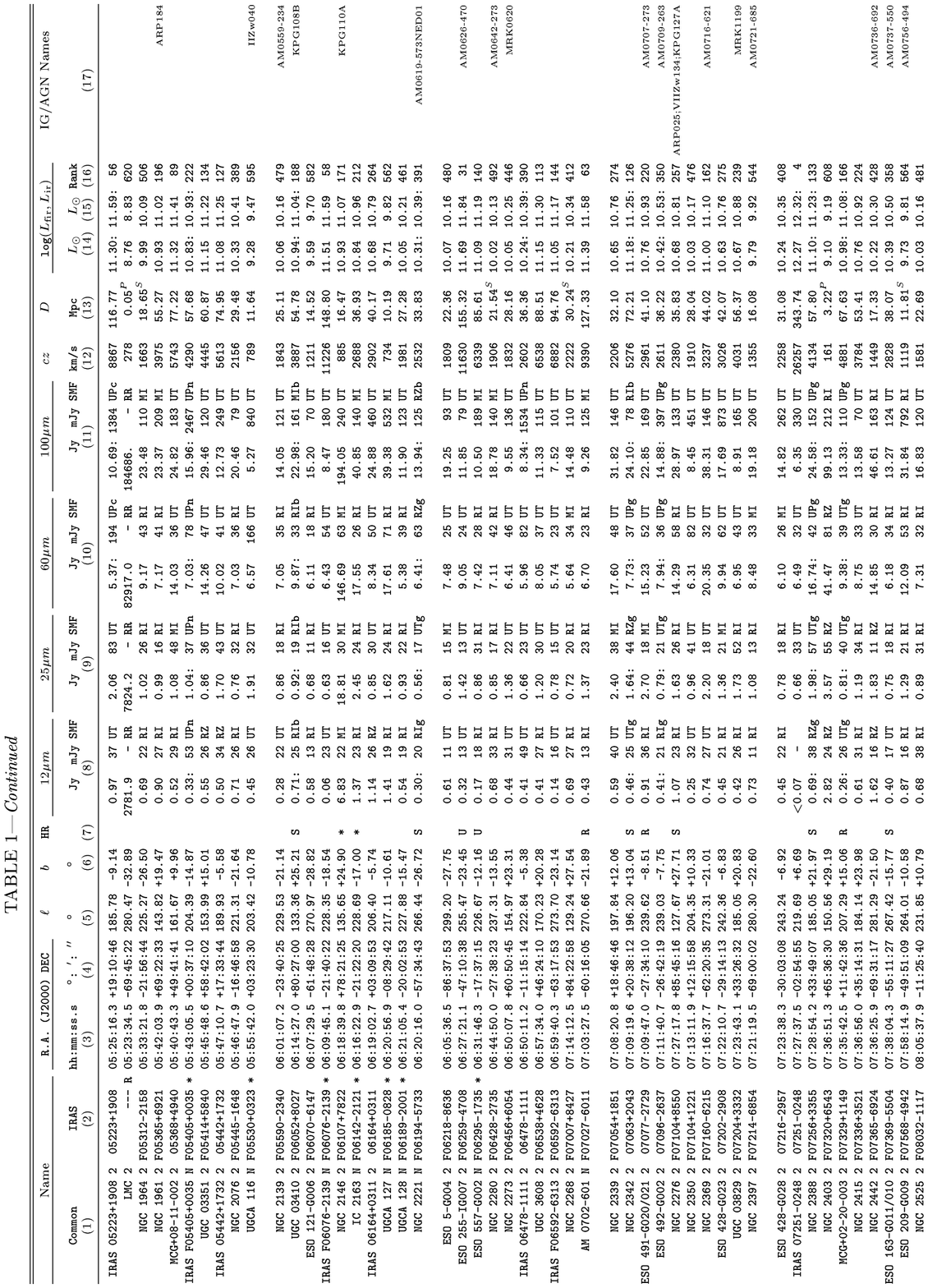}
\end{figure}
\clearpage
\begin{figure}[!t]
\includegraphics[scale=0.9,angle=180]{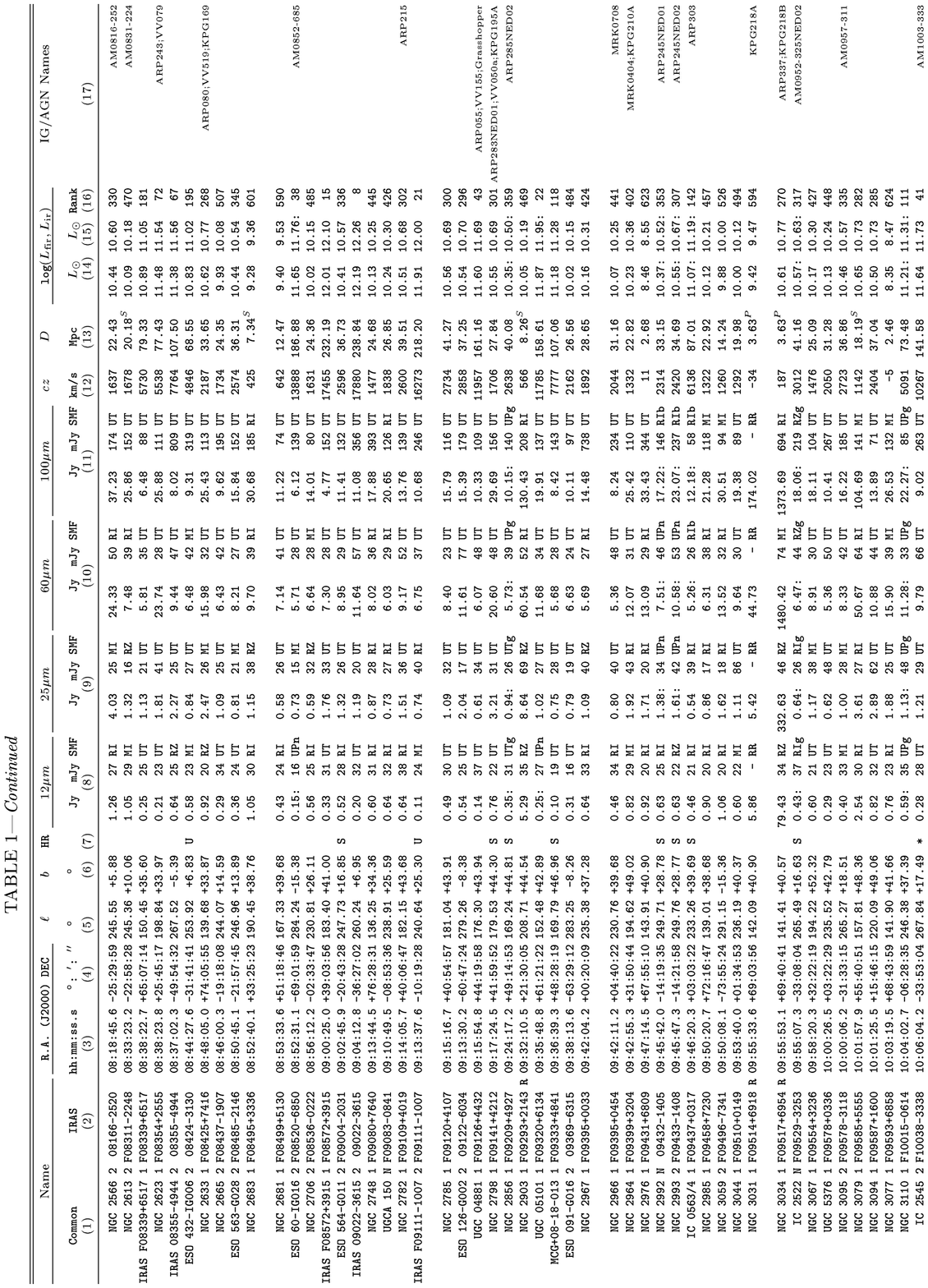}
\end{figure}
\clearpage
\begin{figure}[!t]
\includegraphics[scale=0.9,angle=180]{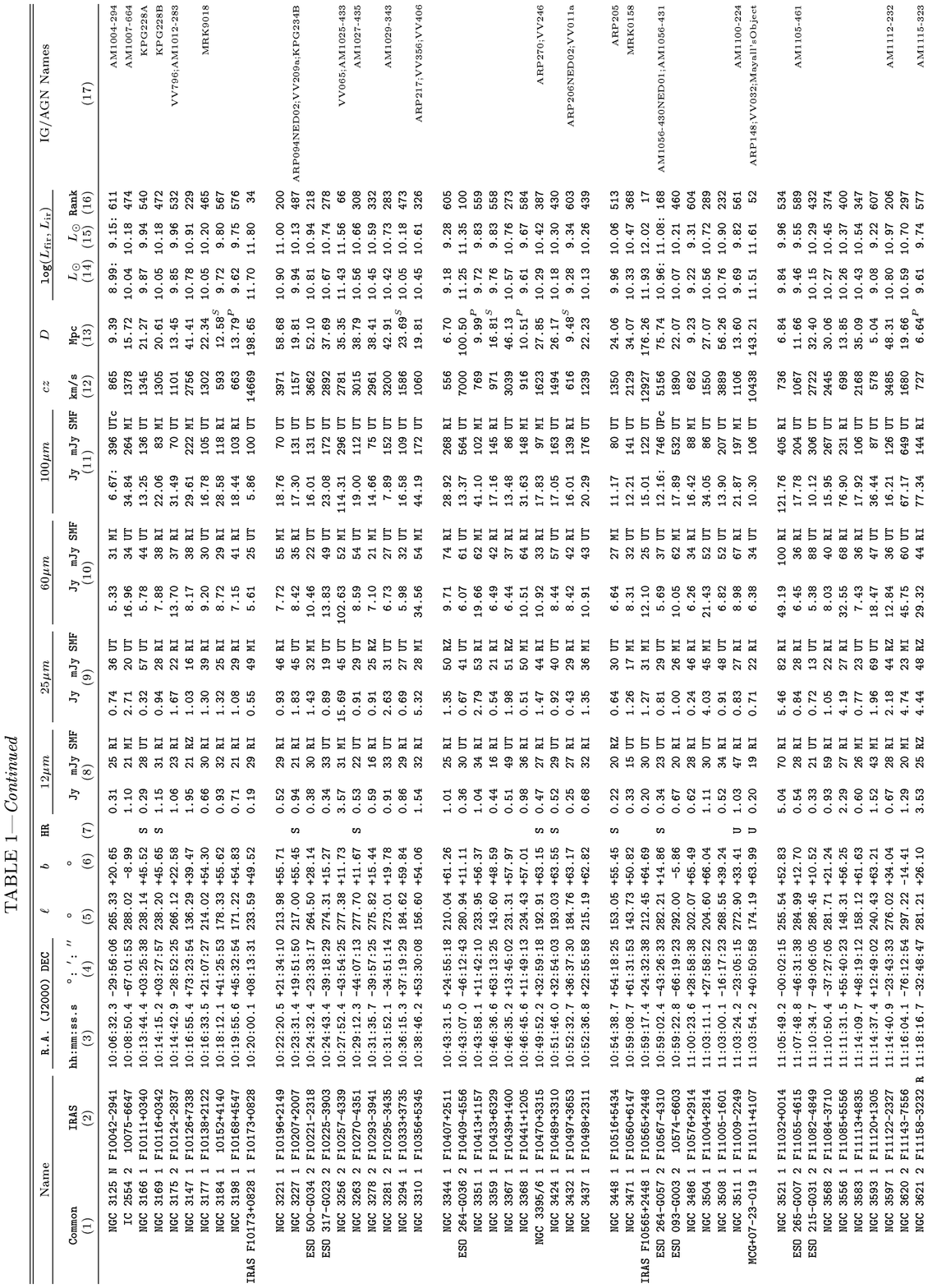}
\end{figure}
\clearpage
\begin{figure}[!t]
\includegraphics[scale=0.9,angle=180]{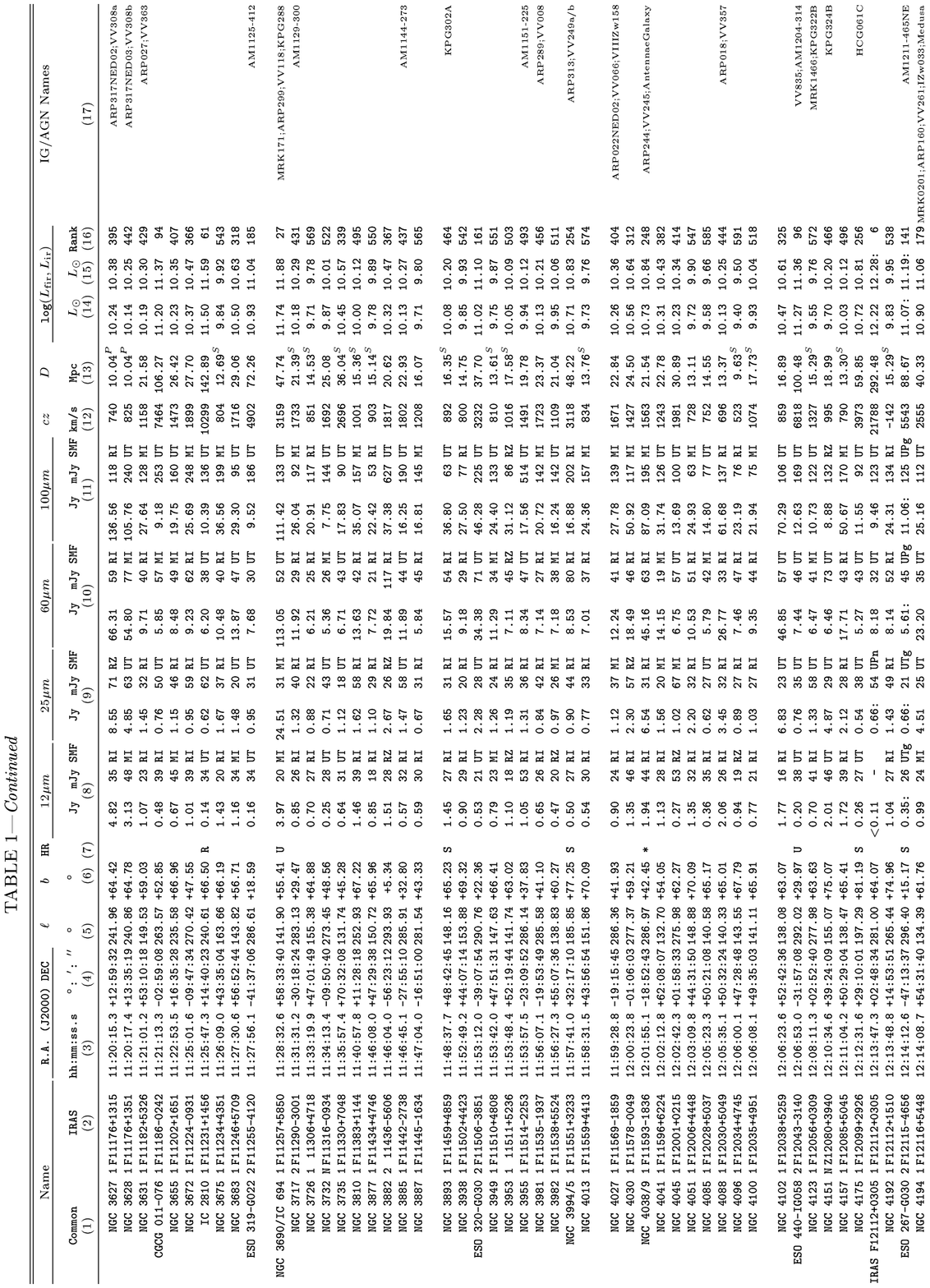}
\end{figure}
\clearpage
\begin{figure}[!t]
\includegraphics[scale=0.9,angle=180]{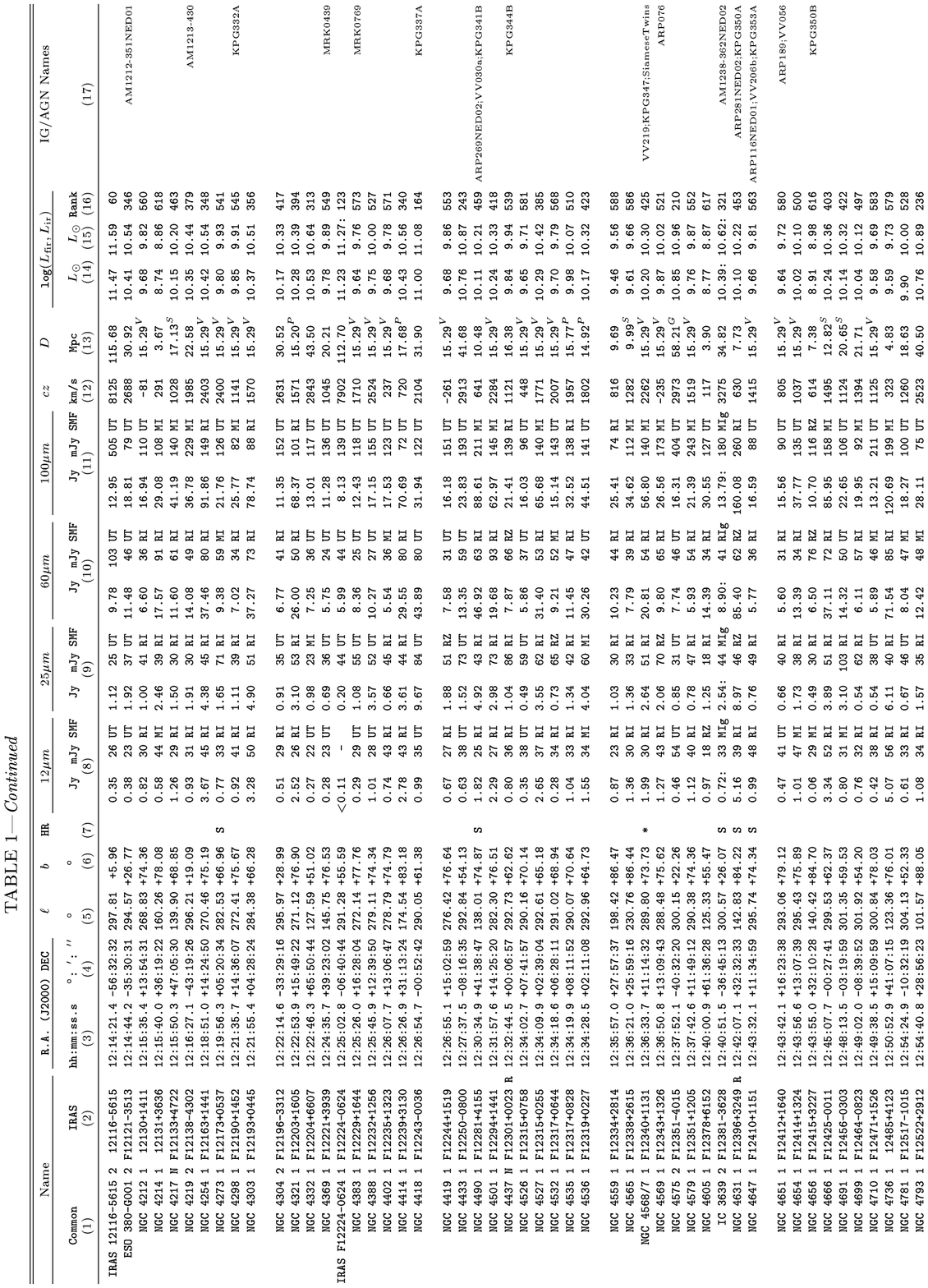}
\end{figure}
\clearpage
\begin{figure}[!t]
\includegraphics[scale=0.9,angle=180]{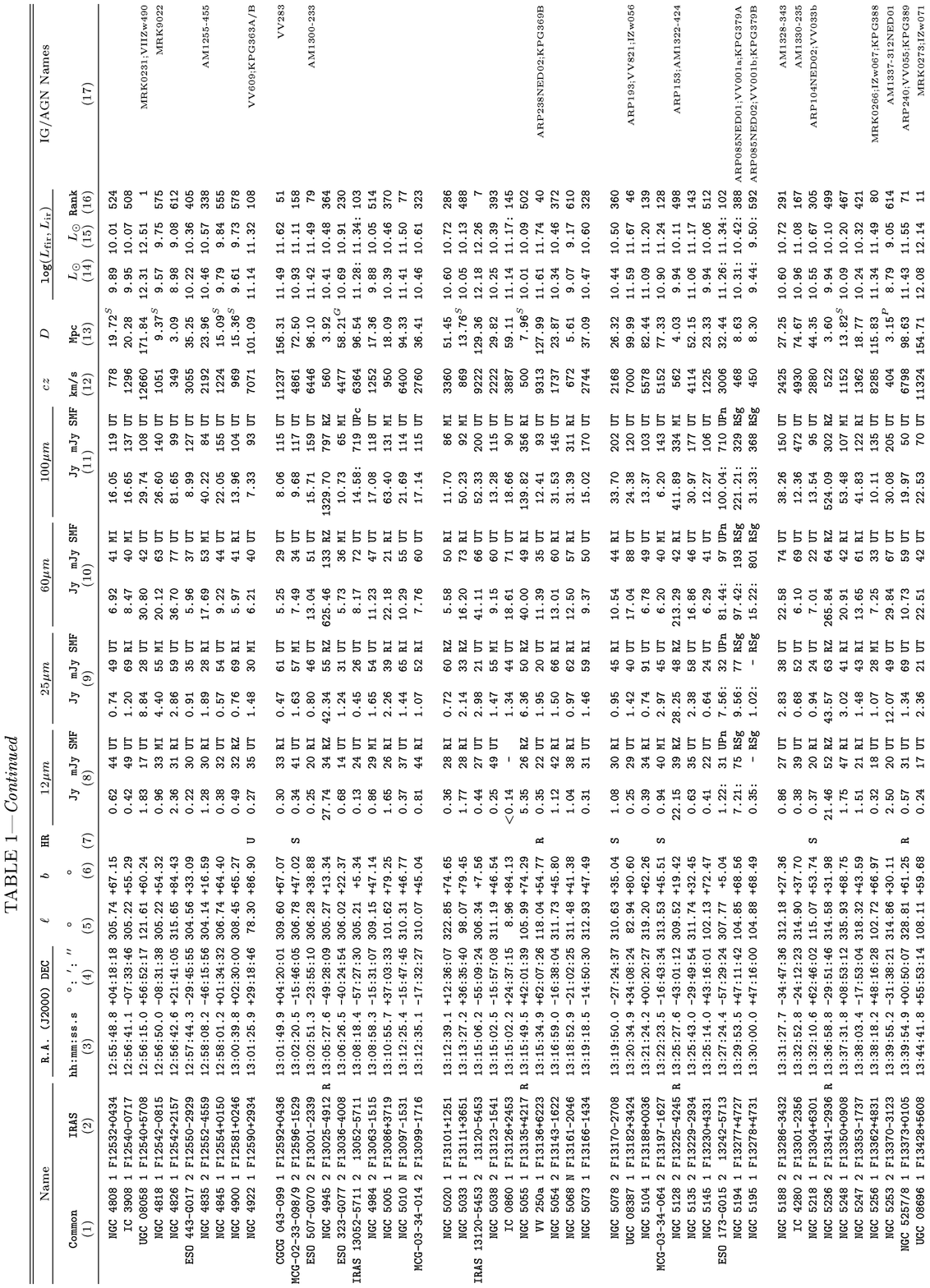}
\end{figure}
\clearpage
\begin{figure}[!t]
\includegraphics[scale=0.9,angle=180]{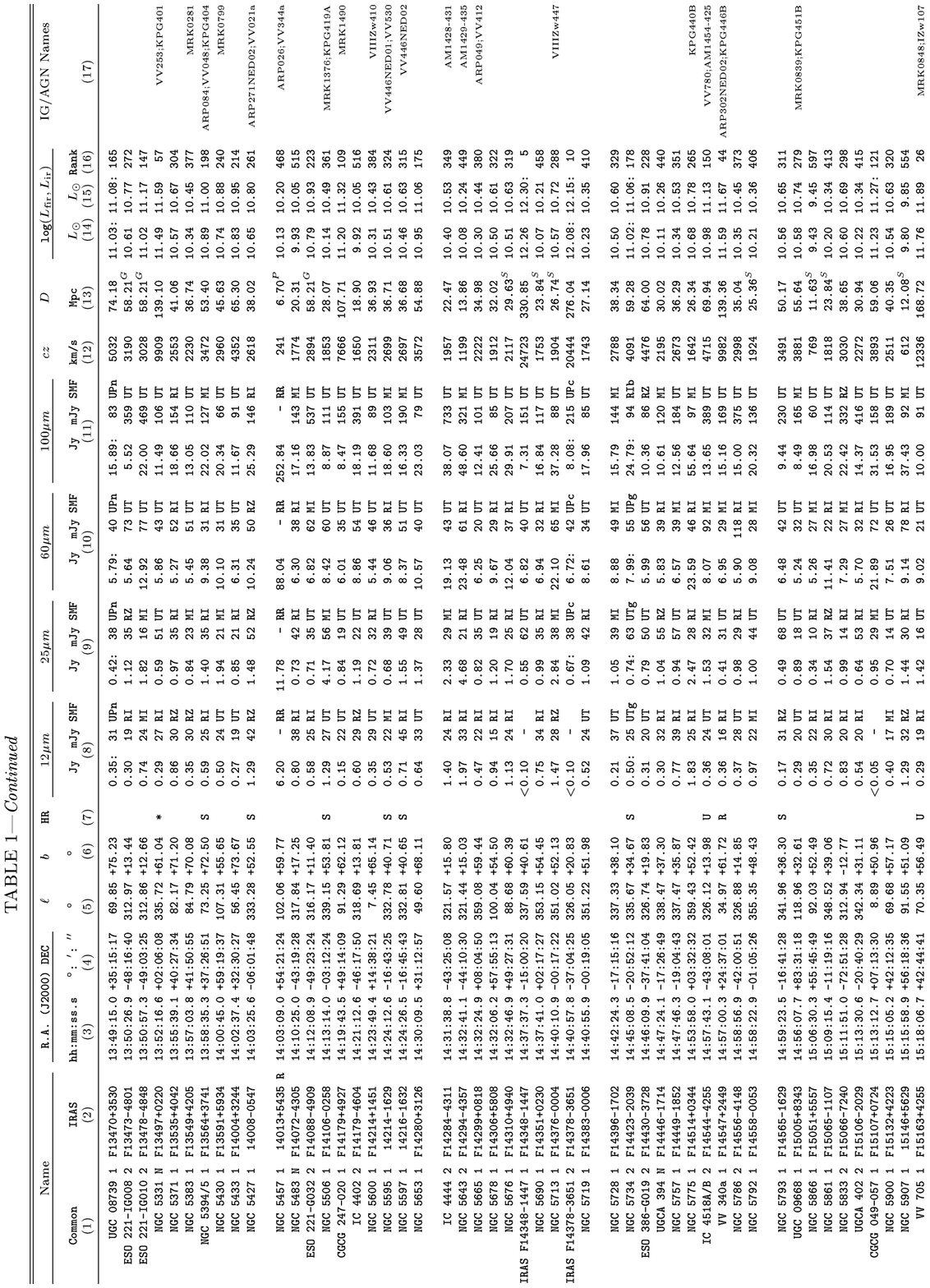}
\end{figure}
\clearpage
\begin{figure}[!t]
\includegraphics[scale=0.9,angle=180]{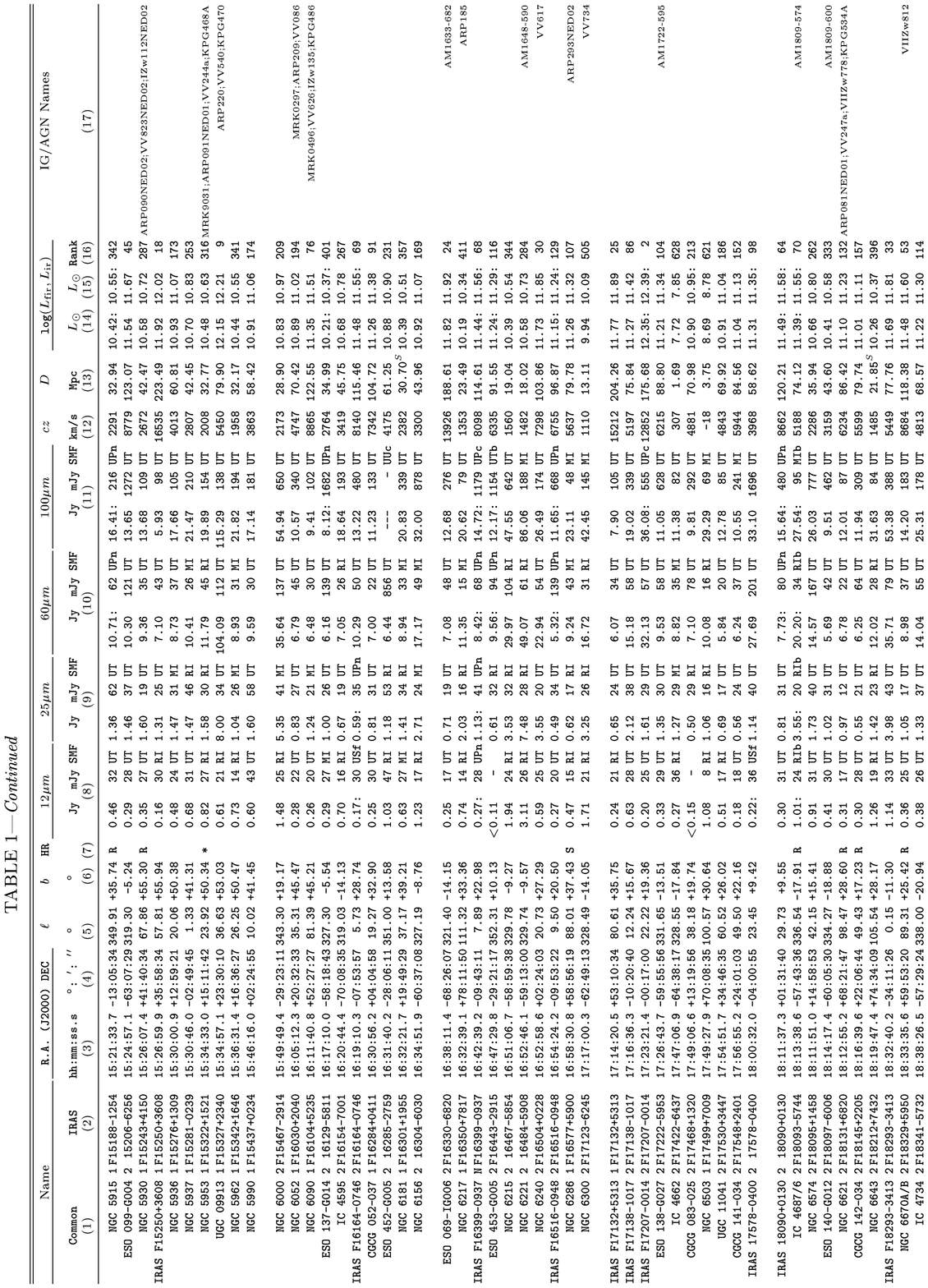}
\end{figure}
\clearpage
\begin{figure}[!t]
\includegraphics[scale=0.9,angle=180]{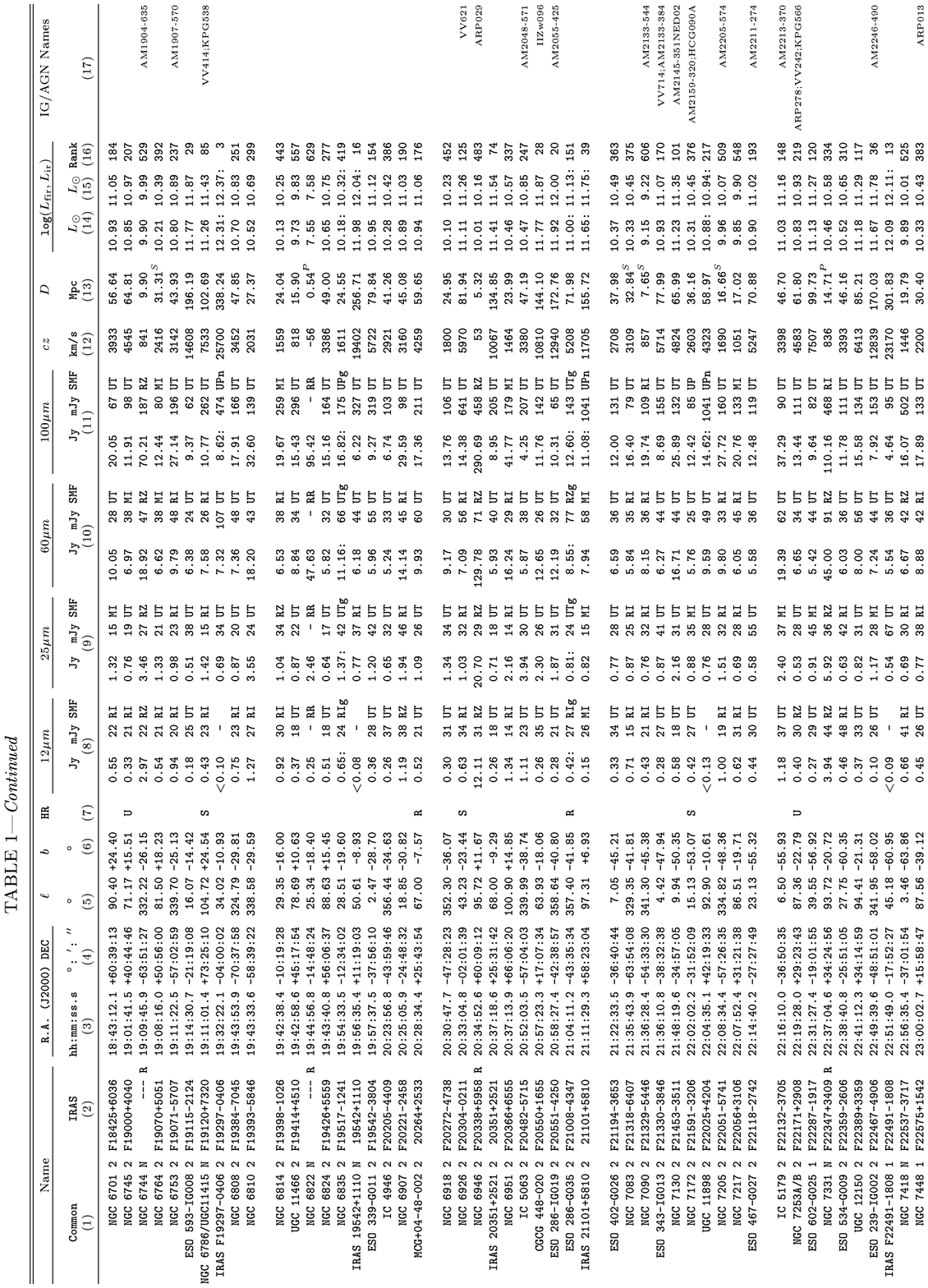}
\end{figure}
\clearpage
\begin{figure}[!t]
\includegraphics[scale=0.9,angle=180]{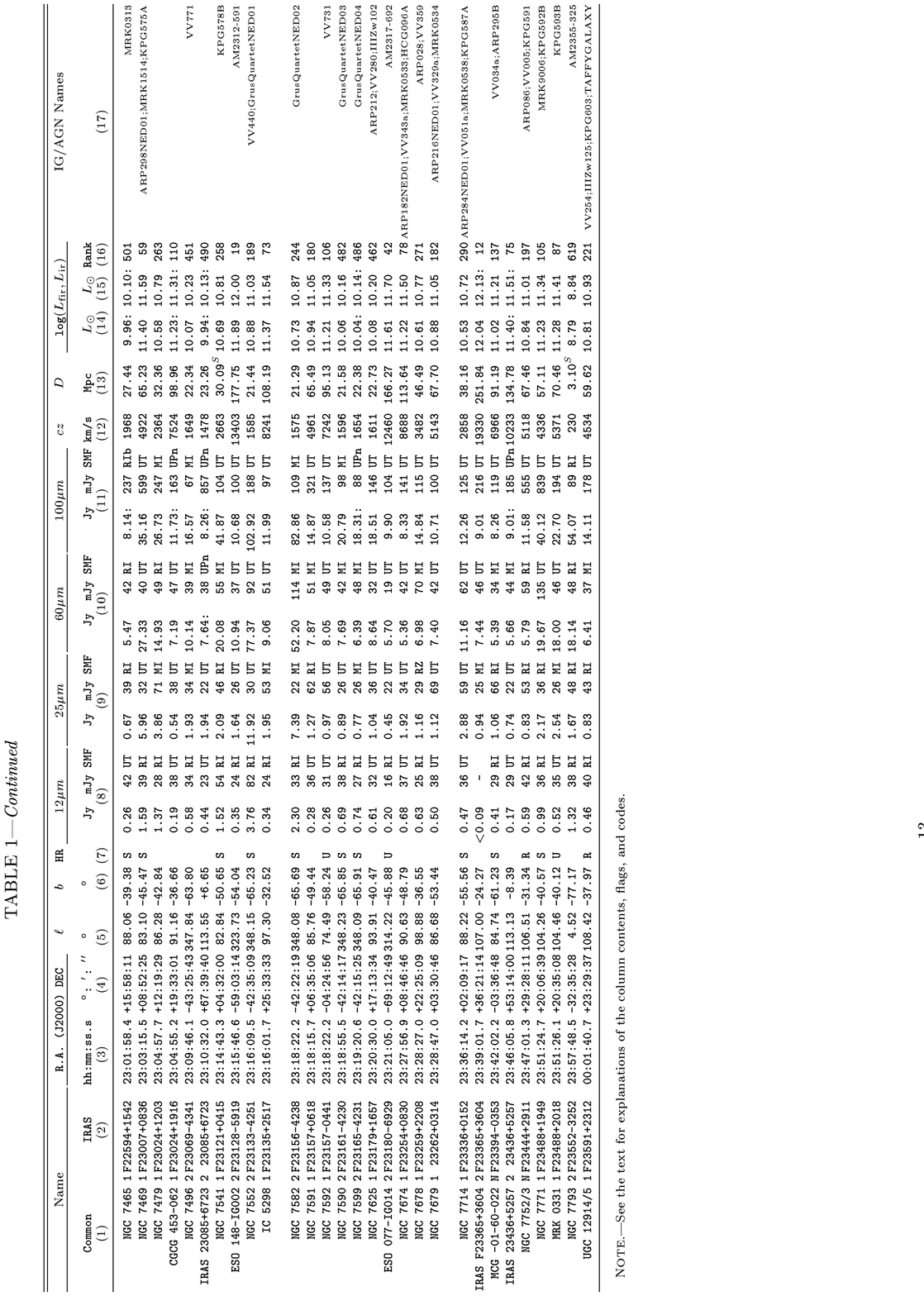}
\end{figure}
\clearpage
} 
\ifnum\Mode=2 \twocolumn \fi 

\fi 

\subsection{Annotated Images from the Digitized Sky Survey}

Figure 1 presents images extracted from the Digitized Sky Survey (DSS1)\footnote{
The Digitized Sky Survey was produced at the Space Telescope Science Institute
under U.S. Government grant NAG W-2166. The images of these surveys are based
on photographic data obtained using the Oschin Schmidt Telescope on Palomar
Mountain and the UK Schmidt Telescope. The plates were processed into the
present compressed digital form with the permission of these institutions.}
for all objects in the RBGS.  They are intended to quickly visualize the
optical morphology in context with the angular and metric size of each
object. Horizontal bars on the bottom of each plot show the angular scale
labeled in arcminutes, and vertical bars on the lower right side of each
plot show the metric scale labeled in kiloparsecs. Metric sizes are derived
using the source distance estimates listed in Table 1. The ellipses represent
3-sigma uncertainties in the \IRAS positions from the FSC and PSC.
Note that for the LMC and SMC, no ellipse is shown because there is no
point source from the FSC or PSC corresponding to the ``center'' 
of these large, diffuse galaxies. In general there is very good agreement
between the \IRAS catalog source position and the coordinates of the optical galaxy.
A notable exception is the position of \IRAS 10574-6603 taken from the PSC, 
which has a large offset from the optical galaxy ESO 093-G003.
The fact that this offset is in the direction of the major axis of
the \IRAS positional uncertainty ellipse (PA = 150\deg), and that SCANPI shows a
peak in the \IRAS emission that is coincident with the position
of the optical galaxy, confirms this cross-identification.
In some cases the IRAS source position is located between components
of optical pairs or groups of galaxies. This is a clear indication
that more than one galaxy contributes significantly to the \IRAS
emission.  Such objects (e.g., AM 0702-601, ESO 297-G011/12, IC
563/4, NGC 7465) are a subset of the pairs and groups investigated
using HIRES processing of the IRAS data (Surace, Sanders \& Mazzarella 2003),
which are flagged by ``(H)'' following the object names in Figure 1.

\ifnum\Mode=0 
\vskip 0.3in
\fbox{\bf Figure 1 (26 pages) goes here.}
\vskip 0.3in
\else
\ifnum\Mode=2 \onecolumn \fi 
\clearpage
{
\headheight     0.0in
\topmargin     -1.0in
\begin{figure}[!htb]
\figurenum{1}
\plotone{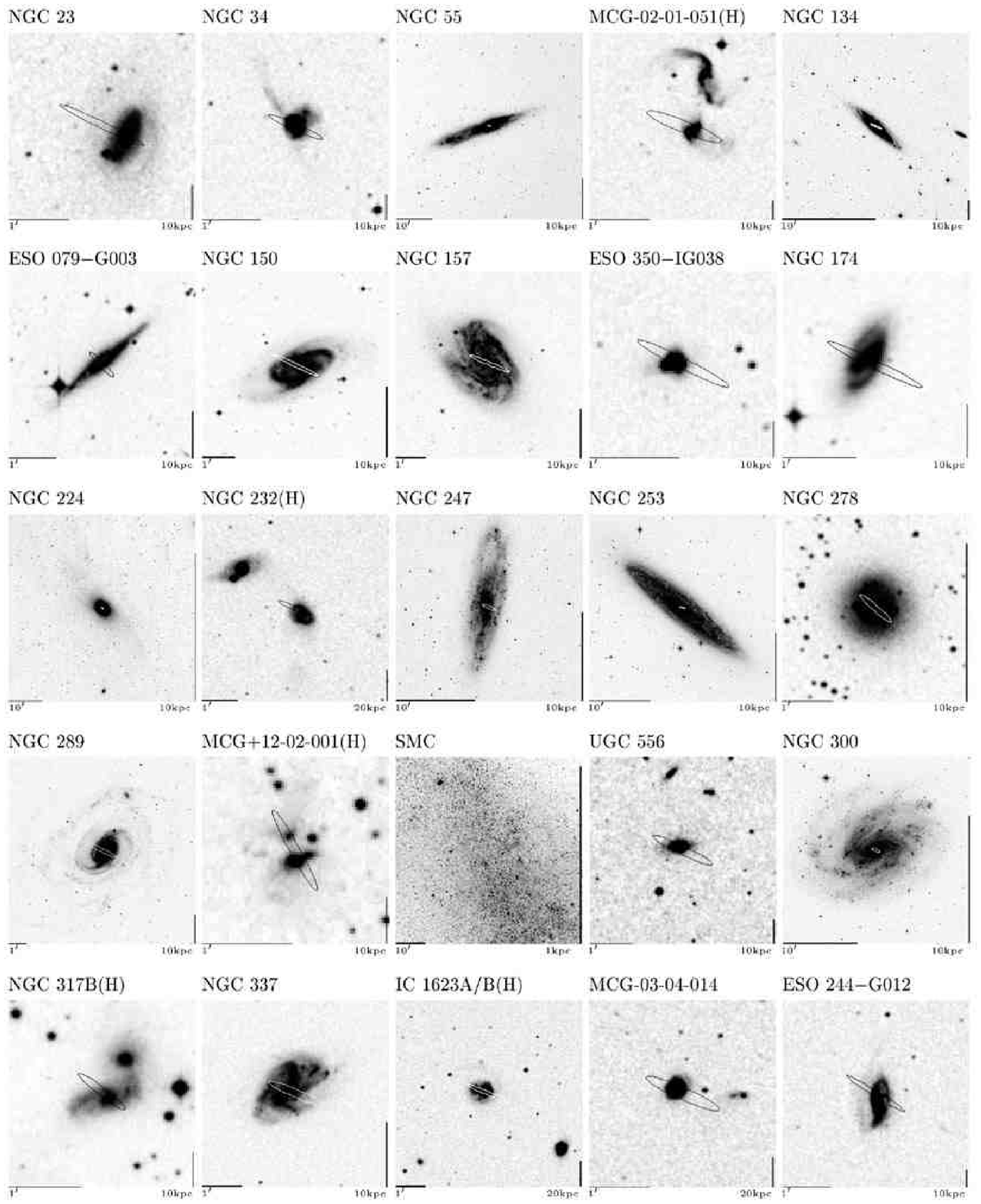}
\caption{\figcapOne}
\end{figure}
\clearpage

\ifnum\InsertAll>0  

\else 
\vskip 5.0in
\fbox{\bf Pages 1-26 of Figure 1:
{\it http://nedwww.ipac.caltech.edu/level5/March03/IRAS\_RBGS/Figures/} (Fig1p*.ps.gz).
}
\clearpage

\fi 

} 

\fi 


\subsection{Revisions from the Original BGS$_1$ and BGS$_2$}

Table 2 is a list of objects from the \BGS\ that have been omitted from the
RBGS because their revised 60{\ts}\um\ flux density is less than or equal
to 5.24 Jy.  The table caption explains why some of these objects were listed
with $S_\nu(60\mu m) \leq 5.24$ Jy in BGS$_1$.

\ifnum\Mode=0 
\vskip 0.3in
\fbox{\bf Table 2 goes here.}
\vskip 0.3in
\else
\tableTwo
\fi 

Table 3 is a list of objects from the \BGS\ which have
revised cross-identifications in various catalogs. These corrected, revised
cross-identifications were determined by visualizing the \IRAS uncertainty
ellipses superimposed on the DSS images (Figure 1), and by using 
current information available in NED.

\ifnum\Mode=0 
\vskip 0.3in
\fbox{\bf Table 3 goes here.}
\vskip 0.3in
\else
\tableThree
\fi 

Table 4 is a summary of the 39 objects which are new members of the RBGS.
These objects were missed during compilation of the \BGS\ for a
variety of reasons. Their presence in the RBGS is the result of comprehensive
searches utilizing NED in conjunction with queries of the \IRAS catalogs,
followed by examination of the \IRAS scan data using SCANPI.

\ifnum\Mode=0 
\vskip 0.3in
\fbox{\bf Table 4 goes here.}
\vskip 0.3in
\else
\tableFour
\fi 

\clearpage
\ifnum\Mode=2 \twocolumn \fi 

\section{Discussion}

The main purpose for compiling the RBGS was to produce a more accurate
and complete list of \IRAS bright galaxies, one that both incorporates the
improved final calibration of the \IRAS Level{\ts}1 Archive and one that
makes use of the best available SCANPI tools to more accurately compute
total \IRAS flux densities in all four \IRAS wavebands. Just as important was
the desire to provide a more accurate assessment of the various sources
of uncertainty (e.g. source confusion, cirrus contamination, etc.) in the
final list of tabulated flux densities.  This was not always clear in the
earlier published versions of the \BGS.  The reader can use the flags in
the RBGS tables to decide when it may be desirable to use the SCANPI tool
available through the Infrared Science Archive (IRSA) at IPAC \footnote{See
{\it http://irsa.ipac.caltech.edu/}.} for a direct visual inspection of
the scan profiles.  Also, rather than leave it to the reader to compute
infrared luminosities for each source, we have attempted to compile the most
recent and accurate redshifts from the literature.  We then list the adopted
source distances which have been used to compute total far-infrared (using
\IRAS bands 3 and 4) and infrared (all 4 \IRAS bands) luminosities according to
standardized prescriptions that are now widely adopted in the literature
(e.g., Sanders \& Mirabel 1996).

We begin the discussion of the RBGS data by first providing a detailed
comparison of how the ``new" \IRAS flux densities compare with the ``old"
previously published values.  This is followed by a discussion of the survey
sky coverage and distribution of sources on the sky, plus a discussion of
the completeness of the survey in all four \IRAS bands.  Various properties
of the RBGS are then discussed, ending with the presentation of the new
infrared luminosity function for \IRAS bright galaxies selected at 60{\ts}\um.

\subsection{Comparison Between Revised and Previous Flux Density Measurements}

Figure 2 shows the ratio of the new RBGS total flux density measurements to
the previously published \BGS\ measurements versus the new total flux density
measurements in each \IRAS band.  In general the largest differences occur at
the low end of the range of measured flux densities in all four \IRAS bands,
with the 12{\ts}\um\ and 25{\ts}\um\ bands showing the most dramatic changes, up to a
factor of $\pm$2 in the few most extreme cases.  Much of these differences
can be accounted for simply by more mature data processing, which had the
greatest effect near the survey lower limits in each of the \IRAS bands.
At 60{\ts}\um\ and 100{\ts}\um, where the measured fluxes were often substantially
above the \IRAS FSC survey limits, the maximum change is typically a more
modest factor of $\pm$30\%.  At flux densities more than a factor of 2 above
the \IRAS FSC limits the ``new" and ``old" values differ by typically $<$15\%.
There is a noticeable tendency for the revised flux density measurements to
be systematically higher, on average $\sim$5\%, among objects brighter than
about 35 Jy at 60{\ts}\um, and across the entire range of observed flux densities
at 100{\ts}\um.  This is due to a better understanding of the fact that for many
of the extended sources with high signal-to-noise ratios, some of the flux
extends beyond the previously adopted $f_\nu (t)$ aperture size,
and is better measured by using $f_\nu (z)$. In addition, the original
processing used for \BGS\ based the choice of whether to use $f_{\nu}(t)$
instead of $f_{\nu}(template)$ only on the coadded scan profile widths
(a comparison of FWHM and the width at 25\% peak to nominal values observed 
for pure point sources). However, many galaxies have profile widths which are
not significantly broader than what is expected for a point source, yet
there is extended emission in a faint \lq\lq pedestal\rq\rq\ that can be
reliably measured as a statistically significant excess of the $f_{\nu}(t)$
aperture value over the $f_{\nu}(template)$ point source fitted value;
see the Appendix (Fig. 16) for further details.

\ifnum\Mode=0 
\vskip 0.3in
\fbox{\bf Figure 2 goes here.}
\vskip 0.3in
\else
\ifnum\Mode=2 \onecolumn \fi 
\begin{figure}[!htb]
\figurenum{2}
\ifnum\Mode=2 
  \epsscale{0.5} 
\else
  \epsscale{0.7} 
\fi
\includegraphics[scale=0.65,angle=270]{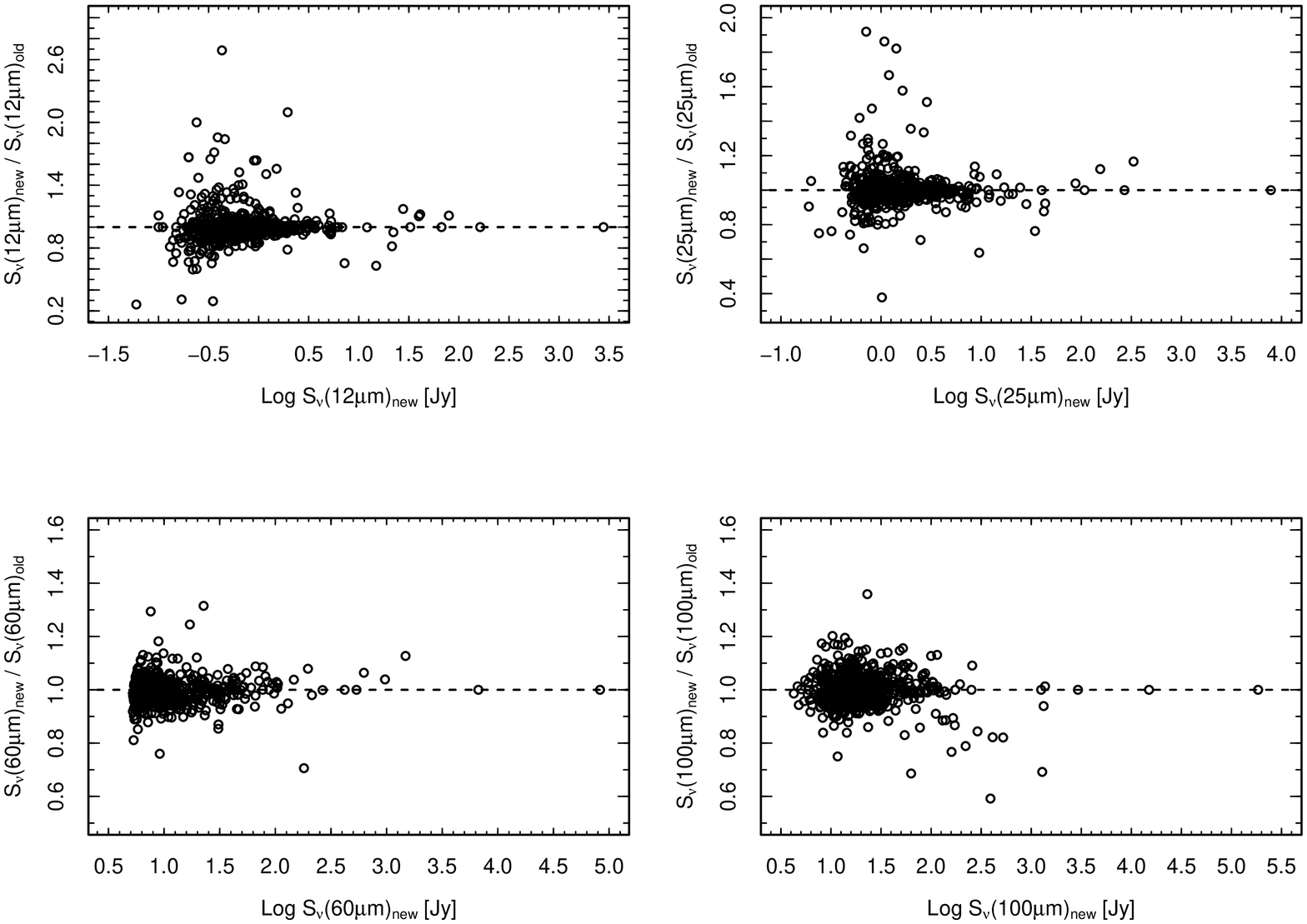}
\caption{\figcapTwo}
\end{figure}
\ifnum\Mode=2 \twocolumn \fi 
\fi 

All sources with extreme $\rm S_{\nu}(new)/S_{\nu}(old)$ flux ratios in Figure
2 were examined in detail, and they are explained by various improvements
in the revised processing. Some objects with $\rm S_{\nu}(new)/S_{\nu}(old)
< 0.8$ are cases where the RBGS flux densities have been estimated by using
SCLEAN  (see Appendix) or peak values to minimize confusion from companion
galaxies in pairs, nearby stars, or cirrus, where in \BGS\ the $f_\nu (t)$
method was used and therefore the flux densities quoted there were contaminated
(over estimated). Examples are NGC 5194 and NGC 5195, where SCLEAN was used
in RBGS to separate the components of this galaxy pair (M 51) at 12{\ts}\um\
and 25{\ts}\um. Another example is IRAS F16516-0948 at 100{\ts}\um\ ($\rm
S_{\nu}(new)/S_{\nu}(old) = 0.75$), where cirrus confusion has been minimized
by using the peak flux estimate in RBGS, and $f_\nu (t)$ was contaminated
(resulting in an over estimated flux density) in BGS$_2$.  Most objects
with $\rm S_{\nu}(new)/S_{\nu}(old) > 1.2$ are cases where the RBGS flux
selection algorithm resulted in the choice of $f_\nu (t)$ or $f_\nu (z)$ over
$f_{\nu}(template)$, where in \BGS\ a lower flux density estimate was made for
reasons explained in the previous paragraph (also see the Appendix). Examples
are NGC 3147 at 12{\ts}\um\ ($\rm S_{\nu}(new)/S_{\nu}(old) = 2.10$) and
{$\rm MCG +07-23-019$} at 25 {\ts}\um\ ($\rm S_{\nu}(new)/S_{\nu}(old) = 1.92$).
Other extreme ratios are due simply to differences between the $f_\nu (t)$
results obtained using the final (PASS3) \IRAS archive calibration versus
the earlier versions utilized in \BGS; an example is NGC 4565 at 60{\ts}\um\
($\rm S_{\nu}(new)/S_{\nu}(old) = 0.79$). Most of the remaining outliers in
Figure 2 are explained by the use of the SCANPI $f_\nu(z)$ measurement for all
objects smaller than 25 arcminutes in size, where in \BGS\ flux measurements
from Rice et al. (1983; 1988) were always used when available. Examples are
NGC 134 at 60{\ts}\um\ ($\rm S_{\nu}(new)/S_{\nu}(old) = 1.32$) and NGC 4631
at 100{\ts}\um\ ($\rm S_{\nu}(new)/S_{\nu}(old) = 0.77$).  As discussed in
Section 2, comparison of the Rice et al. measurements with SCANPI $f_\nu
(z)$ measurements for galaxies with optical sizes less than 25 arcminutes
showed relatively uniform scatter in the residuals, indicating that the use
of SCANPI $f_\nu (z)$, when significantly larger than $f_\nu(t)$, is the
best choice for these objects to insure uniformity and consistency in the
calibration with the rest of the RBGS objects.

Figure 3 shows the ratio of total flux density to the peak flux density in
the coadded scans at 12{\ts}\um, 25{\ts}\um, 60{\ts}\um\ and 100{\ts}\um.
The $f_{\nu}(peak)$ value is used here rather than $f_{\nu}(template)$ because
the latter measurement does not exist for objects in which the point source
template (PSF) fit failed, while for pure point sources $f_{\nu}(total)\ $\app\ 
$f_{\nu}(template)$ and thus the ratio $\rm S_{\nu}(total)/S_{\nu}(peak)$ is very
close to unity.  This figure illustrates the degree to which point-source
fitted measurements in the \IRAS PSC and \IRAS FSC underestimate the total
flux densities for objects in the RBGS.  There are likely numerous errors in
the literature concerning the infrared flux densities and infrared colors
of galaxies, due to the fact that some users of \IRAS data have not fully
appreciated the fact that most bright infrared galaxies in the local universe,
as represented here in the RBGS, are marginally extended or resolved by IRAS.

\ifnum\Mode=0 
\vskip 0.3in
\fbox{\bf Figure 3 goes here.}
\vskip 0.3in
\else
\begin{figure}[!hbt]
\figurenum{3}
\ifnum\Mode=2 
  \includegraphics[scale=0.6,angle=270]{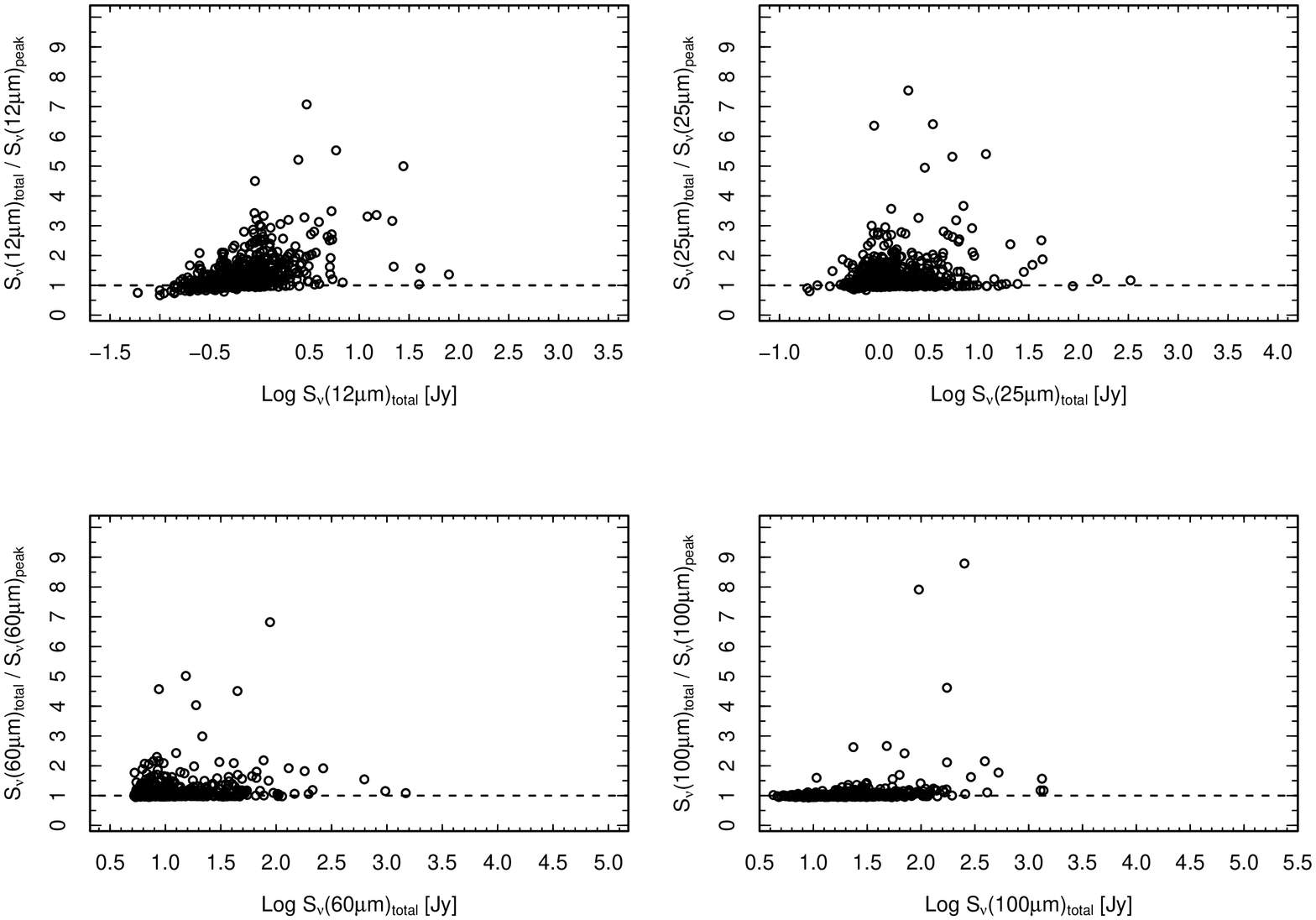}
\else
  \includegraphics[scale=0.65,angle=270]{TotalToPeak.ps}
\fi
\caption{\figcapThree}
\end{figure}
\fi 

A summary of the percentages of sources that were found to be extended in
each of the \IRAS bands in given in Table 5.  The most notable result is that
at 60{\ts}\um\ and 100{\ts}\um, where S/N is highest and distinctions can be reliably made, 
there are significantly more
resolved or marginally resolved objects than previously thought:\
   48\% in the RBGS versus 45\% as previously reported in the \BGS\
   at 60{\ts}\um, and 30\% in the RBGS versus 23\% as previously reported in
   the \BGS\ at 100{\ts}\um.
This is due to a more careful definition of resolved or marginally
extended objects as those having significantly more flux between the
baseline zero-crossings, $f_\nu(z)$, than within the nominal detector size,
$f_\nu(t)$, in combination with a comparison of W25 and W50 measurements to
point-source values.  We should emphasize that the 
\BGS\ used only
W25 and W50 to determine whether sources are resolved, and always chose $f_\nu(t)$
to estimate the flux for the R and M (U$+$) objects. 
The revised processing has resulted in significantly fewer objects with
underestimated total fluxes in the RBGS compared to \BGS.

\ifnum\Mode=0 
\vskip 0.3in
\fbox{\bf Table 5 goes here.}
\vskip 0.3in
\else
\tableFive
\fi 

\clearpage

\ifnum\Mode=2 \onecolumn \fi 

\subsection{\IRAS Flux Densities -- Completeness}

With the exception of 19 sources which were not detected at 12{\ts}\um, the
objects in the RBGS were detected in all four \IRAS bands.  Therefore, despite
the fact that the RBGS was selected only on the basis of 60{\ts}\um\ flux, the
flux distributions and infrared colors involving all four of the \IRAS bands
(see Table 1) are a fair representation of the true distributions of the
\IRAS properties of galaxies selected at 60{\ts}\um.

The distribution of fluxes in Table 1 can be compared with the distribution
$N(S_\nu) \propto S_\nu^{-1.5}$ expected for a complete sample of objects in
a non-evolving Euclidean universe that should be a reasonable approximation
for the relatively small redshift range covered by the objects in the RBGS.
Figure 4 shows the integral and differential log{\ts}$N$--log{\ts}$S_\nu$
plots for each \IRAS band.  The apparent turn over in the fainest bin of the
differential source counts suggests a possible incompleteness near the 60{\ts}\um\
sample flux limit.  However, this interpretation is based on the assumption
that the volume shell containing the bulk of these objects is as uniformly
filled with galaxies as the shells containing the brighter galaxies; studies
of large-scale structure indicate that such uniformity is not actually present.
In addition, the error bars plotted in Figure 4 are merely statistical
uncertainties, $\sqrt N$; they do not account for other possible errors.
The relatively constant slope of the number of sources versus
flux density at 60{\ts}\um\ down to the selection limit, with a power law
fit of $-1.48\pm 0.13$ in the integral counts, shows that at 60{\ts}\um\
the RBGS is reasonably complete to the selection limit of 5.24{\ts}Jy.

\ifnum\Mode=0 
\vskip 0.3in
\fbox{\bf Figure 4 goes here.}
\vskip 0.3in
\else
\begin{figure}[!htb]
\figurenum{4}
\ifnum\Mode=2 
  \includegraphics[scale=0.5,angle=-90]{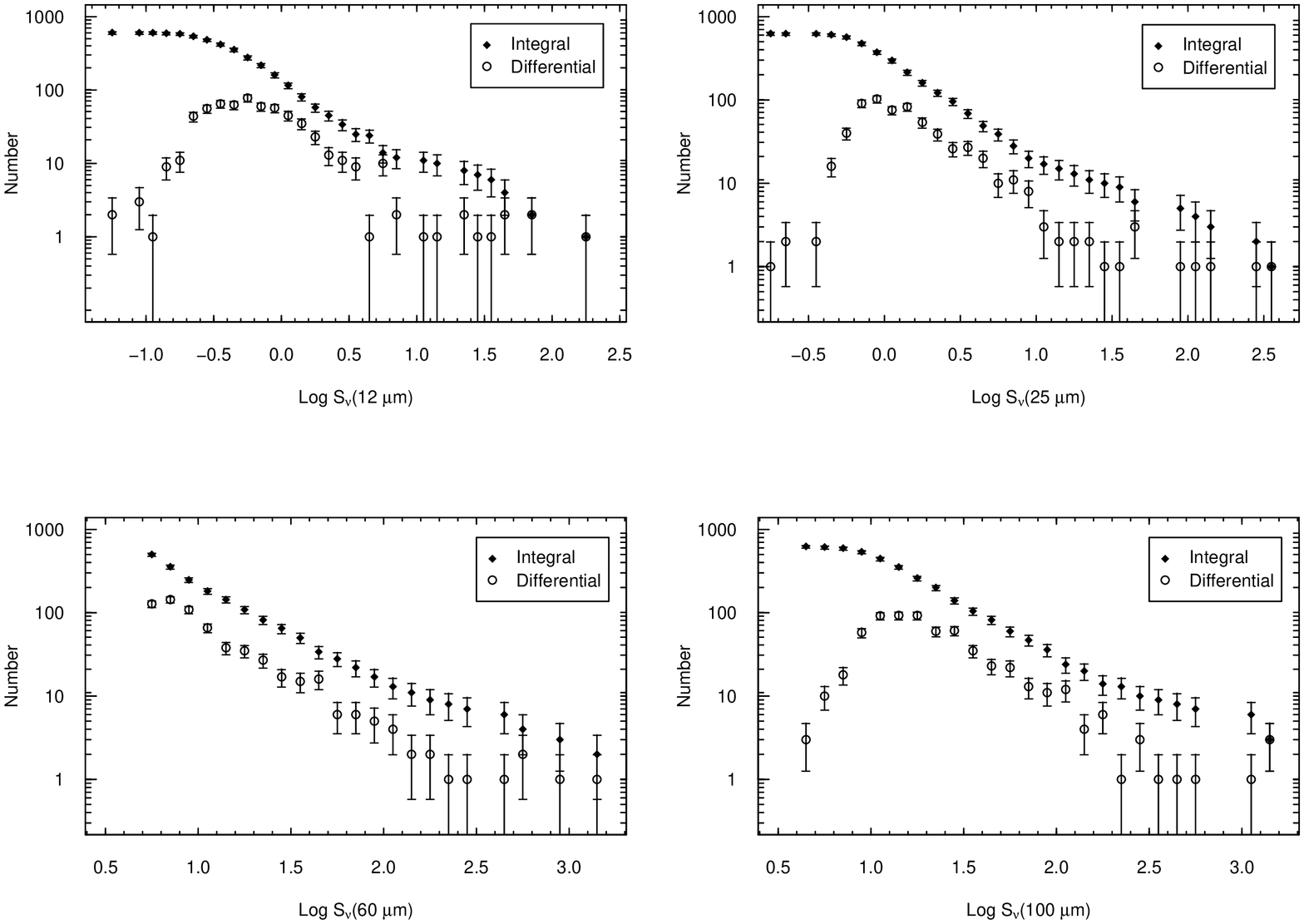}
\else
  \includegraphics[scale=0.6,angle=-90]{logNlogS.ps}
\fi
\caption{\figcapFour}
\end{figure}
\ifnum\Mode=2 \twocolumn \fi 
\fi 

At 12{\ts}{\ts}\um, 25{\ts}\um, and 100{\ts}\um, there is a portion of the
log{\ts}$N$--log{\ts}$ S_\nu$ plot that follows the $-1.5$ power-law relation,
suggesting that the RBGS sample contains a complete flux-density-limited
sample at those wavelengths to the turnover point in the plots (i.e. near
${\rm log}{\ts}S_\nu =$ $-0.1$, 0.0, and 1.1 at 12{\ts}\um, 25{\ts}\um, and
100{\ts}\um\ respectively).  The turnover point represents the flux-density
value beyond which a significant population of sources is being lost as a
result of the 60{\ts}\um\ selection criterion.  At 100{\ts}\um\ the turnover
is at a value $S_\nu \sim${\ts}16{\ts}Jy, nearly 10 times the completeness
limit of the PSC at 100{\ts}\um\ (\IRAS Explanatory Supplement 1988), and is
consistent with the rare occurrence of ``extremely cold" galaxies with $S_{100}
/ S_{60} > 3.5$.  At 25{\ts}\um\ the turnover point at 0.8{\ts}Jy is only
a factor of $\sim${\ts}2.2 above the PSC completeness limit at 25{\ts}\um,
and the turnover reflects a true loss of ``warm" \IRAS galaxies with $S_{25} /
S_{60} > 0.15$ from the RBGS.  Similarly at 12{\ts}\um\ where the turnover
at $\sim${\ts}0.75{\ts}Jy is again approximately twice the PSC completeness
limit at 12{\ts}\um, there is a true loss of objects with $S_{12} / S_{60}
> 0.1$ from the RBGS.

\subsection{Sky Coverage}

The final RBGS covered all of the sky surveyed by \IRAS except for a thin
strip within $\vert b \vert < 5$ degrees of the Galactic Plane.  Figure 5 displays a
Hammer-Aitoff projection in Galactic coordinates of all source positions
in the RBGS.  Also indicated in Figure 5 are three small regions suffering
heavy contamination from nearby Galactic molecular clouds (Orion: 
${\rm 5^h0^m < R.A. < 6^h40^m, -20^\circ < b < -10^\circ}$;
Ophiuchus: ${\rm 16^h15^m <}$ R.A. ${\rm < 16^h30^m, -26^\circ}$ $< \delta <
-20^\circ$)  and an area of confusion near the Large Magallenic Cloud (LMC:
${\rm 5^h01^m <}$ R.A. ${\rm < 6^h 11^m}$, $-72^\circ45^\prime < \delta <
-66^\circ45^\prime$).   These three regions were excluded from the BGS$_2$
(see Figure 1 of Sanders et al. 1995).  However, with the more thorough data
processing carried out for the RBGS, and the fact that the aerial density
of sources in these regions is not significantly different from regions
of high Galactic latitude, we now believe that most if not all \IRAS
bright galaxies within these regions have been identified.  Sources that
lie within these regions are included in Table 1 along with a flag
(an asterisk following the \IRAS source name in Column 2) indicating that they 
are located in a region of high Galactic foreground confusion.

\ifnum\Mode=0 
\vskip 0.3in
\fbox{\bf Figure 5 goes here.}
\vskip 0.3in
\else
\begin{figure}[!ht]
\figurenum{5}
\rotate
\ifnum\Mode=2 
  \includegraphics[scale=0.35,angle=0]{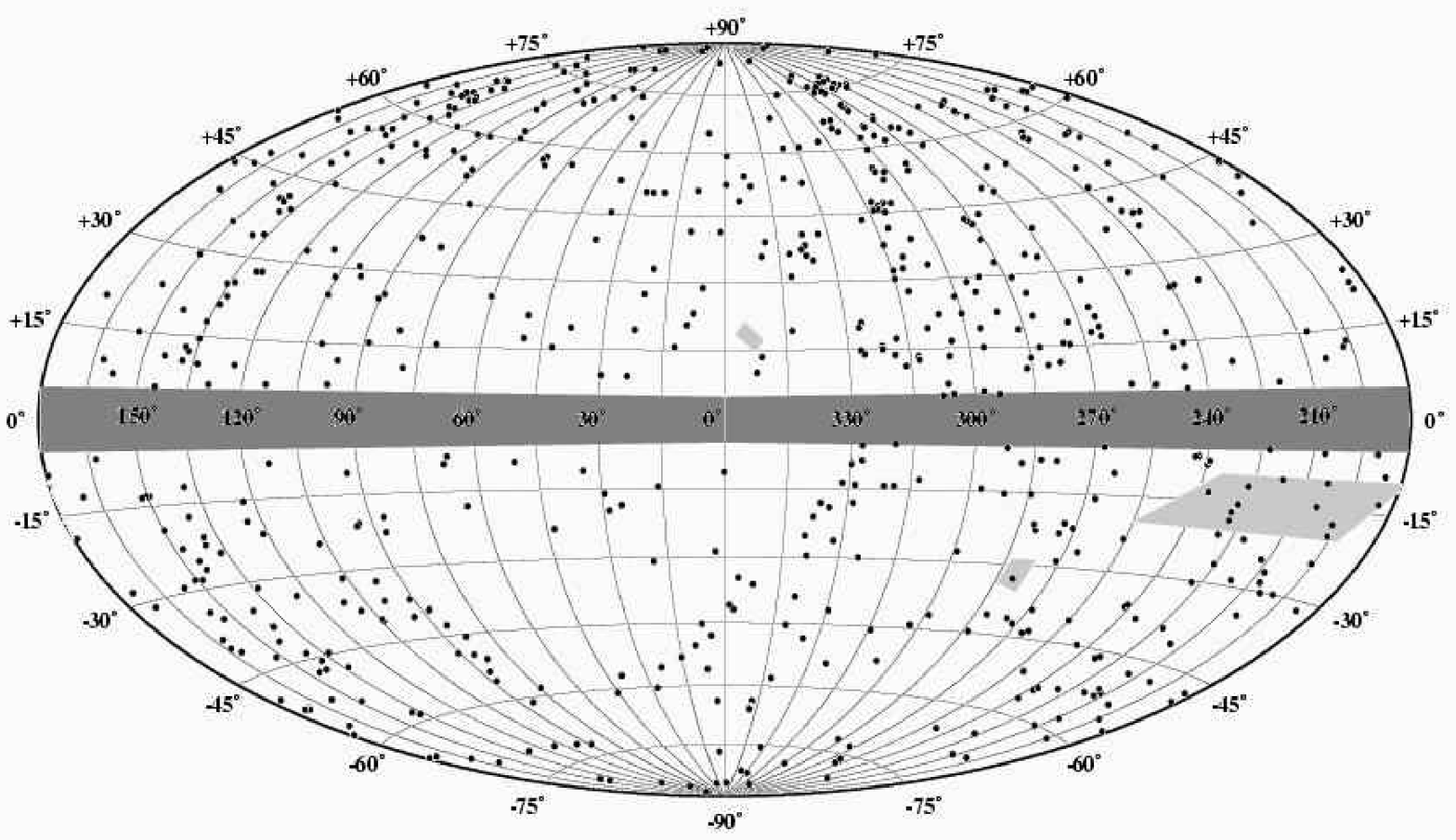}
\else
  \includegraphics[scale=0.55,angle=0]{hammer.ps}
\fi
\caption{\figcapFive}
\end{figure}
\fi 

In the BGS$_2$, Sanders et al. (1995) explored the effect that decreasing
Galactic latitude might have on our ability to detect all of the real
60{\ts}\um\ bright galaxies present.  On the basis of the fact that there was
no statistically significant decrease in the density of sources detected
versus $\vert b \vert$, it was concluded that the BGS$_2$ was essentially
complete down to the limit of $\pm 5^\circ$.   We have re-examined this
issue using the new RBGS data.  Figure 6 shows the surface density of RBGS
sources versus Galactic latitude.  Although the three bins (bin width $\Delta b =
5^\circ$) with the lowest surface density of sources all lie at $b <
40^\circ$, there is again no observed systematic decrease in source
density with Galactic latitude, and the three lowest points are not that
extreme.  However, we do not rule out the possibility that a few sources have
been missed at lower Galactic latitude, and in particular, that the
systematic low values at $b = -35^\circ$ to $-5^\circ$ may reflect this
fact.  What is most noticeable in the data of Figure 6 is the
large excess surface density of sources in the northern hemisphere
compared to the south. This reflects the effect of the Local
Supercluster (e.g., Tully 1982).

\ifnum\Mode=0 
\vskip 0.3in
\fbox{\bf Figure 6 goes here.}
\vskip 0.3in
\else
\begin{figure}[!htb]
\figurenum{6}
\ifnum\Mode=2 
  \epsscale{0.95}
\else
  \epsscale{0.6}
\fi
\plotone{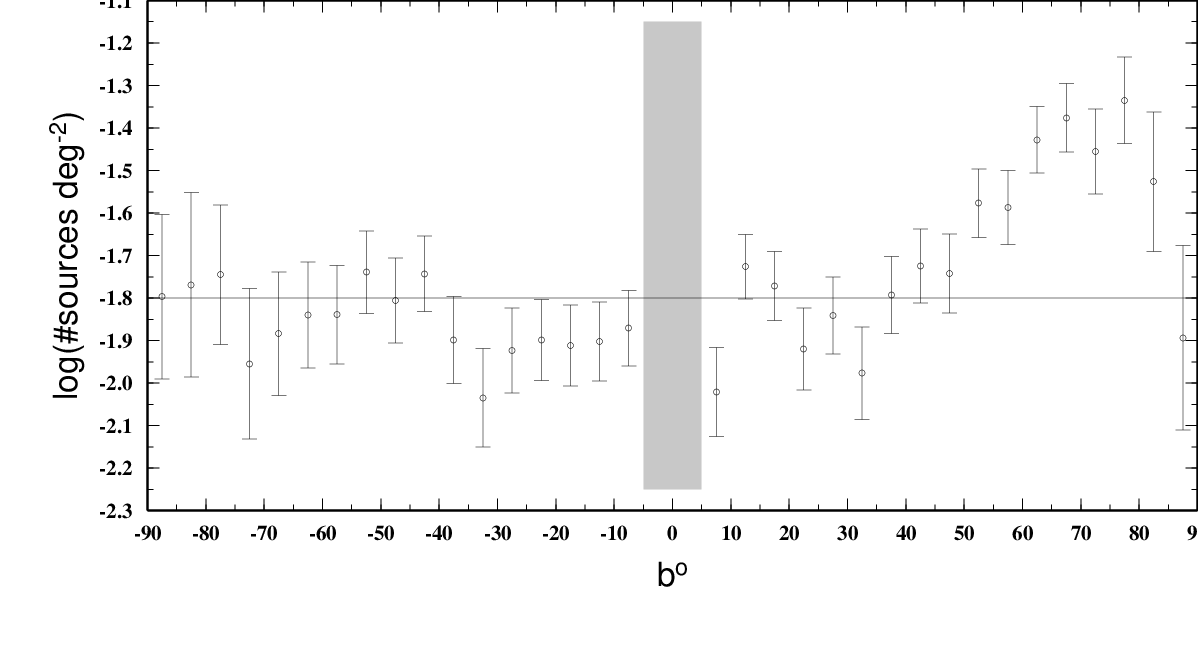}
\caption{\figcapSix}
\end{figure}
\fi 

\subsubsection{Clustering}

In the original discussion of the properties of \IRAS bright galaxies (i.e.
the 324 galaxies in the original BGS$_1$ -- Soifer et al. 1987) significant attention
was given to the effect of the nearby Virgo cluster, which was found to
be reflected in the BGS$_1$ sample as an ``over density" of sources due to
$\sim$30 galaxies within a $\sim 20^\circ$ radius around M87 (taken to be
the center of the Virgo cluster).  These ``Virgo galaxies" were shown to be
gas-rich spirals with relatively modest luminosity, i.e.
log{\ts}($L_{\rm ir}/L_\odot) \sim 9.8-10.4$.  While
nearly all of these galaxies remain in the RBGS, the overall effect of
Virgo on the full RBGS sample is less.  This is partly due to the $\sim
2\times$ increase in the total number of galaxies in the RBGS, but is
also due to the realization that Virgo is simply one (albeit the largest)
of several galaxy concentrations within the Local Supercluster (e.g. Tully
1982; Tully \& Shaya 1984).  We have chosen not to formally separate out
the properties of Virgo galaxies from the overall properties of the RBGS,
but instead will simply point out the effect of Virgo on the distribution
of RBGS galaxy properties where appropriate.

In the Aitoff all-sky plot shown in Figure 5, one of the most obvious features
is the asymmetry in the relative number of \IRAS bright galaxies found in the
north (370) versus the south (259).  Tully (1982) has previously shown that
this asymmetry is also found for optically selected galaxies, and that it
can be accounted for by a relatively small number of ``concentrations" within
the Local Supercluster, which lies almost entirely at high Galactic latitude.
The cumulative affect of the Local Supercluster produces the prominent broad
peak in the surface density distribution of \IRAS bright galaxies versus Galactic
latitude ($b \sim +50^\circ ~{\rm to} +85^\circ$) in Figure 6.  

\subsection{\IRAS Colors}

\ifnum\Mode=0 
\vskip 0.3in
\fbox{\bf Figure 7 goes here.}
\vskip 0.3in
\else
\begin{figure}[!htb]
\figurenum{7}
\ifnum\Mode=2 
  \epsscale{0.95}
\else
  \epsscale{0.6}
\fi
\plotone{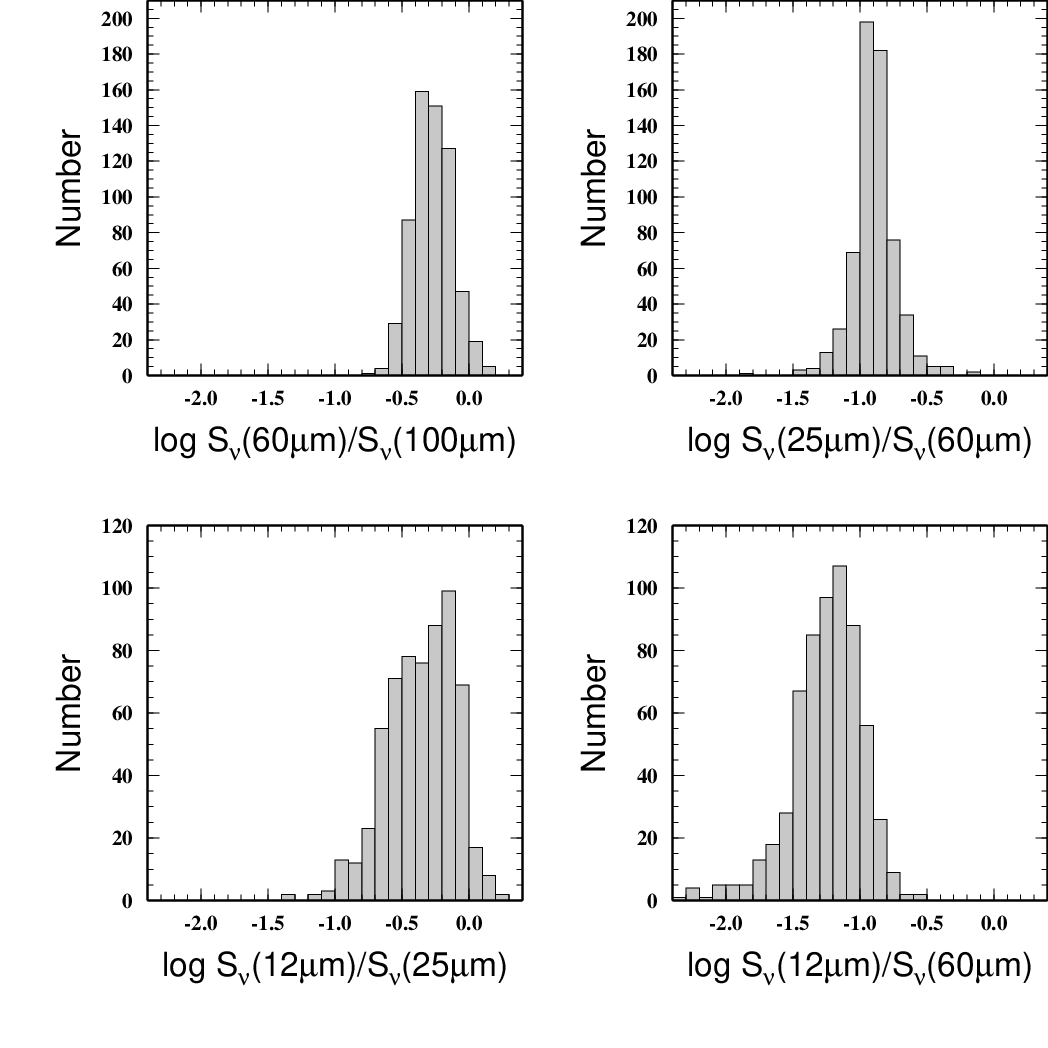}
\caption{\figcapSeven}
\end{figure}
\fi 

A full discussion of the \IRAS colors for this sample is beyond the scope
of this paper. However, it is worth pointing out here that the primary
statistical trends reported previously in BGS$_1$ (Soifer et al. 1989)
have not changed substantially when using the more accurate RBGS fluxes.
We discuss the \IRAS colors of the RBGS objects briefly below.

Figure 7 shows the distributions of the RBGS \IRAS flux density ratios
observed in individual sources. The largest range is found for the
ratios 12\um/25\um\ and 12\um/60\um\ ($\sim${\ts}1.5{\ts}dex), while the ratios
60\um/100\um\ and 25\um/60\um\ cover a narrower range ($\sim${\ts}0.8{\ts}dex).

{\it All} of the galaxies in the RBGS exhibit SEDs whose infrared emission
increases between 25{\ts}\um\ and 60{\ts}\um\ with the bulk of the sample having
$S_{60}/S_{25}$ in the range $\sim 5-15$.  This simply reflects the
dominance of thermal emission from relatively ``cool" dust (i.e. $T_{\rm
dust} \sim 25 - 70 K$) in the infrared SEDs of the 60{\ts}\um\ selected \IRAS
bright galaxies.  This range of dust temperature is also consistent with the range
of observed $S_{100}/S_{60}$ ratios ($\sim 4 - 0.5$).  Several
authors have now shown that this latter ratio is correlated with infrared
bolometric luminosity, with the most luminous \IRAS sources having the
largest values of $S_{60}/S_{100}$ (see Sanders \& Mirabel 1996 for references).

\ifnum\Mode=0 
\vskip 0.3in
\fbox{\bf Figure 8 goes here.}
\vskip 0.3in
\else
\begin{figure}[!htb]
\figurenum{8}
\ifnum\Mode=2 
  \epsscale{0.95}
\else
  \epsscale{0.4}
\fi
\plotone{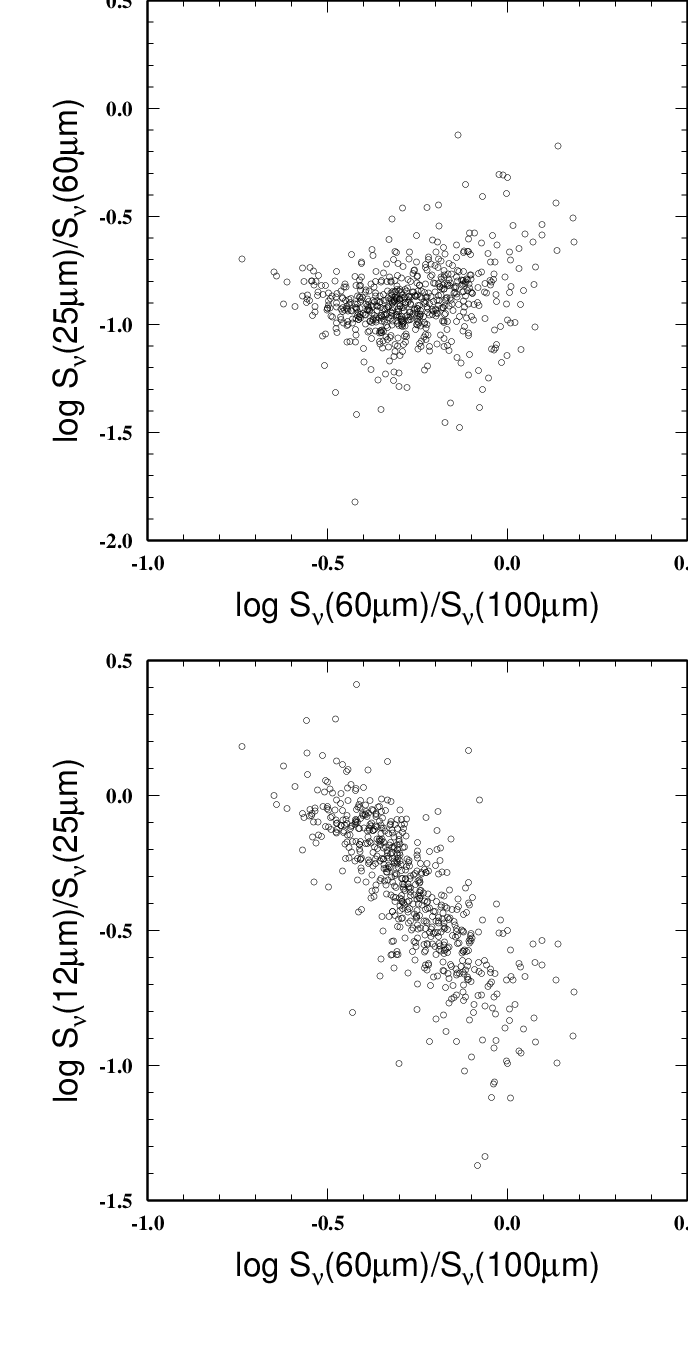}
\caption{\figcapEight}
\end{figure}
\fi 

Not all of the observed \IRAS flux ratios can be simply understood in terms of a single 
dust temperature component.
For example, the relatively flat distribution found for $S_{25}/S_{60}$
versus $S_{60}/S_{100}$ (Figure 8, top panel) likely reflects the combined
effects of more than one prominent dust temperature component. 
Although most galaxies have a dominant ``cool" dust component as discussed above, a
significant fraction also show a secondary ``warm" dust component that typically peaks near
25{\ts}\um\ and that appears to be associated with the presence of
Seyfert/AGN activity (e.g. Miley et al. 1984; deGrijp et al. 1985).

One of the more surprising results concerning the \IRAS colors of 
galaxies in the RBGS is seen in the inverse correlation between the
60\um/100\um\ ratio and the  12\um/25\um\ ratio (Figure 8, bottom panel).  
Soifer \& Neugebauer (1991) previously showed that this ratio was also strongly correlated with
galaxy infrared luminosity (see their Figs. 5 and 6).  One plausible ``simple" interpretation 
that has been proposed is that the component of ``hot" ($T \sim
100-200${\ts}K) dust found in the SEDs of most ``normal" spiral disks decreases due 
to the increasing destruction rate of small grains as the intensity of the nuclear radiation field 
increases with increasing total galaxy infrared luminosity.

\subsection{Luminosity Function}

An appropriate quantity for comparing the RBGS galaxies with other classes
of extragalactic objects selected at other wavelengths is 
the infrared luminosity, $L_{\rm ir} (8-1000\mu m)$, computed using all four
\IRAS bands (Soifer et al. 1987; Sanders \& Mirabel 1996).  Soifer et
al. (1987) first used the ``infrared bolometric luminosity" to compare \IRAS BGS
galaxies with the total bolometric luminosity for several optically selected galaxy samples 
(Seyferts, starbursts, QSOs, etc).  
Here we reconstruct the infrared bolometric
luminosity function using the new \IRAS measurements listed in Table 1 for the RBGS
galaxies.  All of the RBGS objects have measured redshifts, and all have
measured flux densities in all four \IRAS bands (except for a very small percentage of 
objects with upper limits at 12{\ts}\um). 
The space density of the galaxies, $\rho$, is the number of objects per
cubic megaparsec per unit absolute magnitude interval.
The units of $\rho$ are ${\rm Mpc}^{-3}\ M_{\rm ir}^{-1}$, where
$\rm M_{ir}$ signifies infrared absolute magnitude bins computed
using logarithmic intervals in which
each luminosity bin boundary is a factor of $10^{0.4}$ larger than the previous one
\footnote{\footnotesize
This effectively converts intervals of infrared luminosity ($\rm L_{ir}/L_{\odot}$)
to equivalent intervals of absolute magnitude ($\rm M_{ir}$); an alternate way
to express the units of $\rho$ is $\rm Mpc^{-3}\ [0.4*log_{10}(L_{ir}/L_{\odot})]^{-1}$.}.

Figure 9 plots the distribution of heliocentric radial velocities
($c*z$) for the complete RBGS using the redshifts tabulated in Table
1 taken from the references given in Table 7.  The sharp peak in the
1000-2000{\ts}km{\ts}s$^{-1}$ redshift bin is largely due to the Virgo cluster.
Otherwise, the redshift distribution for the RBGS shows a relatively smooth high 
redshift tail out to a cut-off near $cz \sim 26,000${\ts}km{\ts}s$^{-1}$.

\ifnum\Mode=0 
\vskip 0.3in
\fbox{\bf Figure 9 goes here.}
\vskip 0.3in
\else
\begin{figure}[!htb]
\figurenum{9}
\ifnum\Mode=2 
  \epsscale{0.95}
\else
  \epsscale{0.4}
\fi
\plotone{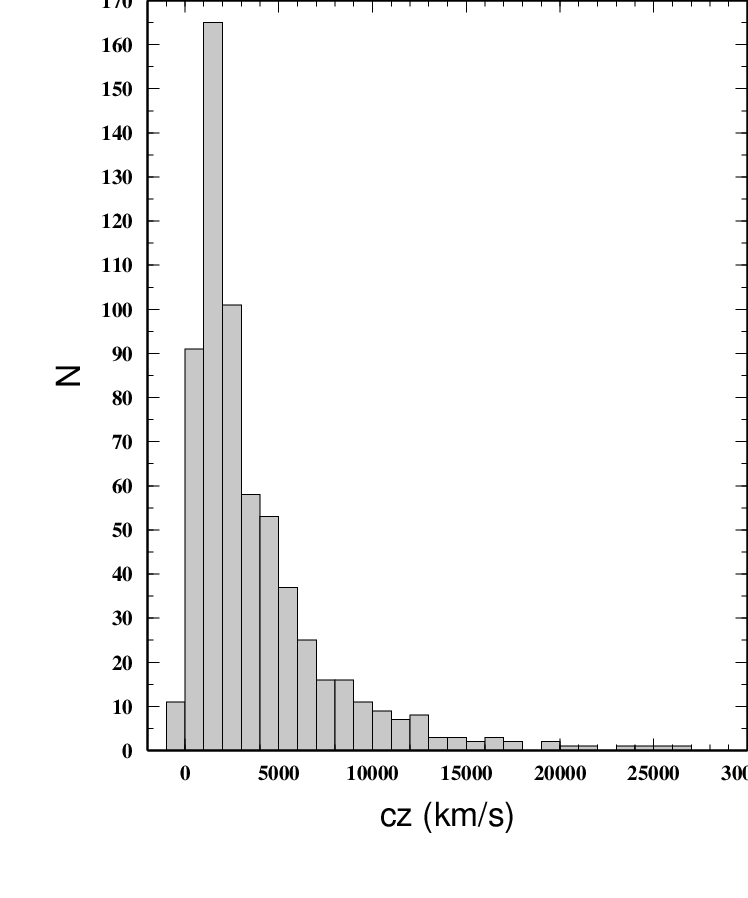}
\caption{\figcapNine}
\end{figure}
\fi 

Distances for the RBGS galaxies have been computed using the new cosmic
attractor flow model outlined in Appendix A of Mould et al. (2000), assuming
$H_{\rm o} = 75${\ts}km{\ts}s$^{-1}$Mpc$^{-1}$ and adopting a flat
cosmology, $\Omega_{\rm M} = 0.3$ and $\Omega_\Lambda = 0.7$.  Figure 10
plots the distribution of distances (Mpc) as tabulated in Table 1.  Again,
the effect of the Virgo cluster can be seen as affecting the strength of
the peak in the 10-20{\ts}Mpc bin (assuming our adopted distance to Virgo of 15.3{\ts}Mpc).

\ifnum\Mode=0 
\vskip 0.3in
\fbox{\bf Figure 10 goes here.}
\vskip 0.3in
\else
\begin{figure}[!htb]
\figurenum{10}
\ifnum\Mode=2 
  \epsscale{0.95}
\else
  \epsscale{0.5}
\fi
\plotone{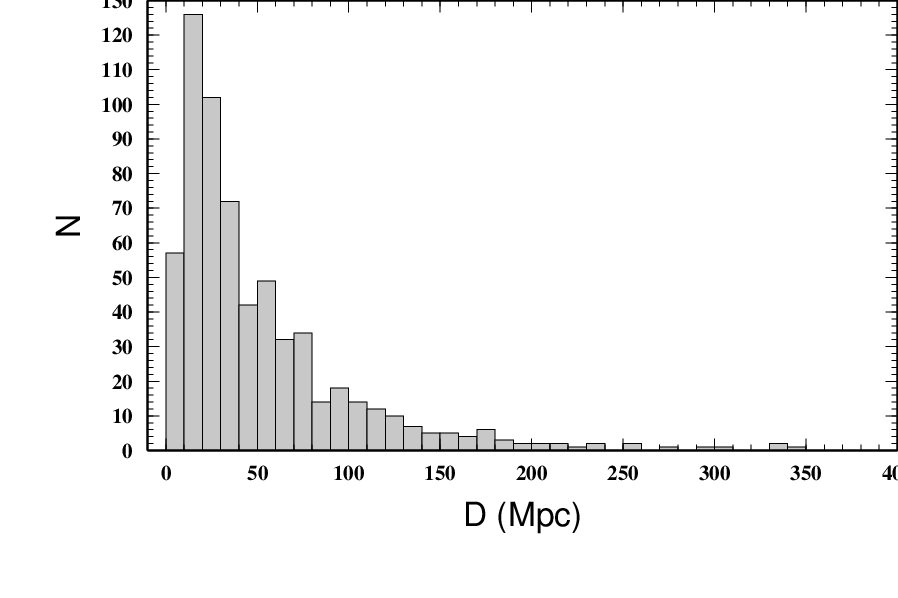}
\caption{\figcapTen}
\end{figure}
\fi 

The resulting distribution of infrared luminosities is shown in Figure 11.
The prescription and references used for computing $L_{\rm ir}$ are given
in the column notes to Table 1. Except for a modest excess of objects at
$L_{\rm ir} \sim 10^{10} L_\odot$ (largely due to Virgo) the distribution
shows a  relatively broad peak over the luminosity range log{\ts}($L_{\rm
ir}/L_\odot) \sim 9.8 - 11.4$ (half-power).  The median observed luminosity,
log{\ts}($L_{\rm ir}/L_\odot) \sim 10.65$, is somewhat larger than the total
bolometric luminosity of the Milky Way, and the maximum observed luminosity in
the sample, log{\ts}($L_{\rm ir}/L_\odot) = 12.51$ (Mrk 231), 
is nearly 100 times larger than the median.

\ifnum\Mode=0 
\vskip 0.3in
\fbox{\bf Figure 11 goes here.}
\vskip 0.3in
\else
\begin{figure}[!htb]
\figurenum{11}
\ifnum\Mode=2 
  \epsscale{0.95}
\else
  \epsscale{0.4}
\fi
\plotone{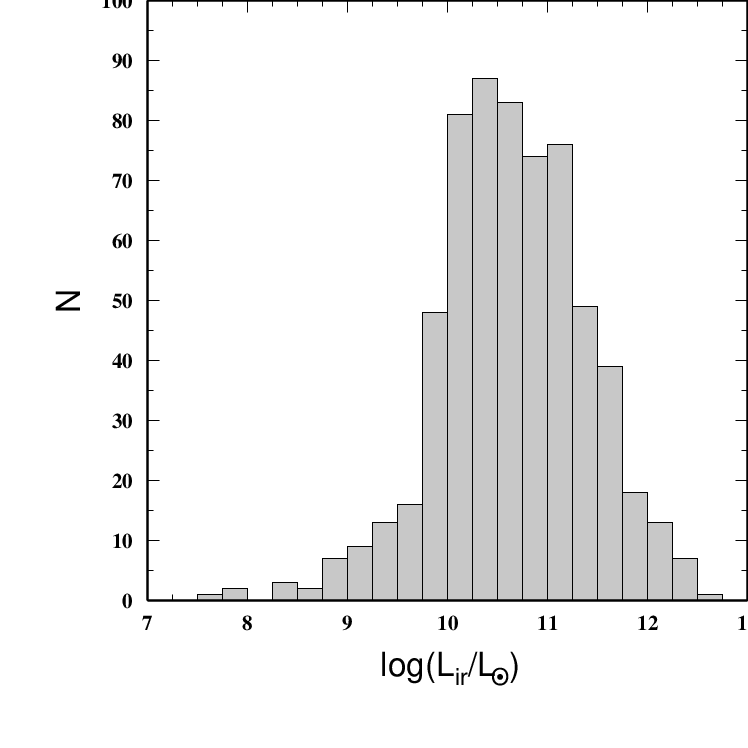}
\caption{\figcapEleven}
\end{figure}
\fi 

The luminosities plotted in Figure 11 were used to compute the infrared
bolometric luminosity function for the RBGS (Figure 12), using the $1/V_{\rm
max}$ method (Schmidt 1968).  The computed values are listed in Table 6.
The ``double power-law" shape of the luminosity function for \IRAS bright
galaxies is similar to that derived earlier for the BGS$_1$ (e.g. Soifer
et al.  1987), except for improved statistics at both low and high infrared
luminosities, plus the decreased influence of the Virgo cluster in the
all-sky sample as compared to its effect in the smaller BGS$_1$ survey.
The best fit power-laws, $\phi (L) \propto L^\alpha$, give $\alpha = -0.6\
(\pm 0.1)\ {\rm and}\ \alpha = -2.2\ (\pm 0.1)$ below and above $L_{\rm ir}
\sim 10^{10.5} L_\odot$ respectively.

\ifnum\Mode=0 
\vskip 0.3in
\fbox{\bf Figure 12 goes here.}
\vskip 0.3in
\else
\begin{figure}[!htb]
\figurenum{12}
\ifnum\Mode=2 
  \epsscale{0.95}
\else
  \epsscale{0.75}
\fi
\plotone{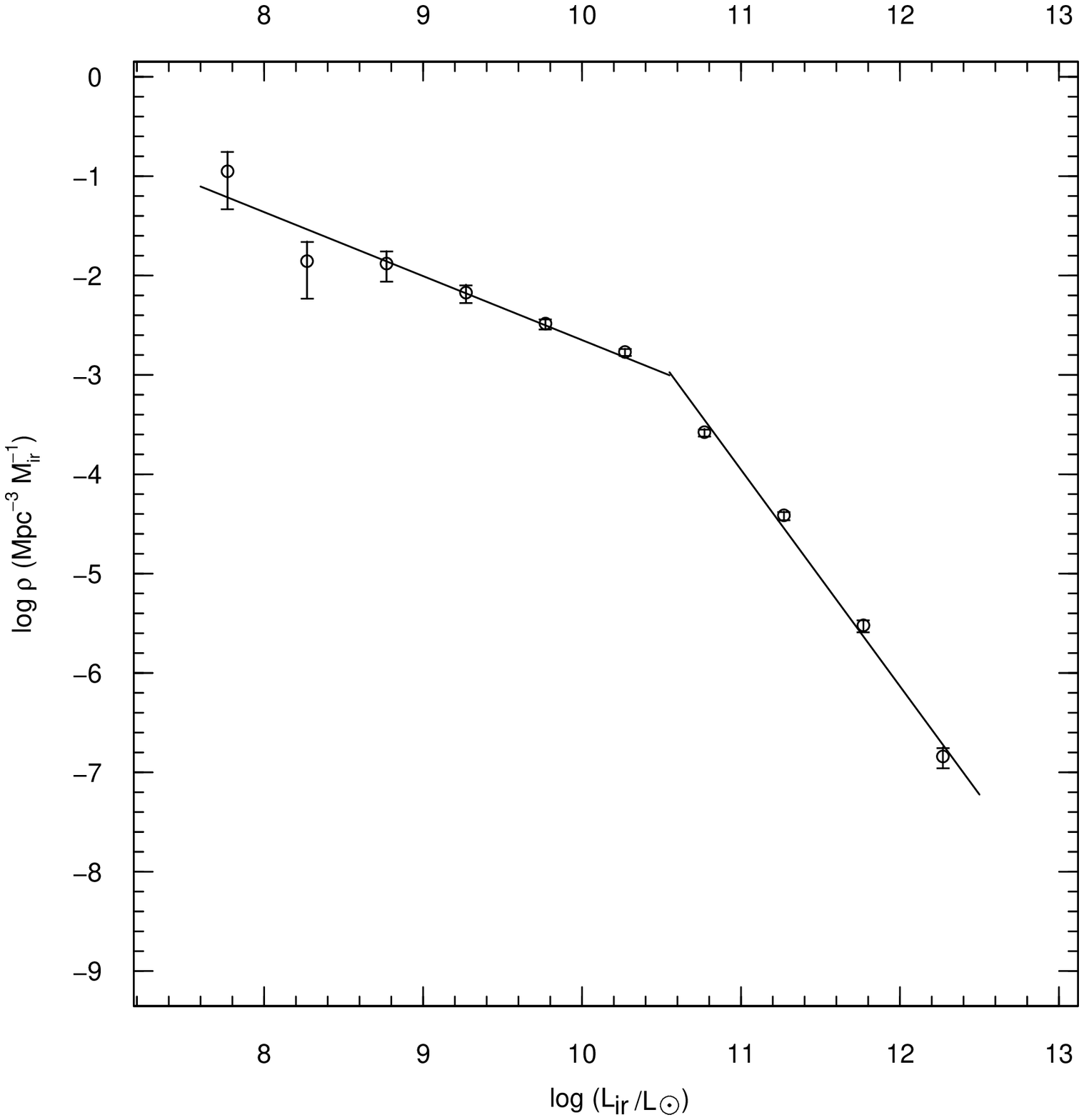}
\caption{\figcapTwelve}
\end{figure}
\fi 

\ifnum\Mode=0 
\vskip 0.3in
\fbox{\bf Table 6 goes here.}
\vskip 0.3in
\else
\tableSix
\fi 

\ifnum\Mode=1
 \clearpage
\fi
\section{Summary}

This paper presents the complete list of objects in the \IRAS Revised Bright
Galaxy Sample, a sample of extragalactic objects selected at 60$\mu$m
from the \IRAS all-sky survey.  The observed slope of $-1.49\pm0.10$ in the
log{\ts}$N$--log{\ts}$S_\nu$ relation at 60{\ts}{\ts}\um\ confirms that the RBGS
is statistically complete down to the limit of $S_{60} = 5.24${\ts}Jy.
The properties of the RBGS objects were computed using the final release
of the \IRAS Level 1 Archive and thus the RBGS replaces the earlier BGS
samples which were determined using older versions of the \IRAS data products.
The RBGS contains 39 objects which were not present in the \BGS, and 28 objects
from the \BGS\ have been dropped from the RBGS because their revised 60{\ts}\um\ flux
densities are not greater than 5.24{\ts}Jy.  Improved methods were used to
measure the total \IRAS flux densities of individual sources, resulting in typical
changes of 5-25\% when compared to previous values reported for the \BGS,
with changes of up to a factor of 2 for the faintest sources at 12{\ts}\um\ and
25{\ts}\um. Better procedures for resolving position uncertainties and resolving
cross-identifications with other galaxy catalogs resulted in name changes for
$\sim$7\% of the previous \BGS\ compilations. This work presents the
most accurate estimates to date of the total \IRAS flux densities and derived
infrared luminosities of galaxies in the local Universe.
Basic properties of the RBGS galaxies are summarized below.

\begin{enumerate}

\item
The RBGS sample contains a total of 629 galaxies with
$S_{60}{\ts}>{\ts}5.24${\ts}Jy in an area of 37,657.5 square degrees (91.3\%
of the sky) covering the entire sky surveyed by \IRAS down to Galactic latitude
$\vert b \vert = 5^\circ$;

\item
Extended flux ($>${\ts}0.77$^\prime$) at 12{\ts}\um, ($>${\ts}0.78$^\prime$)
at 25{\ts}\um, ($>${\ts}1.44$^\prime$) at 60{\ts}\um, and ($>${\ts}2.94$^\prime$)
at 100{\ts}\um\ is detected in 61\%, 54\%, 48\%, and 30\% of the galaxies
respectively. 

\item
The mean and median redshift for the entire RBGS sample is $z = 0.0126$ ($cz
= 3777${\ts}km{\ts}s$^{-1}$) and $z = 0.0082$ ($cz = 2458${\ts}km{\ts}s$^{-1}$),
respectively.  The object with the highest redshift is IRAS~$07251-0248 (z=0.0876)$,
and the object with the largest computed infrared luminosity in this local sample 
is Mrk 231 ($L_{\rm ir} = 3.2 \times 10^{12} L_\odot$).

\item
The bolometric infrared luminosity function, $\phi (L_{\rm ir})$, for infrared
bright galaxies in the Local Universe remains best fit by a double power
law, $\phi (L) \propto L^\alpha$, with $\alpha = -0.6 \pm 0.1, {\rm and}\
\alpha = -2.2 \pm 0.1$ below and above $L_{\rm ir} \sim 10^{10.5} L_\odot$,
respectively.

\end{enumerate}


\acknowledgments

We thank George Helou for helpful discussions, and the anonymous
referee for comments that improved the presentation of this material.
This research made extensive use of the \ifnum\Mode=2 \\ \fi 
NASA/IPAC Extragalactic Database (NED), which is operated by
the Jet Propulsion Laboratory, California Institute of Technology, under
contract with the National Aeronautics and Space Administration. Queries of
the \IRAS catalogs and scan coadd processing using SCANPI were supported by
the NASA/IPAC Infrared Science Archive (IRSA), which is operated by the Jet
Propulsion Laboratory, California Institute of Technology, under contract with
the National Aeronautics and Space Administration.  DBS acknowledges support
from a Senior Award from the Alexander von Humboldt-Foundation and from the
Max-Planck-Institut fur extraterrestrische Physik as well as support from NASA
grant NAG{\ts}90-1217. JMM, DCK, and JS were supported by the Jet Propulsion
Laboratory, California Institute of Technology, under contract with NASA.
DCK is grateful for financial support from KOSEF grant No. R14-2002-058-01000-0
and from the BK21 project of the Korean Government.
JS and BTS are supported by the SIRTF Science Center at the California
Institute of Technology; SIRTF is carried out at JPL, under contract with NASA.

\clearpage

\appendix

\centerline{APPENDIX}
\vskip 0.2in

\centerline{Data Reduction and Detailed SCANPI Measurements}
\vskip 0.2in

Many sources in the RBGS are near enough such that they appear resolved or
marginally extended in one or more of the \IRAS detector bands, while others
are unresolved. Therefore, an objective and consistent procedure had to
be developed to select the best estimate of the total flux density for
each object in each of the four \IRAS bands.  Table 7 lists the \IRAS SCANPI
measurements for all sources in the RBGS; these data were used along with
the coadded scan plots to determine the ``best" flux density estimates as
listed in Table 1. These measurements are from the SCANPI median (1002)
method of \IRAS scan coaddition (Helou et al. 1988).  Table 7 includes the
coaddition results from all four SCANPI methods (``zc", ``tot", ``template",
``peak") for each source in the RBGS.  Our automated processing methods
selected the final flux densities listed in Table 1 based on the relative
values from these four coadd methods, plus the use of important additional
information concerning source extent, and possible confusion due to blended
sources, companions, Galactic cirrus, or excessive noise.  Examples of how
these choices were made are illustrated in Figures 14-16, and
more thoroughly discussed in the captions to these figures.

The column entries in Table 7 are as follows:

({\bf 1}) {\it  Name} -- The Common name as listed in Table 1.

({\bf 2}) {\it  Redshift Reference} -- 19 digit reference code from NED for
the redshift listed in Table 1.

({\bf 3}) {\it  N/O} -- Ratio of the new \IRAS flux density estimate to the
old flux density estimate published previously in BGS$_1$ or BGS$_2$ at 12{\ts}\um; objects
new to the RBGS have missing values\\ (`` ---") in this and following columns.

({\bf 4}) {\it  zc} -- flux density from SCANPI's zero-crossing measurement,
$f_\nu (z)$ (Jy) at 12{\ts}\um.

({\bf 5}) {\it  tot} -- flux density from SCANPI's in-band total measurement,
$f_\nu (t)$ (Jy) at 12{\ts}\um.

({\bf 6}) {\it  temp} -- flux density from SCANPI's template amplitude
measurement, tmpamp (Jy) at 12{\ts}\um.

({\bf 7}) {\it  peak} -- flux density from SCANPI's peak measurement, peak (Jy) at 12{\ts}\um.

({\bf 8}) {\it  W25} -- scan profile full width (arcminutes) measured at 25\% of the peak signal at 12{\ts}\um.

({\bf 9}) {\it  W50} -- scan profile full width (arcminutes) measured at 50\% of the peak signal at 12{\ts}\um.

({\bf 10--16})  -- Same measurements as in columns (3) -- (9), but at $25{\mu}m$.

({\bf 17--23})  -- Same measurements as in columns (3) -- (9), but at $60{\mu}m$.

({\bf 24--30})  -- Same measurements as in columns (3) -- (9), but at $100{\mu}m$.

\ifnum\Mode=2 \onecolumn \fi 
\ifnum\Mode=0 
\vskip 0.3in
\fbox{\bf Table 7 goes here.}
\vskip 0.3in
\else
\clearpage
{
\headheight     0.0in
\topmargin     -1.0in
\begin{figure}[!t]
\includegraphics[scale=0.9,angle=180]{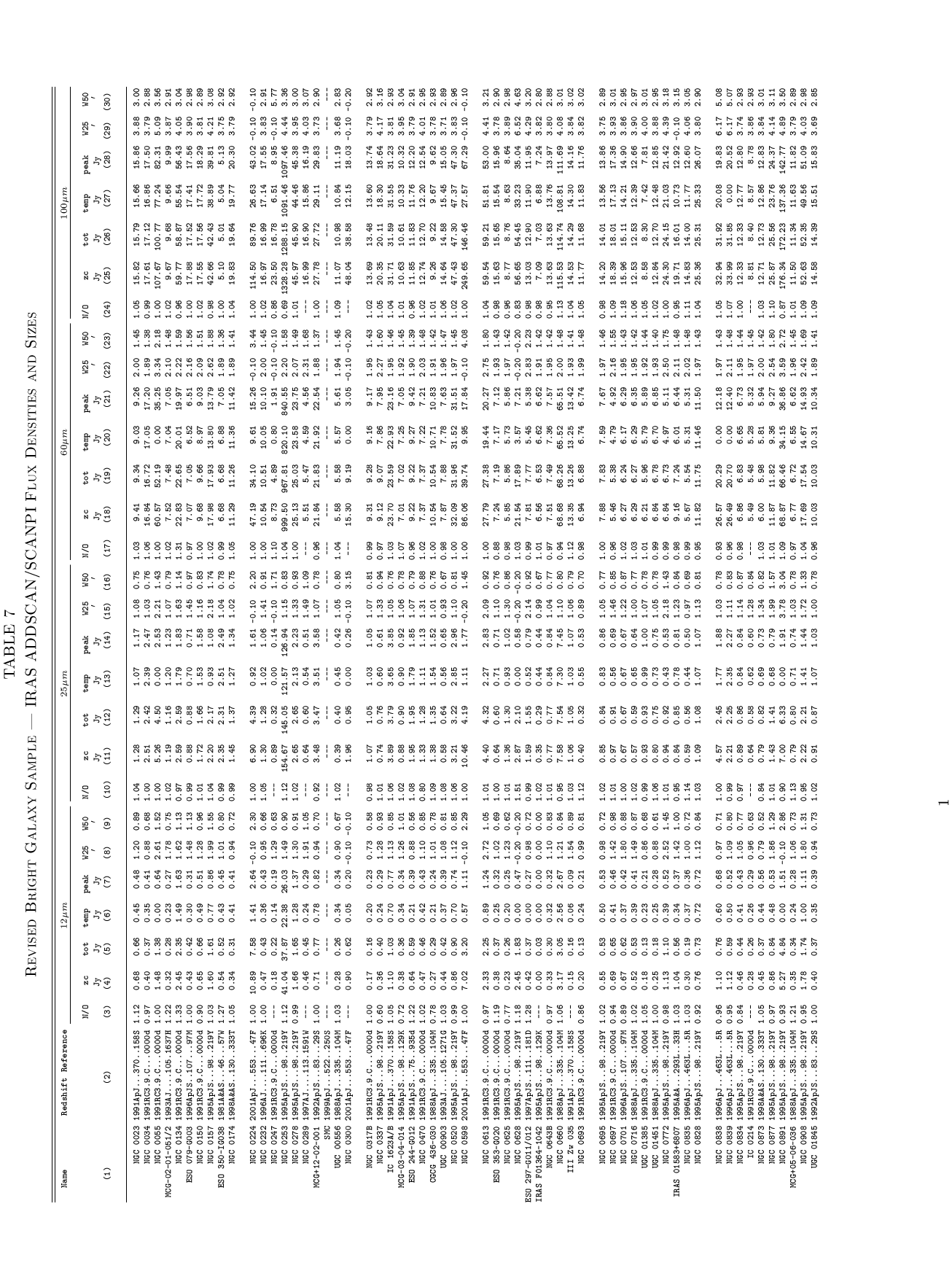}
\end{figure}
\clearpage
%
%
%
%
%
%
%
%
%
%

\vskip 5.0in
\fbox{\bf Pages 1-11 of Table 7:
{\it http://nedwww.ipac.caltech.edu/level5/March03/IRAS\_RBGS/Figures/Table7.ps.gz.}
}
\clearpage

} 
\fi 

For some objects, values of {\it temp} are 0.00 or values of {\it  W25}
and {\it  W50} are negative in Table 7. These are indications that the point
source template amplitude fit failed, which occurred for some very extended
objects best measured using the $f_{\nu}(z)$ flux estimator; see the example
of NGC 1532 at 25{\ts}\um\ as plotted in panel (a) of Figure 15.

Perhaps of most immediate interest to those familiar with the previous \BGS\
data are the {\it N/O} values given in columns (3), (10), (17) and (24) of Table 7.
Part of the differences in the ``new" versus ``old" flux densities is simply
that due to the improved ``Pass 3" calibration adopted for the final release
of the \IRAS Level{\ts}1 Archive.  However, a significant effect is that due
to the improved methods of estimating the total flux, in particular the
use of $f_\nu (z)$ when this flux measurement was significantly larger
than the other coadd methods due to extended emission not captured by
$f_\nu (t)$.  In addition, as mentioned in Section 4.1,
many galaxies have profile widths which are not significantly broader than
what is expected for a point source, yet there is extended emission in a
faint \lq\lq pedestal\rq\rq\ that can be easily seen in the profiles, and
reliably measured as a statistically significant excess of the $f_{\nu}(t)$
aperture value over the $f_{\nu}(template)$ point source fitted value.
Figure 13 shows the distribution of the ratio $f_{\nu}(t)/f_{\nu}(template)$
at 60{\ts}\um; also plotted are Gaussian fits intended to model the distribution
expected solely from noise in the relative $f_{\nu}(t)$ and $f_{\nu}(template)$
measurements for unresolved objects, data for a sample of comparably bright
stars, and 60{\ts}\um\ profiles for representative RBGS objects. This information
was used as follows to establish the threshold for when $f_{\nu}(t)$ could be
selected as a reliable, confident indicator of extended emission in excess
of the value measured by the point source template fit.

A sample of candidate stars was selected from the \IRAS Point Source Catalog
(PSC) using the joint criteria of $f_{\nu}(60\mu m) > 5.24~Jy$, a high point
source fit correlation coefficient ($99$\%), and positional association
with objects in one or more star catalogs. The resulting candidate list (195
objects) was further filtered by cross-identification of each \IRAS source
with known stars using information available in SIMBAD.  Some are planetary
nebulae or unknown object types and were therefore omitted for this purpose
of identifying a large sample of pure 60{\ts}\um\ point sources.  The remaining
sample of confirmed stars were then processed with SCANPI using the same
procedure as the RBGS objects. The distribution of the measured ratios of
$f_{\nu}(t)/f_{\nu}(template)$ for 121 confirmed bright stars with high
quality 60{\ts}\um\ \IRAS scans (e.g., not confused by cirrus, excessive noise,
or companions) is plotted in the same bins as the RBGS objects in Figure 13
(asterisks).  The presence of stars with ratios greater than $\sim$1.05 was
unexpected, and this complicated the goal of building a comparison sample of
pure \IRAS 60{\ts}\um\ point sources; close inspection of the data showed that each
of these objects have 60{\ts}\um\ scan profiles similar to the RBGS galaxy profiles
plotted in Figure 13, with faint \lq\lq pedestals\rq\rq\ of faint emission
under a dominating point source. Larger ratios correspond to higher
or more extended pedestals well above the background noise in the scans.
These are clear candidates for stars embedded in extended circumstellar dust
disks, shells or nebulae; some objects have published data that support this
interpretation, including spectral classifications such as carbon stars,
emission-line stars, and stars with known OH/IR envelopes. These objects are
not considered further in this paper, but their presence required omitting
stars outside the range 0.95 - 1.05 to form a Gaussian fit representative
of unresolved stars measured by SCANPI.  This fitted distribution of stars
(dotted line, with a mean of $1.000 \pm 0.004$) was then scaled to match the
peak of the Gaussian fit to the RBGS objects (solid line), and plotted as a
dashed line in Figure 13. 

\ifnum\Mode=0 
\vskip 0.3in
\fbox{\bf Figure 13 goes here.}
\vskip 0.3in
\else
\setcounter{figure}{13}
\begin{figure}[!htb]
\figurenum{13}
\ifnum\Mode=2 
  \epsscale{0.90}
  \plotone{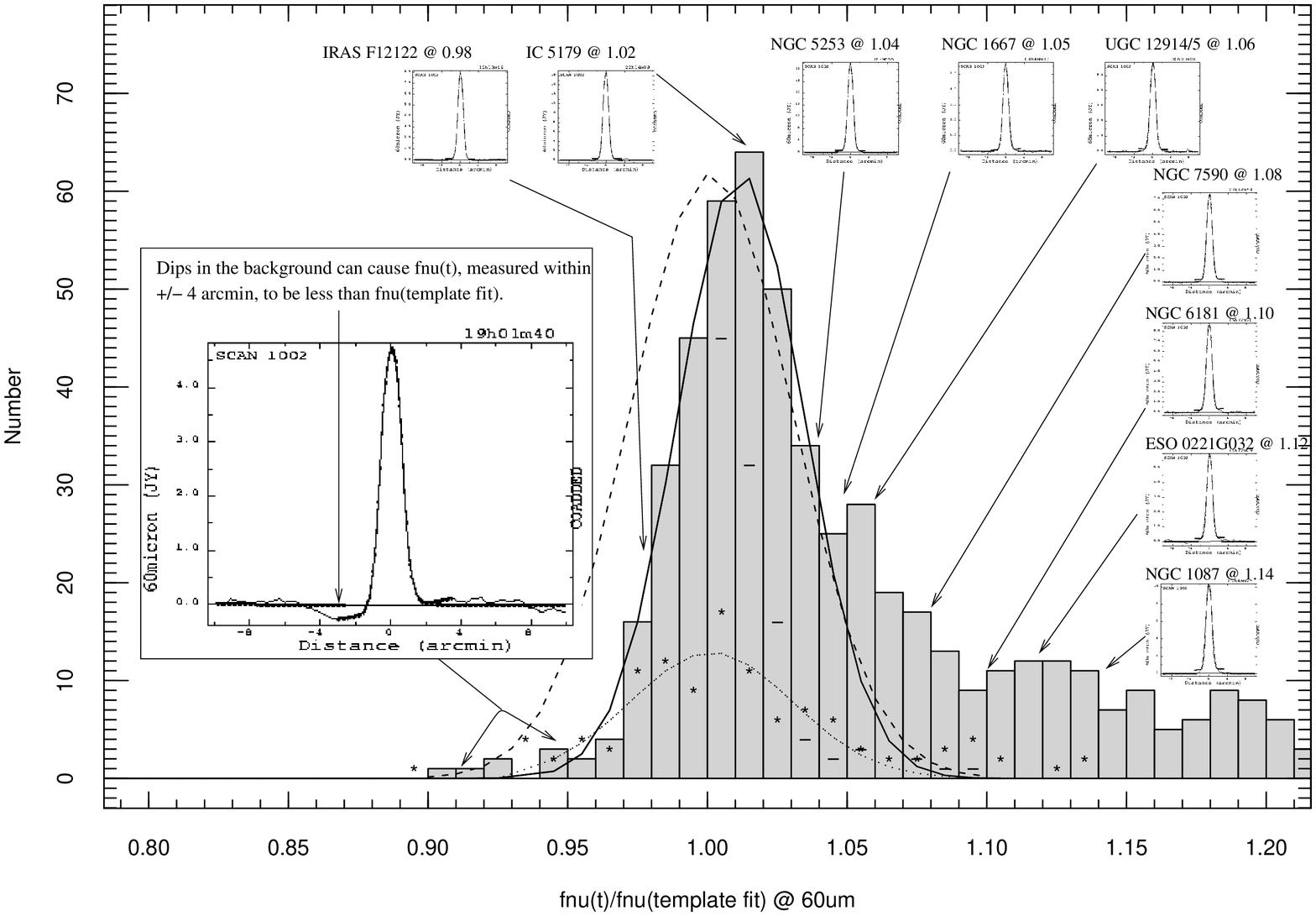}
  \caption{\figcapThirteen}
  \end{figure}
  \clearpage
\else
  \epsscale{0.85}
  \plotone{threshold6.ps}
  \caption{\figcapThirteen}
  \end{figure}
  \clearpage
\fi
\fi 


Another approach to predicting the expected distribution for unresolved
galaxies with\\ $f_{\nu}(t)/f_{\nu}(template) > 1.0$ is based on the assumption
that small differences between the two measurements (whether negative or
positive) are due solely to noise in the coadded scans and uncertainties
in the \IRAS point source template fits.  That is, if all galaxies with
$f_{\nu}(t)/f_{\nu}(template)$ between 1.00 and 1.10 were unresolved by
\IRAS at 60{\ts}\um, we would expect the observed RBGS histogram bins over this
range to match a reflection of the distribution over the range 0.90 -
1.00, where differences between $f_{\nu}(t)$ and $f_{\nu}(template)$ are
clearly not physical and therefore due solely to noise. These expected
counts are shown as horizontal line segments drawn within the bins with
$f_{\nu}(t)/f_{\nu}(template)$ values between 1.0 and 1.10.  Using this
noise symmetry argument to predict the $f_{\nu}(t)/f_{\nu}(template)$
ratios expected for truly unresolved objects over the range 1.0 - 1.1,
there is a clear excess of galaxies with ratios as small as 1.04 -- 1.05
that likely have real (but weak) extended components; roughly 50\% of the
RBGS objects in these bins are in this category. However, since we cannot
distinguish, using visual inspection of the coadded scan profiles, between
galaxies that have true extended emission and those which have $f_{\nu}(t) >
f_{\nu}(template)$ due only to noise among these objects, we cannot reliably
use a threshold ratio smaller than 1.05 to classify specific objects
as marginally extended. Although the bins with $f_{\nu}(t)/f_{\nu}(template)$
in the range 1.05 -- 1.10 have counts that suggest some of these objects may
be explained by the Gaussian fits described above for unresolved objects,
visual inspection of the SCANPI profiles shows that every object in this range
(and of course larger values) have clear, obvious extended emission as shown
in the example profiles inset in Figure 13. Finally, the reality of extended
emission for RBGS objects with ratios $f_{\nu}(t)/f_{\nu}(template) > 1.05$
is visualized in Figure 13 through the progressive increase in the height or
spatial extent of the pedestals corresponding to an increase in the flux ratio,
all clearly defined well above the background noise in the coadded scans.

The threshold ratio of $f_{\nu}(t)/f_{\nu}(template) > 1.05$ (5\% excess flux
over the point source template fit) was therefore used for defining marginally
extended objects (M). Visual examination of the coadded scan profiles widths in
conjunction with the flux ratio distribution in Figure 13 lead to selection
of a threshold of $f_{\nu}(t)/f_{\nu}(template) > 1.20$ (20\% flux excess
over the point source template fit) to flag a source as fully resolved (R),
even if there is no additional excess flux measured by the $f_{\nu}(z)$
method and therefore the $f_{\nu}(t)$ value is selected. The final algorithm
chosen for selecting the best SCANPI method for estimating the total flux
density and for assigning \IRAS source size codes in each \IRAS band is as follows:

\begin{enumerate}

\item If the condition 
$[f_{\nu}(t) >= 1.05*f_{\nu}(template)~{\rm AND}~f_{\nu}(t) >= f_{\nu}(template)+3*sigma]$
is true OR the condition $[W50 >= W50_{psf}~{\rm OR}~W25 >= W25_{psf}]$ is true,
$f_{\nu}(t)$ is selected and the source is classified as marginally extended (M).
The $sigma$ above and in other conditions that follow is the standard
deviation of the coadded data measured by SCANPI in the background noise
outside the signal range; these values are tabulated in columns (8) -- (11)
of Table 1 in units of mJy. $W50_{psf}$ and $W25_{psf}$ are the thresholds
used to establish when a galaxy profile's width at 50\% peak (FWHM) or 25\%
peak are significantly larger than observed among unresolved point sources.
The $W50_{psf}$ and $W25_{psf}$ thresholds adopted, rather conservatively,
are the computed mean plus 3 times the rms dispersion observed in each \IRAS
band for a large sample of point sources:
1.04\arcm and 1.40\arcm at 12{\ts}\um, 
1.00\arcm and 1.38\arcm at 25{\ts}\um, 
1.52\arcm and 2.06\arcm at 60{\ts}\um, and
3.22\arcm and 4.32\arcm at 100{\ts}\um. 
For reference, the nominal FWHM of
the \IRAS detectors is 0.77\arcm (12{\ts}\um), 0.78\arcm (25{\ts}\um), 1.44\arcm (60{\ts}\um),
and 2.94\arcm (100{\ts}\um). 

\item If the condition 
$[f_{\nu}(t) >= 1.20*f_{\nu}(template)~{\rm AND}~f_{\nu}(t) >= f_{\nu}(template)+3*sigma]$
is true OR the condition $[W50 >= W50_{psf}~{\rm AND}~W25 >= W25_{psf}]$ is true, 
$f_{\nu}(t)$ is selected and the object is classified as resolved (R).

\item If the condition 
$[f_{\nu}(z) >= f_{\nu}(t)+3*sigma]$ is true, AND the condition 
$[f_{\nu}(z) >= f_{\nu}(peak)]$ is true, AND the condition
$[f_{\nu}(z) >= 1.05*f_{\nu}(t)]$ is true,
$f_{\nu}(z)$ is selected and the source is classified as resolved (R).

\item If none of the above conditions are satisfied, the $f_{\nu}(template)$
method is selected and the source is classified as unresolved (U).

\item For coadded scans that display confusion from nearby sources, blends
of close pairs, excessive noise, cirrus contamination, etc., based on visual
inspection made for all of the data, selections resulting from the algorithm above
were overridden using flags specified in the input data. For example, the peak
flux value is a better estimate (flagged with method code ``P'' in Table 1)
when the $\rm f_{\nu}(t)$ value is tainted by a bad point source template
fit due to a nearby confusing source or cirrus (e.g., IRAS $02572+7002$
at 60{\ts}\um; see Fig. 15.). Another example is when the SCLEAN algorithm was
used in an attempt to de-blend components of a pair (e.g., NGC 5194/95 $=$
M 51; see Fig. 15.).

\end{enumerate}

Figures 14 -- 16 display coadded \IRAS scan profiles that illustrate the meaning
of the source size codes (S), SCANPI flux density methods (M), and uncertainty flags
(F) as listed for each source in Table 1.

\ifnum\Mode=0 
\vskip 0.3in
\fbox{\bf Figure 14 goes here.}
\vskip 0.3in
\else
\setcounter{figure}{14}
\begin{figure}[!htb]
\figurenum{14}
\ifnum\Mode=2 
  \epsscale{0.35}
  \plotone{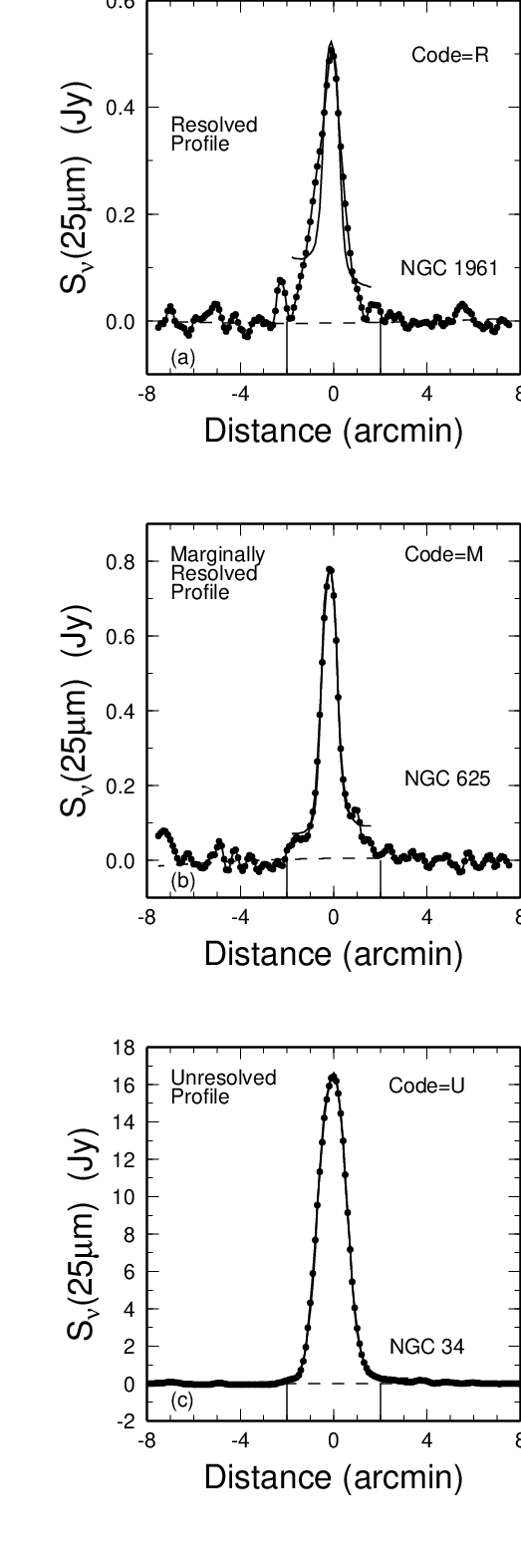}
  \caption{\figcapFourteen}
  \end{figure}
  \clearpage
\else
  \epsscale{0.35} 
  \plotone{SizeCodes.ps}
  \caption{\figcapFourteen}
  \end{figure}
  \clearpage
\fi
\fi 

\ifnum\Mode=0 
\vskip 0.3in
\fbox{\bf Figure 15 goes here.}
\vskip 0.3in
\else
\setcounter{figure}{15}
\begin{figure}[!htb]
\figurenum{15}
\ifnum\Mode=2 
  \epsscale{0.55}
  \plotone{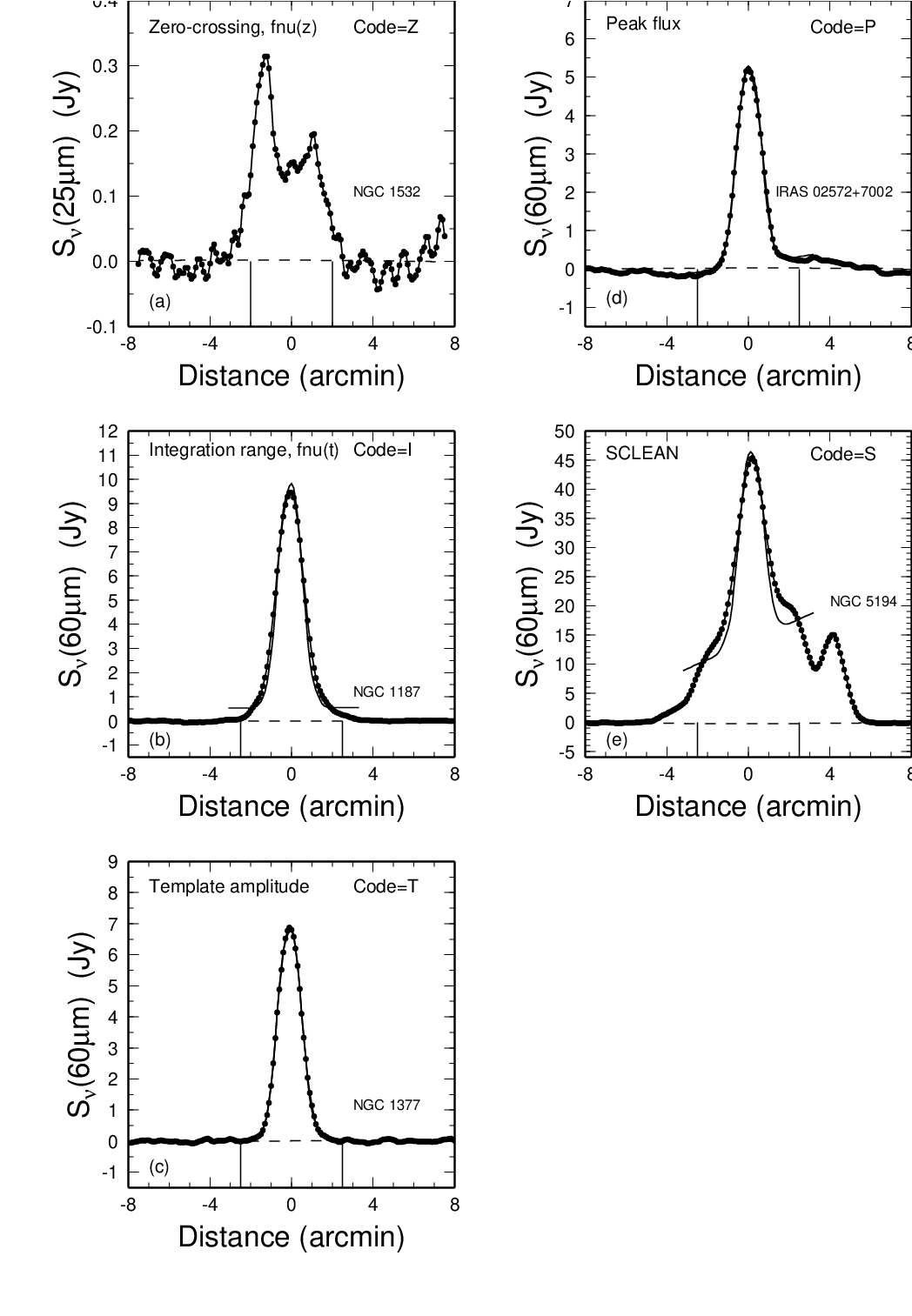}
  \caption{\figcapFifteen}
  \end{figure}
  \clearpage
\else
  \epsscale{0.55}
  \plotone{MethodCodes.ps}
  \caption{\figcapFifteen}
  \end{figure}
  \clearpage
\fi
\fi 

\ifnum\Mode=0 
\vskip 0.3in
\fbox{\bf Figure 16 goes here.}
\vskip 0.3in
\else
\setcounter{figure}{16}
\begin{figure}[!htb]
\figurenum{16}
\ifnum\Mode=2 
  \epsscale{0.65}
  \plotone{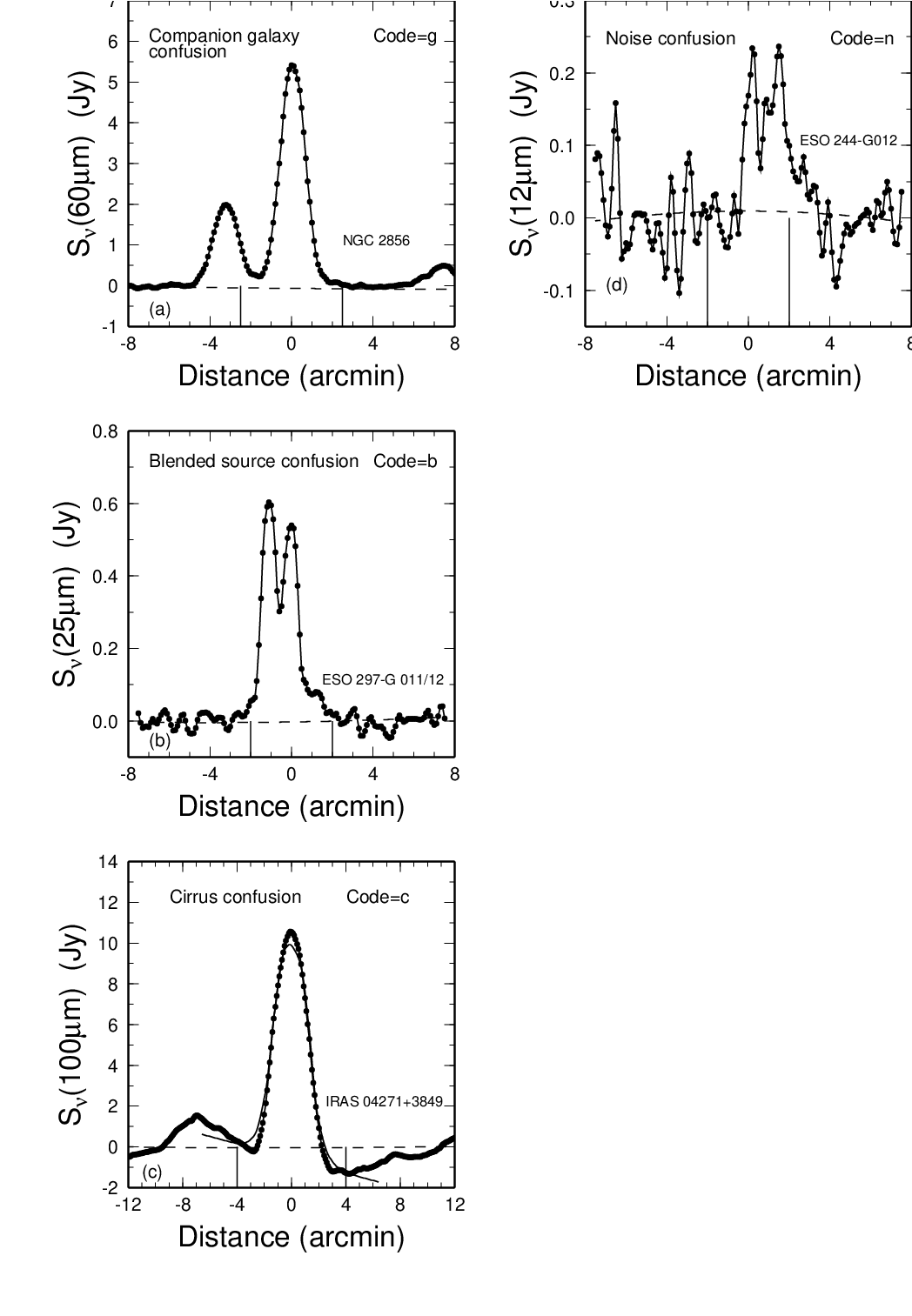}
  \caption{\figcapSixteen}
  \end{figure}
  \clearpage
\else
  \epsscale{0.65}
  \plotone{FlagCodes.ps}
  \caption{\figcapSixteen}
  \end{figure}
  \clearpage
\fi
\fi 


The flux estimates chosen by the final processing are indicated
by the ``Method'' codes following the flux densities quoted in Table 1: 
Z = ``zero crossing" (``zc" in Table 7); 
I = ``in-band total" (``tot" in Table 7); 
T = ``template fit" (``temp" in Table 7); 
P = ``peak value" (``peak" in Table 7); 
S = ``deconvolution with SCLEAN"\footnote{SCLEAN is a simple routine that
fits an \IRAS point-source template at an input position and subtracts (``cleans'')
the fit from the 1-D coadded profile. This allows the user to estimate
the flux remaining in a source which is not accounted for by point
source component(s). SCLEAN was used to estimate the flux densities for
components of pairs and a number of confused objects, as indicated in Table 1.}; 
R = from Rice et al. (1988). For objects with ``R'' listed as the Method code in
Table 1 (objects larger than $\sim$25 arcminutes) the SCANPI measurements 
in were not used; they are included in Table 7 just for reference.

Table 5 (Section 4.1) lists the distribution among the size codes (U,M,R) for
the sources at each wavelength, and reflects primarily the changing angular
resolution of the \IRAS detectors as a function of wavelength, although the
increased sensitivity of the \IRAS detectors at 60{\ts}\um\ partially compensates
for the larger size of the 60{\ts}\um\ detectors compared to the smaller angular
resolution of the detectors at 12{\ts}\um\ and 25{\ts}\um.  
The number of resolved or marginally resolved objects
(i.e. size codes ``M" or ``R" respectively in Table 1) is 61\% at 12{\ts}\um, 54\%
at 25{\ts}\um, 48\% at 60{\ts}\um, and drops to 30\% at 100{\ts}\um, as listed in Table 5.


\clearpage

\ifnum\Mode>0 \end{document} \fi 

\centerline{FIGURE CAPTIONS}
\vskip 0.3in

\figcaption[Fig1p01.ps]{
\figcapOne
}

\figcaption[NewVsOld.ps]{
\figcapTwo
}

\figcaption[TotalToPeak.ps]{
\figcapThree
}

\figcaption[logNlogS.ps]{
\figcapFour
}

\figcaption[hammer.ps]{
\figcapFive
}

\figcaption[GlatCounts.eps]{
\figcapSix
}

\figcaption[FIRColorHistograms.ps]{
\figcapSeven
}

\figcaption[S12S25vsS60S100.ps]{
\figcapEight
}

\figcaption[zhist.ps]{
\figcapNine
}

\figcaption[Dmpc.ps]{
\figcapTen
}

\figcaption[Lir.ps]{
\figcapEleven
}

\figcaption[RBGSlumFunc.ps]{
\figcapTwelve
}

\figcaption[SizeCodes.ps]{
\figcapThirteen
}

\figcaption[MethodCodes.ps]{
\figcapFourteen
}

\figcaption[FlagCodes.ps]{
\figcapFifteen
}

\figcaption[threshold6.ps]{
\figcapSixteen
}



\clearpage
\vskip 0.3in
\fbox{\bf Table 1 goes here.
It has 13 landscape pages included in a separate file.
}
\vskip 0.3in

\clearpage
\tableTwo

\clearpage
\tableThree

\clearpage
\tableFour

\clearpage
\tableFive

\clearpage
\tableSix

\clearpage
\vskip 0.3in
\fbox{\bf Table 7 goes here.
It has 11 landscape pages included in a separate file.
}
\vskip 0.3in


\begin{thebibliography}{}

\bibitem[]{}Aaronson, M., \EA\ 1982a, \apj, 258, 64.

\bibitem[Aaronson et al. (1982b)]{1982ApJS...50..241A} Aaronson, M., \EA\ 1982b, \apjs, 50,
241.

\bibitem[Aaronson \& Mould (1983)]{1983ApJ...265....1A} Aaronson, M., \& Mould, J. 1983,
\apj, 265, 1.

\bibitem[Aaronson, Mould, \& Huchra (1980)]{1980ApJ...237..655A} Aaronson, M., Mould, J., 
\& Huchra, J. 1980, \apj, 237, 655

\bibitem[Aumann, Fowler, \& Melnyk(1990)]{1990AJ.....99.1674A} Aumann,
H.~H., Fowler, J.~W., \& Melnyk, M.\ 1990, \aj, 99, 1674

\bibitem[]{c1} Cataloged Galaxies and Quasars Detected in the \IRAS Survey
1985, prepared by  C. J. Lonsdale, G. Helou, J. C. Good, \& W. L. Rice
(JPL D1932).

\bibitem[da Costa et al.(1991)]{1991ApJS...75..935D} da Costa, L.~N., 
Pellegrini, P.~S., Davis, M., Meiksin, A., Sargent, W.~L.~W., \& Tonry, 
J.~L.\ 1991, \apjs, 75, 935. 

\bibitem[Dale et al. (2001)]{2001ApJ...549..215D} Dale, D.~A., Helou, G., Contursi, A., 
Silbermann, N.~A., \& Kilhatkar, S. 2001, \apj, 549, 215.

\bibitem[de Carvalho, Ribeiro, Capelato, \& 
Zepf(1997)]{1997ApJS..110....1D} de Carvalho, R.~R., Ribeiro, A.~L.~B., 
Capelato, H.~V., \& Zepf, S.~E.\ 1997, \apjs, 110, 1. 

\bibitem[deGrijp et al. (1985)]{1985Natur.314..240D} de{\ts}Grijp, M.~H.~K., Miley, G.~K., 
Lub, J., \& de{\ts}Jong, T. 1985, Nature, 314, 240.

\bibitem[de Vaucouleurs et al.(1991)]{1991RC3.9.C...0000d}
de Vaucouleurs, G., de Vaucouleurs, A., Corwin, JR.,~H.~G., Buta, R.~J.,
Paturel, G., \& Fouque, P.\ 
1991, Third Reference Catalogue of Bright Galaxies,
Austin: University of Texas Press.

\bibitem[de Vaucouleurs, de Vaucouleurs, \& 
Corwin(1976)]{1976RC2...C......0D} de Vaucouleurs, G., de Vaucouleurs, A., 
\& Corwin, J.~R.\ 1976, Second Reference Catalogue of Bright Galaxies, 
Austin: University of Texas Press.

\bibitem[Dey, Strauss, \& Huchra(1990)]{1990AJ.....99..463D} Dey, A., 
Strauss, M.~A., \& Huchra, J.\ 1990, \aj, 99, 463. 

\bibitem[Fajardo-Acosta, Stencel \& Backman (1997)]{1997ApJ...487L.151F} 
Fajardo-Acosta, S.~B., Stencel, R.~E. \& Backman, D.~E. 1997,
\apjl, 487, 151.

\bibitem[Ferrarese et al. (2000)]{2000ApJS..128..431F} Ferrarese, L. et al. 2000, 
\apjs, 128, 431.

\bibitem[Freedman et al.(2001)]{2001ApJ...553...47F} Freedman, W.~L.~et 
al.\ 2001, \apj, 553, 47. 

\bibitem[Giovanelli \& Haynes(1993)]{1993AJ....105.1271G} Giovanelli, R.~\& 
Haynes, M.~P.\ 1993, \aj, 105, 1271. 

\bibitem[Giovanelli et al.(1997)]{1997AJ....113...22G} Giovanelli, R., 
Haynes, M.~P., Herter, T., Vogt, N.~P., Wegner, G., Salzer, J.~J., da 
Costa, L.~N., \& Freudling, W.\ 1997, \aj, 113, 22. 

\bibitem[Giovanelli, Avera, \& Karachentsev(1997)]{1997AJ....114..122G} 
Giovanelli, R., Avera, E., \& Karachentsev, I.~D.\ 1997, \aj, 114, 122. 

\bibitem[Gudehus(1976)]{1976ApJ...208..267G} Gudehus, D.~H.\ 1976, \apj, 
208, 267. 

\bibitem[Haynes et al.(1997)]{1997AJ....113.1197H} Haynes, M.~P., 
Giovanelli, R., Herter, T., Vogt, N.~P., Freudling, W., Maia, M.~A.~G., 
Salzer, J.~J., \& Wegner, G.\ 1997, \aj, 113, 1197. 

\bibitem[Haynes et al.(1998)]{1998AJ....115...62H} Haynes, M.~P., van Zee, 
L., Hogg, D.~E., Roberts, M.~S., \& Maddalena, R.~J.\ 1998, \aj, 115, 62. 

\bibitem[]{h1} Helou, G., Kahn, I. R., Malek, L., \& Boehmer, L. 1988,
\apjs, 68, 151.
  
\bibitem[Huchra, Davis, Latham, \& Tonry(1983)]{1983ApJS...52...89H} 
Huchra, J., Davis, M., Latham, D., \& Tonry, J.\ 1983, \apjs, 52, 89. 

\bibitem[Huchra et al.(1993)]{1993AJ....105.1637H} Huchra, J., Latham, 
D.~W., da Costa, L.~N., Pellegrini, P.~S., \& Willmer, C.~N.~A.\ 1993, \aj, 
105, 1637. 

\bibitem[Huchra et al.(1992)]{1992ZCAT..M...0000H}
Huchra, J.,~P.~et al.\ 1992, CfA Redshift Catalog (ZCAT).

\bibitem[Huchra, Geller, \& Corwin(1995)]{1995ApJS...99..391H} Huchra, 
J.~P., Geller, M.~J., \& Corwin, H.~G.\ 1995, \apjs, 99, 391. 

\bibitem[Huchra, Vogeley, \& Geller(1999)]{1999ApJS..121..287H} Huchra, 
J.~P., Vogeley, M.~S., \& Geller, M.~J.\ 1999, \apjs, 121, 287. 

\bibitem[]{i1} \IRAS Catalogs and Atlases -- Explanatory Supplement 1985
(NASA RP-1190).

\bibitem[]{i2} \IRAS Catalogs and Atlases: Small Scale Structure Catalog
1988, eds. G. Helou \&  D. Walker (Washington, DC: GPO) (SSS).

\bibitem[]{i3} \IRAS Catalogs and Atlases: Point Source Catalog 1988,
(Washington, DC: GPO)  (PSC Version 2).

\bibitem[]{1992ifss.book.....M} \IRAS Faint Source Survey, Explanatory Supplement version 2,
1992,  eds. M. Moshir, G. Kopman, T. Conrow (Pasadena: IPAC).


\bibitem[Keel(1996)]{1996AJ....111..696K} Keel, W.~C.\ 1996, \aj, 111, 696. 

\bibitem[Keel(1996)]{1996ApJS..106...27K} Keel, W.~C.\ 1996, \apjs, 106, 27. 

\bibitem[Kim et al.(1995)]{1995ApJS...98..129K} Kim, D.-C., Sanders, D.~B., 
Veilleux, S., Mazzarella, J.~M., \& Soifer, B.~T.\ 1995, \apjs, 98, 129. 

\bibitem[Kirshner(1977)]{1977ApJ...212..319K} Kirshner, R.~P.\ 1977, \apj, 
212, 319. 

\bibitem[]{k1} Kirshner, R. P., Oemler, A., \& Schechter, P. L. 1978,
\aj, 83, 1549.

\bibitem[Lauberts \& Valentijn(1989)]{1989spce.book.....L} Lauberts, A.~\& 
Valentijn, E.~A.\ 1989, ``The Surface Photometry Catalogue of the
ESO-Uppsala Galaxies," (Garching: European Southern Observatory). 

\bibitem[Lebofsky \&  Rieke (1979)]{} Lebofsky, M.~J. ~\& 
Rieke, G.~H.\ 1979, \apj, 229, 111.

\bibitem[Lu et al.(1990)]{1990ApJ...357..388L} Lu, N.~Y., Dow, M.~W., 
Houck, J.~R., Salpeter, E.~E., \& Lewis, B.~M.\ 1990, \apj, 357, 388. 

\bibitem[Lu et al. (1993)]{1993ApJS...88..383L} Lu, N.~Y., Hoffman, G.~L., 
Groff, T., Roos, T., \& Lamphier, C.\ 1993, \apjs, 88, 383. 

\bibitem[Mathewson, Ford, \& Buchhorn (1992)]{1992ApJS...81..413M} 
Mathewson, D.~S., Ford, V.~L., \& Buchhorn, M.\ 1992, \apjs, 81, 413. 

\bibitem[Madore \& Freedman (1998)]{MF98} Madore, B. F., \& Freedman, W. L.\ 1998,
in ``Stellar Astrophysics for the Local Group: 
VIII Canary Islands Winter School of Astrophysics."
Edited by A. Aparicio, A. Herrero, \& F. Sanchez. 
Cambridge; New York: Cambridge University Press, p.263

\bibitem[Mathewson \& Ford (1996)]{1996ApJS..107...97M} Mathewson, D.~S.~\& 
Ford, V.~L.\ 1996, \apjs, 107, 97. 

\bibitem[Miley et al. (1984)]{1984ApJ...278L..79M} Miley, G., Neugebauer, G., 
Soifer, B.~T., Clegg, P.~E., Harris, S., Rowan-Robinson, M., \& Young, E. 1984, \apj, 278, L79

\bibitem[Mirabel \& Sanders (1988)]{1988ApJ...335..104M} Mirabel, I.~F.~\& 
Sanders, D.~B.\ 1988, \apj, 335, 104. 

\bibitem[Moshir et al. (1992)]{FSC} Moshir, M., Kopan, G., Conrow, T.,
McCallon, H., Hacking, P., Gregorich, D., Rohrbach, G., Melnyk, M.,
Rice, W., Fullmer, L., White, J., \& Chester, T. 1992, 
Explanatory Supplement to the \IRAS Faint Source Survey, Version 2, JPL D-10015 8/92
(Pasadena: JPL)

\bibitem[Mould et al. (1991)]{1991ApJ...383..467M} Mould, J. R. et al. 1991, \apj, 383, 467

\bibitem[Mould et al. 2000]{2000ApJ...529..786M} Mould, J.~R. et al. 2000, \apj, 529, 786

\bibitem[Murphy et al.(2001)]{2001AJ....121...97M} Murphy, T.~W., Soifer, 
B.~T., Matthews, K., Armus, L., \& Kiger, J.~R.\ 2001, \aj, 121, 97. 

\bibitem[Neugebauer et al. 1984]{1984ApJ...278L...1N} 
Neugebauer, G., et al. 1984, \apjl, 278, L1.

\bibitem[Nordgren, Chengalur, Salpeter, \& 
Terzian (1997)]{1997AJ....114...77N} Nordgren, T.~E., Chengalur, J.~N., 
Salpeter, E.~E., \& Terzian, Y.\ 1997, \aj, 114, 77. 

\bibitem[Nordgren, Chengalur, Salpeter, \& 
Terzian (1997)]{1997AJ....114..913N} Nordgren, T.~E., Chengalur, J.~N., 
Salpeter, E.~E., \& Terzian, Y.\ 1997, \aj, 114, 913.
 
\bibitem[Palumbo, Tanzella-Nitti, \& Vettolani(1983)]{1983QB857.P34......} 
Palumbo, G.~G.~C., Tanzella-Nitti, G., \& Vettolani, G.\ 1983, 
``Catalogue of Radial Velocities of Galaxies," New York, 
Gordon and Breach Science Publishers, 592 pp. 

\bibitem[Pantoja, Altschuler, Giovanardi, \& 
Giovanelli(1997)]{1997AJ....113..905P} Pantoja, C.~A., Altschuler, D.~R., 
Giovanardi, C., \& Giovanelli, R.\ 1997, \aj, 113, 905. 

\bibitem[]{} Perault, M. 1987, Structure et Evolution des Nuages Moleculaires, 
PhD thesis, Univ. Paris

\bibitem[Ribeiro et al.(1996)]{1996ApJ...463L...5R} Ribeiro, A.~L.~B., de 
Carvalho, R.~R., Coziol, R., Capelato, H.~V., \& Zepf, S.~E.\ 1996, \apjl, 
463, L5. 

\bibitem[Rice 1993]{1993AJ....105...67R} Rice, W. 1993, AJ, 105, 67.

\bibitem[]{r1} Rice, W. L., Lonsdale, C. J., Soifer, B. T., Neugebauer,
G., Kopan, E. L., Lloyd,  L. A., deJong, T., \& Habing, H. 1988, \apjs,
68, 91.

\bibitem[Rubin, Kenney, \& Young(1997)]{1997AJ....113.1250R} Rubin, V.~C., 
Kenney, J.~D.~P., \& Young, J.~S.\ 1997, \aj, 113, 1250. 

\bibitem[Sandage, Bell, \& Tripicco(1999)]{1999ApJ...522..250S} Sandage, 
A., Bell, R.~A., \& Tripicco, M.~J.\ 1999, \apj, 522, 250. 

\bibitem[]{s1} Sandage, A., \& Tammann, G. A. 1981, A Revised
Shapley-Ames Catalog of Bright Galaxies  (Carnegie Institute, Washington,
DC) (RSA).

\bibitem[Sanders \& Mirabel 1996]{1996ARAA..34..749S} Sanders, D.~B., \& Mirabel, I.~F.
1996, ARA\& A, 34, 749.

\bibitem[Sanders et al.(1995)]{1995AJ....110.1993S} Sanders, D.~B., Egami, 
E., Lipari, S., Mirabel, I.~F., \& Soifer, B.~T.\ 1995, \aj, 110, 1995 (BGS$_2$). 

\bibitem[Sanders, Scoville, \& Soifer(1991)]{1991ApJ...370..158S} Sanders, 
D.~B., Scoville, N.~Z., \& Soifer, B.~T.\ 1991, \apj, 370, 158. 

\bibitem[Schmidt 1968]{1968ApJ...151..393S} Schmidt, M. 1968, \apj, 151, 393

\bibitem[Schneider, Thuan, Magri, \& Wadiak(1990)]{1990ApJS...72..245S} 
Schneider, S.~E., Thuan, T.~X., Magri, C., \& Wadiak, J.~E.\ 1990, \apjs, 
72, 245. 

\bibitem[Schneider, Thuan, Mangum, \& Miller(1992)]{1992ApJS...81....5S} 
Schneider, S.~E., Thuan, T.~X., Mangum, J.~G., \& Miller, J.\ 1992, \apjs, 
81, 5. 

\bibitem[Shier \& Fischer(1998)]{1998ApJ...497..163S} Shier, L.~M.~\& 
Fischer, J.\ 1998, \apj, 497, 163. 

\bibitem[Shostak(1975)]{1975ApJ...198..527S} Shostak, G.~S.\ 1975, \apj, 
198, 527. 

\bibitem[]{s4} Soifer, B. T., Boehmer, L., Neugebauer, G., \& Sanders, D.
B. 1989, \aj, 98, 766 (BGS$_1$). 

\bibitem[]{s5} Soifer, B. T., Sanders, D. B., Madore, B. F., Neugebauer,
G., Danielson, G. E.,  Elias, JU. H., Lonsdale, C. J., \& Rice, W. L.
1987, \apj, 320, 238.

\bibitem[]{s6} Soifer, B. T., Sanders, D. B., Neugebauer, G., Danielson,
G. E., Lonsdale, C. J.,  Madore, B. F., \& Persson, S. E. 1986, \apjl,
303, L41.

\bibitem[Solomon, Downes, Radford, \& Barrett(1997)]{1997ApJ...478..144S} 
Solomon, P.~M., Downes, D., Radford, S.~J.~E., \& Barrett, J.~W.\ 1997, 
\apj, 478, 144. 

\bibitem[Storchi-Bergmann et al. (1996)]{1996ApJ...472...83S} 
Storchi-Bergmann, T., Rodriguez-Ardila, A., Schmitt, H.~R., Wilson, A.~S., 
\& Baldwin, J.~A.\ 1996, \apj, 472, 83. 

\bibitem[Strauss et al. (1992)]{1992ApJS...83...29S} Strauss, M.~A., Huchra, 
J.~P., Davis, M., Yahil, A., Fisher, K.~B., \& Tonry, J.\ 1992, \apjs, 83, 
29. 

\bibitem[]{}Surace, J. A., Sanders, D. B., \& Mazzarella, J. M. 2003, \apjs, submitted.

\bibitem[Tift \& Cocke (1988)]{1988ApJS...67....1T} Tift, W.~G.~\& Cocke, 
W.~J.\ 1988, \apjs, 67, 1. 

\bibitem[Tinney, Scoville, Sanders, \& Soifer (1990)]{1990ApJ...362..473T} 
Tinney, C.~G., Scoville, N.~Z., Sanders, D.~B., \& Soifer, B.~T.\ 1990, 
\apj, 362, 473. 

\bibitem[Tully (1982)]{1982ApJ...257..389T} Tully, R.~B. 1982, \apj, 257, 389.

\bibitem[Tully \& Shaya (1984)]{1984ApJ...281...31T} Tully, R.~B., \& Shaya, E. 1984, \apj,
281, 31.

\bibitem[Walsh, Staveley-Smith, \& Oosterloo(1997)]{1997AJ....113.1591W} 
Walsh, W., Staveley-Smith, L., \& Oosterloo, T.\ 1997, \aj, 113, 1591. 

\bibitem[Young et al. (1995)]{1995ApJS...98..219Y} Young, J.~S.~et al.\ 
1995, \apjs, 98, 219. 

 \bibitem[]{z1} Zwicky, F., \EA\ (1961-68), Catalog of Galaxies and
Clusters of Galaxies, (California Institute of Technology, Pasadena).
 
\bibitem[]{z2} Zwicky, F., \& Zwicky, M. A. 1971, Catalog of Selected
Galaxies and Post-Eruptive Galaxies, (Zurich: Offsetdruck L. Speich).

\end{thebibliography}
\end{document}